\documentclass[a4paper,10pt]{article}
\pdfoutput=1 

\usepackage{jheppubmod} 

\usepackage[T1]{fontenc} 
\usepackage{lmodern} 

\usepackage{xcolor}
\usepackage{amsmath,amssymb}
\usepackage{mathrsfs,wasysym}
\usepackage{booktabs}
\usepackage{multirow}
\usepackage{slashed}
\usepackage{cancel}
\usepackage{subcaption}
\usepackage{xspace}
\usepackage{placeins}

\newcommand{\Cb}{\bar C}

\newcommand{\adani}{\href{https://github.com/niclaurenti/adani}{\texttt{adani}}\xspace}

\newcommand{\h}{h}
\newcommand{\li}{g,q,\bar q}
\newcommand{\he}{\h,\bar\h}
\newcommand{\lh}{\li,\he}

\newcommand{\as}{\alpha_s}

\newcommand{\Ord}{\mathcal{O}}

\newcommand{\Mell}{\mathcal{M}}

\newcommand{\xB}{x_{\rm Bj}}
\newcommand{\muh}{\mu_{\h}}
\newcommand{\mQ}{\frac{m^2}{Q^2}}
\newcommand{\Qm}{\frac{Q^2}{m^2}}
\newcommand{\mmu}{\frac{m^2}{\mu^2}}
\newcommand{\mmuh}{\frac{m^2}{\muh^2}}
\newcommand{\mumuh}{\frac{\mu^2}{\muh^2}}
\newcommand{\muQ}{\frac{\mu^2}{Q^2}}

\newcommand{\gammae}{\gamma_{\scriptscriptstyle E}}

\let\originalleft\left
\let\originalright\right
\renewcommand{\left}{\mathopen{}\mathclose\bgroup\originalleft}
\renewcommand{\right}{\aftergroup\egroup\originalright}

\makeatletter
\newcommand{\vast}{\bBigg@{3}}
\newcommand{\Vast}{\bBigg@{4}}
\makeatother

\def\beq{\begin{equation}}  
\def\eeq{\end{equation}}
\def\({\left(}
\def\){\right)}
\def\[{\left[}
\def\]{\right]}

\let\oldsubsection\subsection
\renewcommand\subsection[2][\subsectiontoc]{%
  \def\subsectiontoc{#2}%
  \oldsubsection[#1]{\boldmath #2}%
}

\let\oldsubsubsection\subsubsection
\renewcommand\subsubsection[2][\subsubsectiontoc]{%
  \def\subsubsectiontoc{#2}%
  \oldsubsubsection[#1]{\boldmath #2}%
}

\allowdisplaybreaks

\newcommand{\zmax}{z_{\rm max}}
\newcommand{\ord}[1]{\mathcal{O} (\as^{#1})}

\title{\boldmath Implementation of DIS at N$^3$LO for PDF determination}

\author[a]{Andrea Barontini,}
\author[b]{Marco Bonvini,}
\author[a]{Niccol\`o Laurenti}
\affiliation[a]{Tif Lab, Dipartimento di Fisica, Universit\`a di Milano and
INFN, Sezione di Milano,\\ Via Celoria 16, I-20133 Milano, Italy}
\affiliation[b]{INFN, Sezione di Roma,\\ Piazzale Aldo Moro~5, 00185 Roma, Italy}

\preprint{ }

\emailAdd{andrea.barontini@unimi.it}
\emailAdd{marco.bonvini@roma1.infn.it}
\emailAdd{niccolo.laurenti@unimi.it}

\abstract{%
  In this work we construct an accurate description of Deep-Inelastic Scattering (DIS)
  at third order in perturbative QCD that is valid at all energy scales.
  We do so by assembling massless and massive results in a variable flavour number scheme,
  performing a careful power counting of the various contributions.
  We also propose an improved approximation for the massive neutral-current DIS coefficient function
  at order $\as^3$, whose construction is validated against lower order results.
  These results are instrumental for state-of-the-art next-to-next-to-next-to-leading order fits of parton distribution functions.
}

\begin{document}

\maketitle


\section{Introduction}
\label{sec:intro}

Parton distribution functions (PDFs) play a key role in high-precision high-energy physics,
as they are a necessary ingredient for any theoretical prediction of processes induced by initial-state hadrons.
A careful treatment of heavy quark mass effects is a crucial aspect for accurate PDF determination.
Indeed, PDFs are mostly determined by deep inelastic scattering (DIS) data,
spanning a wide range of energies crossing both the charm and bottom masses.
When the scale $Q$ is close to a heavy quark mass $m$, all mass dependence of the computation must be
taken into account.
When $Q\gg m$, instead, power corrections proportional to $m^2/Q^2$ are suppressed and can be neglected,
but collinear logarithms of $Q/m$ become large and must be resummed to all orders.

The smooth transition between these two regimes is realised with what is usually called
a variable flavour number scheme (VFNS).
Indeed, resummation of collinear mass logarithms can be easily achieved by promoting
the heavy quark to a dynamical one, i.e.\ by changing the factorization scheme
from one with $n_f$ light (massless) active quark flavours to one with $n_f+1$ active quark flavours.
Here, ``active'' means that the quark flavour participates to DGLAP evolution,
as a consequence of the factorization of collinear logarithms into the PDFs,
thus generating a PDF for the heavy flavour as well.\footnote
{We are not considering the case of intrinsic heavy quark PDFs for simplicity.
Further information on this can be found for instance in Refs.~\cite{Ball:2015tna,Ball:2015dpa}.}
This resummation is most simply implemented by treating the heavy quark as massless,
but power-suppressed mass effects must be restored to obtain a reliable description
in the transition region between the two regimes $Q\lesssim m$ and $Q\gg m$.
A variety of practical implementations of this procedure exist,
such as ACOT~\cite{Aivazis:1993kh, Aivazis:1993pi},
S-ACOT~\cite{Collins:1997sr, Kramer:2000hn},
TR and TR'~\cite{Thorne:1997ga, Thorne:2006qt},
FONLL~\cite{Buza:1996wv, Cacciari:1998it, Forte:2010ta}
and BPT~\cite{Bonvini:2015pxa,Bonvini:2016fgf}.
They are all equivalent to all orders in perturbation theory~\cite{Bonvini:2015pxa,Ball:2015dpa},
but they differ in the way they combine the various ingredients at finite order.

In this work we propose an explicit implementation of a VFNS to achieve
next-to-next-to-next-to-leading order (N$^3$LO) accuracy in DIS.
We follow the approach of BPT~\cite{Bonvini:2015pxa,Bonvini:2016fgf}
which performs a careful counting of all the perturbative ingredients,
enabling us to construct a perturbative expansion that is stable and smooth across the heavy quark threshold
at any order.
The key difference of BPT with respect to other approaches is
the fact that DGLAP evolution factors are included in the perturbative counting.
This implies for instance that the heavy quark (perturbative) PDF is effectively of $\Ord(\as)$
at scales close to the heavy quark mass, in contrast with light quark and gluon PDFs
that are of non-perturbative origin and thus count as $\Ord(\as^0)$.
This has clear implications in the way the various perturbative ingredients
are combined order by order, resulting in different behaviours close to heavy quark threshold.

Achieving N$^3$LO accuracy requires also a currently unknown ingredient, namely
the $\Ord(\as^3)$ contribution to heavy quark production in DIS
with full heavy quark mass dependence.\footnote
{While completeing this work, the exact result at this order for charged-current DIS has been computed~\cite{Caola:2026kvl}.
  The exact neutral-current DIS massive result at $\Ord(\as^3)$ remains as yet unknown.}
For this, in the neutral-current case, we make use of the approximation proposed in Ref.~\cite{Kawamura:2012cr},
that we improve by using the most recent and accurate determinations of the various ingredients
and by combining them in a novel way that better describes the various kinematic regions.
We also provide predictions for the longitudinal structure function at this order for the first time.

This paper is organised as follows.
In section~\ref{sec:VFNS} we introduce our notation and describe the general ingredients and the construction of a VFNS.
In section~\ref{sec:dis-accuracy} we focus on the power counting and how to achieve N$^3$LO accuracy,
discussing our approach based on BPT and comparing with other approaches.
In section~\ref{sec:approx} we present our novel construction of the approximate $\mathcal{O}(\as^3)$ massive coefficient functions.
Finally, in section~\ref{sec:VFNS-N3LO} we show our results for DIS structure functions up to N$^3$LO.
We conclude in section~\ref{sec:conclusions}, and collect additional material in the appendices.

\section{Variable flavour number schemes}
\label{sec:VFNS}

In the context of PDF fitting, DIS data are an essential ingredient, even though 
modern PDF determination typically include thousands of datapoints covering different
processes. Unlike other data, coming for example from the LHC, DIS data span a rather large 
$Q^2$ range, roughly from $\sim2\, \text{GeV}^2$ to $\sim20000\, \text{GeV}^2$,
which crosses both the charm mass ($m_c \sim 1.5\, \text{GeV}$)
and the bottom mass ($m_b \sim 4.9\, \text{GeV}$) scales.
For this reason, DIS predictions
in a single flavour scheme for all the datapoints is not accurate enough and a variable flavour number scheme (VFNS) is needed.

We now review general features of a VFNS, postponing to the next section
the details of an explicit construction up to N$^3$LO.
Let us consider photon-mediated DIS, focusing on the subprocess in which the off-shell photon scatters off the proton,
\beq
p(P) + \gamma^*(q) \to X.
\eeq
Here $X$ represents any final-state radiation (including the remnant of the proton),
and the process is inclusive over $X$.

Note that we are not considering the contributions to the DIS process where the mediators are
weak bosons ($Z$ or $W$).
Indeed, as they are much heavier than the charm and bottom quarks, the power-suppressed mass effects of these quarks,
proportional to powers of $m_c^2/Q^2$ and $m_b^2/Q^2$, are negligible at scales $Q$ where the contributions of weak bosons
to DIS are relevant.\footnote
{This is certainly the case for neutral-current DIS, as at low scales $Q$ where mass effect would be relevant
  the $Z$ contribution (and the $\gamma$-$Z$ interference) are totally negligible in comparison with the dominant photon contribution.
  In charged-current DIS, instead, the $W$ contribution is the only one, and so at low scales it is not suppressed
  in comparison with anything else. It is still true that the DIS cross section is suppressed at these scales in comparison with higher scales,
  but if an accurate description is required the mass effect should be taken into account.
  Even though we focus on neutral-current DIS in this work, the construction of the VFNS presented here
  is applicable also in the charged-current case.}
Therefore, for weak mediators (and for $\gamma$-$Z$ interference) we can simply treat
both the charm and bottom quarks as massless,\footnote
{This would not be true for top quarks, whose mass is larger than those of $Z$ and $W$.
  However, top contributions at the energies of DIS
  experiments currently available are not significant, and the resummation of mass
  logarithms due to top quarks is certainly not needed at these scales.}
thus avoiding the complications of the matching procedure that we are going to discuss for photon mediated DIS.

In the photon-mediated case, the DIS cross section can be written in terms of two structure functions
\beq
\frac{d\sigma}{d\xB\, dQ^2}=\frac{2\pi\alpha_{\rm em}^2}{\xB Q^4}\Big[\(1+(1-y)^2\)F_2(\xB, Q^2)-y^2F_L(\xB, Q^2)\Big],
\eeq
where $\alpha_{\rm em}$ is the electromagnetic coupling and we have defined
\beq
Q^2=-q^2,\qquad
\xB=\frac{Q^2}{2 P \cdot q}, \qquad
y = \frac{P \cdot q}{P \cdot k},
\eeq
being $k$ the momentum of the incoming lepton probe.
We will focus on the two structure functions $F_2$ and $F_L$ from now on.

For definiteness, we now consider a single heavy quark $h$ of mass $m$, and
we assume that there are $n_f$ light flavours that are treated as massless.
Each structure function can thus be written as
\beq\label{eq:SF}
F_a(\xB,Q^2) = \xB \sum_{i=\li}
\int_{\xB}^1 \frac{dx}x\,C_{a,i}^{[n_f]}\(\frac{\xB}{x},\mQ,\muQ\) f_i^{[n_f]}(x,\mu^2),
\qquad a=2,L,
\eeq
in terms of partonic coefficient functions $C_{a, i}$ and PDFs $f_i$.
The integration variable $x$ represents the proton momentum fraction carried by the parton,
namely $p=xP$ being $p$ the parton momentum,
and the sum is extended over all light partons,
namely the gluon $g$ and the light quarks denoted schematically as $q,\bar q$.
We are emphasising here the dependence on the various energy scales:
the hard scale $Q$, the mass scale $m$, and the factorization scale $\mu$.
Note that the first argument of the coefficient function is
\beq
z\equiv \frac{\xB}{x}=\frac{Q^2}{2p\cdot q},
\eeq
which is the parton-level analog of $\xB$.
The label $^{[n_f]}$ indicates that only the light flavours are considered active,
namely their collinear divergences are factorized into the PDFs and they thus participate to DGLAP evolution.
The heavy quark instead can only be produced perturbatively via gluon splittings in the coefficient functions.

In principle we could also consider the case of intrinsic heavy quark PDF.
In this case the sum would extend over the heavy quark as well,
and there would exist a (static) PDF for the intrinsic heavy quark.
This possibility has been studied extensively,
see e.g.~Refs.~\cite{Brodsky:1980pb,Aivazis:1993kh,Aivazis:1993pi,Vogt:1994zf,Collins:1998rz,Pumplin:2005yf,Ball:2015tna,Ball:2015dpa}.
As this complicates the treatment without changing the big picture,
we assume that there is no intrinsic heavy quark PDF in our work.

Let us go back to Eq.~\eqref{eq:SF}.
Since we are including the heavy quark contribution in the coefficient function perturbatively,
mass logarithms are present order by order in perturbation theory.
Therefore, in the high-scale limit $Q\gg m$, these logarithms
become large and invalidate the fixed-order perturbative expansion,
calling for their resummation.
If we had assumed the heavy quark to be massless, which is a reasonable approximation in the $Q\gg m$ limit,
there would be no logarithms in the final result,
but collinear singularities would have appeared in the computation requiring their factorization,
and making the heavy quark an active quark.
Using a schematic representation of the Mellin convolution (and consistently omitting the first argument from the functions), we can wite
\begin{align}\label{eq:main}
  \frac1{\xB}F_a(\xB,Q^2)
  &= \sum_i^{n_f} C_{a,i}^{[n_f]}\(\mQ,\muQ\) \otimes f_{i}^{[n_f]}(\mu^2)\\
  &= \sum_k^{n_f+1} C_{a,k}^{[n_f+1]}\(0,\muQ\) \otimes f_{k}^{[n_f+1]}(\mu^2) + \Ord\(\mQ\) \nonumber\\
  &= \sum_i^{n_f} \sum_k^{n_f+1} C_{a,k}^{[n_f+1]}\(0,\muQ\) \otimes K_{ki}^{[n_f+1]\leftarrow [n_f]}\(\mmu\) \otimes f_{i}^{[n_f]}(\mu^2) + \Ord\(\mQ\), \nonumber
\end{align}
where we have introduced a symbolic expression for the sums
\begin{align}
  \sum_i^{n_f} &\equiv  \sum_{i=\li}\;,
  &
  \sum_k^{n_f+1} &\equiv  \sum_{k=\lh}
\end{align}
to streamline the notation.
The first line of Eq.~\eqref{eq:main} reports the $n_f$-scheme expression Eq.~\eqref{eq:SF},
and in the second line we have introduced the coefficient functions and PDFs in the $n_f+1$ scheme,
namely the scheme where the heavy quark $h$ is active and thus summed over,
emphasising that the coefficient function
is computed assuming a massless heavy quark (first argument equal to
zero).
These two lines represent two different but legitimate ways of computing the structure functions,
and are thus equal (to all orders), up to the explicitly missing power
corrections in the $n_f+1$ scheme.
In the last line of Eq.~\eqref{eq:main} we have introduced
the scheme-change matching function $K_{ki}^{[n_f+1]\leftarrow [n_f]}\(\mmu\)$ through the equation
\beq\label{eq:fsc}
f_{k}^{[n_f+1]}(\mu^2) = \sum_i^{n_f} K_{ki}^{[n_f+1]\leftarrow [n_f]}\(\mmu\) \otimes f_{i}^{[n_f]}(\mu^2).
\eeq
Note that, while we do not write it explicitly, the PDFs in the $n_f+1$ scheme do depend on the heavy quark mass.
In fact, it is exactly this dependence that implements the resummation of the collinear mass logarithms.
Indeed, the matching functions contain the same mass logarithms of the $n_f$-scheme coefficient functions,
appearing in the form of $\log\mmu$. They can be resummed by introducing suitable DGLAP evolution factors $U$,
such that the scheme change happens at a scale $\muh\sim m$, where the logs are small and thus harmless.
Namely, we better write
\beq\label{eq:fres}
f_{k}^{[n_f+1]}(\mu^2) = \sum_s^{n_f+1}\sum_j^{n_f}\sum_i^{n_f}
U_{ks}^{[n_f+1]}(\mu^2,\muh^2) \otimes
K_{sj}^{[n_f+1]\leftarrow [n_f]}\(\mmuh\) \otimes
U_{ji}^{[n_f]}(\muh^2,\mu_0^2) \otimes
f_{i}^{[n_f]}(\mu_0^2),
\eeq
where the (non-perturbative) $n_f$-scheme PDFs are evolved from $\mu_0$
up to the matching scale $\muh$, then the scheme change happens,
and the $(n_f+1)$-scheme PDFs are finally evolved from $\muh$ to the factorization scale $\mu$.
The latter step resums the logarithms of $\mu/\muh$ through DGLAP evolution,
corresponding exactly to the sought resummation of the collinear logarithms of $Q/m$
if $\muh\sim m$ and $\mu\sim Q$.
Note that to all orders in perturbation theory Eq.~\eqref{eq:fsc} and Eq.~\eqref{eq:fres} are identical;
equivalently, expanding the evolution factors in Eq.~\eqref{eq:fres} in powers of $\as$ one recovers
Eq.~\eqref{eq:fsc} order by order.

We have now shown how the second line of Eq.~\eqref{eq:main} resums collinear logarithms.
However, because of the missing power corrections, this result is only valid in the $Q\gg m$ limit.
To construct a full VFNS that smoothly describes all $Q^2$ regions we need to restore the
power corrections in the $n_f+1$ scheme.
In other words, we seek the functions $\Delta C_{a,k}^{[n_f+1]}\(\mQ,\muQ\)$ such that the structure functions can
be written as
\begin{align}\label{eq:matched}
  \frac1{\xB}F_a(\xB,Q^2)
  &= \sum_k^{n_f+1} \[C_{a,k}^{[n_f+1]}\(0,\muQ\) + \Delta C_{a,k}^{[n_f+1]}\(\mQ,\muQ\) \] \otimes f_{k}^{[n_f+1]}(\mu^2).
\end{align}
Comparing this equation with the first and last lines of Eq.~\eqref{eq:main} we immediately find
\begin{align}\label{eq:DeltaCdef}
\sum_k^{n_f+1} &\Delta C_{a,k}^{[n_f+1]}\(\mQ,\muQ\) \otimes f_{k}^{[n_f+1]}(\mu^2) \\
&=
\sum_i^{n_f} \[C_{a,i}^{[n_f]}\(\mQ,\muQ\) - \sum_k^{n_f+1} C_{a,k}^{[n_f+1]}\(0,\muQ\) \otimes K_{ki}^{[n_f+1]\leftarrow [n_f]}\(\mmu\) \]
\otimes f_{i}^{[n_f]}(\mu^2).\nonumber
\end{align}
To explicitly write $\Delta C_{a,k}^{[n_f+1]}$
we need to express $f_{i}^{[n_f]}(\mu^2)$ in terms of $f_{k}^{[n_f+1]}(\mu^2)$.
However, the inverse of Eq.~\eqref{eq:fsc} is not unique, because the scheme-change matrix
$K_{ki}^{[n_f+1]\leftarrow [n_f]}$ is rectangular.
This opens up various possibilities, leading to different (all correct) incarnations of a VFNS.
We will consider the simplest one,
in which the inverse is computed on the square submatrix of light entries~\cite{Ball:2015dpa},
corresponding to choosing $\Delta C_{a,k}^{[n_f+1]}=0$ when $k=\he$ is the heavy quark.
In this way, contributions to the structure functions initiated by the heavy quarks are always computed in the massless limit,
avoiding the enormous complication of dealing with initial-state massive quarks~\cite{Collins:1998rz}.
This approach corresponds to the widely used
S-ACOT~\cite{Collins:1997sr, Kramer:2000hn},
TR and TR'~\cite{Thorne:1997ga, Thorne:2006qt},
FONLL~\cite{Buza:1996wv, Cacciari:1998it, Forte:2010ta}
and BPT~\cite{Bonvini:2015pxa,Bonvini:2016fgf}
schemes.

According to this choice, the result in the $n_f+1$ scheme Eq.~\eqref{eq:matched} can be written in a more compact form as
\begin{align}\label{eq:matched2}
  \frac1{\xB}F_a(\xB,Q^2)
  &= \sum_k^{n_f+1} \Cb_{a,k}^{[n_f+1]}\(\mQ,\muQ\) \otimes f_{k}^{[n_f+1]}(\mu^2),
\end{align}
where we have introduced mass-dependent coefficient functions in the $n_f+1$ scheme defined by\footnote
{Note that in principle there is no need to give a different name to the coefficient functions $\Cb$ and $C$,
  as they represent the same object, the latter being the massless limit of the former.
  However, we think that for clarity it is useful to emphasize the difference, also for consistency with traditional notations.}
\begin{subequations}\label{eq:Cbardef}
\begin{align}
  \Cb_{a,k}^{[n_f+1]}\(\mQ,\muQ\) &\equiv C_{a,k}^{[n_f+1]}\(0,\muQ\) + \Delta C_{a,k}^{[n_f+1]}\(\mQ,\muQ\), & k&=g,q,\bar q, \\
  \Cb_{a,k}^{[n_f+1]}\(\mQ,\muQ\) &\equiv C_{a,k}^{[n_f+1]}\(0,\muQ\), & k&=h,\bar h,
\end{align}
\end{subequations}
which obviously tend to the massless coefficients in the limit $Q^2\gg m^2$,
and differ from them only by power suppressed terms (in the light channels only).
Explicit expressions are given in appendix~\ref{app:DeltaC}.

Before concluding the section,
we note that from the comparison of the first and last lines of Eq.~\eqref{eq:main}
one can recognise the collinear factorization of the heavy quark mass logarithms,
\beq\label{eq:Csc}
C_{a,i}^{[n_f]}\(\mQ,\muQ\) = \sum_k^{n_f+1} C_{a,k}^{[n_f+1]}\(0,\muQ\) \otimes K_{ki}^{[n_f+1]\leftarrow[n_f]}\(\mmu\) + \Ord\(\mQ\),
\eeq
which in turn can be used to compute the universal (process-independent) matching functions $K_{ki}^{[n_f+1]\leftarrow [n_f]}\(\mmu\)$.
This equation will be important in section~\ref{sec:approx} to construct the approximation
for the $\Ord(\as^3)$ massive coefficient functions.

\section{DIS at N$^3$LO accuracy}
\label{sec:dis-accuracy}

In this section we focus on the explicit construction of the VFNS for DIS
at each perturbative order up to N$^3$LO, i.e.\ $\Ord(\as^3)$, clarifying what are the various ingredients needed
and how to combine them.

First of all, we separate the structure functions into a light and heavy contributions,
as customary in DIS:
\begin{align}
  F_a(\xB,Q^2) = F_a^{\rm light}(\xB,Q^2) + F_a^{\rm heavy}(\xB,Q^2).
\end{align}
The latter is defined as the contributions coming from the subprocesses
in which the vector boson (the photon in our case) couples only to the heavy quark,
while the former includes everything else.
This definition of the heavy structure function does not match exactly
the experimental definition of heavy quark production in DIS,
where at least one heavy quark is required in the final state.
Indeed, such definition is not infrared safe unless kinematic restrictions
on the momentum of the final-state heavy quark are imposed
(as indeed required to identify the process experimentally),
losing the inclusive nature of the observable.
In fact, the ``theory'' definition of $F_a^{\rm heavy}(\xB,Q^2)$ contains loop induced contributions
which do not lead to heavy quarks in the final state,
and similarly $F_a^{\rm light}(\xB,Q^2)$ can have heavy quarks in the final state
coming from the splitting of gluons attached to a light quark line.
However, such contributions are typically very small.

In what follows we focus on the heavy structure functions only.
Indeed, the mass effects to the light structure functions are very small,
as they start at order $\as^2$ (NNLO) and are thus genuinely suppressed by two powers of $\as$
with respect to the leading terms. Treating the heavy quark contributions
to the light structure function in the massless limit is thus a reasonable approximation.
Note that in practice these contributions are known exactly (with full mass dependence) at NNLO,
so the massless approximation shall be used at N$^3$LO only,
thus leading to a very mild deviation from the unknown exact N$^3$LO result.
For an explicit implementation of a VFNS at NNLO for the light structure functions see e.g.~\cite{Forte:2010ta}.

In the heavy structure functions the situation is different.
Indeed, the leading contribution (in the $n_f$ scheme) is given by a gluon splitting to a heavy quark pair,
one of which interacts with the virtual photon, and it is thus an $\Ord(\as)$ term.\footnote
{For this reason, the $\Ord(\as)$ contribution to the heavy structure functions is often called LO.
  Here, instead, we call it NLO for a more direct comparison with the light structure functions, that start at $\Ord(\as^0)$.
  As we shall see in section~\ref{sec:PCres}, an $\Ord(\as^0)$ contribution to the heavy structure functions
  is effectively generated by the resummation of mass logarithms at high scales,
  making our notation more suitable for identifying each term properly.}
Moreover, mass effects are essential for the proper description of this process,
and mass logarithms appear already in this lowest order diagram.

\subsection{Power counting at fixed-order level}

In the $n_f$ scheme, the heavy structure function can thus be written as an expansion
in powers of $\as$ as\footnote
{In this expansion we can conveniently assume $\as\equiv\as(\mu^2)$,
  i.e.\ the renormalization scale is chosen equal to the factorization scale.
  Of course, the renormalization scale at which $\as$ is computed can be changed to any other value
  provided the corresponding logarithms are correctly retained in the coefficient functions.
  Similarly, it is convenient to imagine that in this expression $\as$ is computed in the $n_f$ scheme;
  other choices are possible provided the coefficient functions are consistently modified.}
\begin{align}\label{eq:FOexp}
  \frac1{\xB} F_a^{\rm heavy}(\xB,Q^2)
  &= \as C_{a,g}^{[n_f](1)}\(\mQ,\muQ\) \otimes f_{g}^{[n_f]}(\mu^2) &&\text{(NLO)}\nonumber\\
  &+ \as^2 \sum_{i}^{n_f} C_{a,i}^{[n_f](2)}\(\mQ,\muQ\) \otimes f_{i}^{[n_f]}(\mu^2) &&\text{(NNLO)}\nonumber\\
  &+ \as^3 \sum_{i}^{n_f} C_{a,i}^{[n_f](3)}\(\mQ,\muQ\) \otimes f_{i}^{[n_f]}(\mu^2) &&\text{(N$^3$LO)}\nonumber\\
  &+\Ord(\as^4),
\end{align}
where $C_{a,i}^{[n_f](n)}$ represents the $\Ord(\as^n)$ contribution to $C_{a,i}^{[n_f]}$
(for ease of notation, we do not include the label ``heavy'' on the coefficient functions,
as from now on we discuss only the heavy structure functions).
The NLO coefficients are known for a long time~\cite{Witten:1975bh,Babcock:1977zv,Shifman:1977yb,Leveille:1978px,Gluck:1979aw,Gluck:1987uk},
and the NNLO coefficients have been computed in Ref.~\cite{Laenen:1992zk,Laenen:1992cc,Riemersma:1994hv,Harris:1995tu,Hekhorn:2018ywm,Klann:2026svr}. 
The exact N$^3$LO coefficients are as yet unknown, and they will be the subject of section~\ref{sec:approx}.

The $n_f$-scheme PDFs are supposed to be all of order 1:
this is the case for the massless quark PDFs at the ``initial'' (fit) scale.\footnote
{If we are considering $n_f=4$, i.e.\ the heavy quark is the bottom, one may question whether
  this assumption is valid for the charm PDF.
  To properly treat the region where mass effects of both charm and bottom are relevant,
  one has to deal with two different non-zero masses at a time~\cite
  {Ablinger:2017err,Ablinger:2017xml,Ablinger:2018brx,Blumlein:2018jfm,Ablinger:2019gpu,Ablinger:2020snj,Bierenbaum:2022biv,Ablinger:2025nnq}.
  The extension of a VFNS from the one-mass to the two-mass case can be achieved as in Ref.~\cite{Barontini:2024xgu}.
  }
However, in Eq.~\eqref{eq:FOexp} the PDFs are computed at the factorization scale,
which is to be close to the hard scale, and so they are given by
DGLAP evolution from the initial scale $\mu_0$ to the hard scale:
\beq
f_{i}^{[n_f]}(\mu^2) = \sum_j^{n_f}
U_{ij}^{[n_f]}(\mu^2,\mu_0^2) \otimes
f_{j}^{[n_f]}(\mu_0^2).
\eeq
DGLAP evolution factors, despite their all-order nature, can also be expanded
in powers of $\as$ as
\begin{align}\label{eq:Uexp}
  U_{ij}^{[n_f]}(\mu^2,\mu_0^2)
  =
  U_{ij}^{[n_f](0)}(\mu^2,\mu_0^2)
  +\as U_{ij}^{[n_f](1)}(\mu^2,\mu_0^2)
  +\as^2 U_{ij}^{[n_f](2)}(\mu^2,\mu_0^2)
  +\as^3 U_{ij}^{[n_f](3)}(\mu^2,\mu_0^2)
  +\ldots,
\end{align}
where $U_{ij}^{[n_f](0)}(\mu^2,\mu_0^2)$ represents the leading-log (LL) solution
to DGLAP equation (namely the one determined by the LO splitting functions)
resumming to all orders powers of $\as\log\frac{\mu^2}{\mu_0^2}$,
then $U_{ij}^{[n_f](1)}(\mu^2,\mu_0^2)$ is the correction due to next-to-leading-log (NLL) terms
(those induced by NLO splitting functions), which is again a function resumming to all orders
powers of $\as\log\frac{\mu^2}{\mu_0^2}$ but with an overall power of $\as$ in front,
and so on.
If $\mu^2\gg\mu_0^2$, then each of the $U_{ij}^{[n_f](n)}$ is of order 1,
and the explicit powers of $\as$ in Eq.~\eqref{eq:Uexp}
determine the size of each contribution.
Taking the expansion Eq.~\eqref{eq:Uexp} into account,
Eq.~\eqref{eq:FOexp} becomes
\begin{align}\label{eq:FOexp2}
  \frac1{\xB} F_a^{\rm heavy}(\xB,Q^2)
  &= \as \sum_{j}^{n_f} C_{a,g}^{[n_f](1)}\(\mQ,\muQ\) \otimes U_{gj}^{[n_f](0)}(\mu^2,\mu_0^2) \otimes f_{j}^{[n_f]}(\mu_0^2)\nonumber\\
  &+ \as^2 \Bigg[\sum_{i,j}^{n_f} C_{a,i}^{[n_f](2)}\(\mQ,\muQ\)  \otimes U_{gj}^{[n_f](0)}(\mu^2,\mu_0^2)\nonumber\\
  &\qquad  +\sum_{j}^{n_f} C_{a,g}^{[n_f](1)}\(\mQ,\muQ\) \otimes U_{gj}^{[n_f](1)}(\mu^2,\mu_0^2)\Bigg]
    \otimes f_{j}^{[n_f]}(\mu_0^2)\nonumber\\
  &+\Ord(\as^3).
\end{align}
While Eq.~\eqref{eq:FOexp2} is formally more correct than Eq.~\eqref{eq:FOexp},
in which the latter includes ``spurious'' higher order interference terms,
in practice it is more difficult to construct a perturbative expansion
as in Eq.~\eqref{eq:FOexp2} (one would need several PDF sets evolved at different orders)
and there would be no much gain in the result, especially at high orders where these
spurious effects become small.

Taking into account the perturbative expansion of the DGLAP evolution factors
is however important if it modifies the starting point of the expansion.
In the case of the fixed-order ($n_f$-scheme) result this does not happen,
because at any scale $\mu$ the evolution is such that each PDF $f_{i}^{[n_f]}(\mu^2)$
is of order 1.
The situation changes in the resummed case.

\subsection{Power counting at resummed level}
\label{sec:PCres}

At resummed level, the $(n_f+1)$-scheme heavy structure function
expanded in powers of $\as$ takes the form\footnote
{Here it is reasonable to consider $\as$ renormalized in the $n_f+1$ scheme.
  As the scheme change in $\as$ is not central for our discussion, we do not change notation here.
  However, we must keep in mind that, when comparing expressions computed with different renormalization schemes,
  the scheme change in $\as$ must be taken into account (see appendix~\ref{app:DeltaC} for further details).
}
\begin{align}\label{eq:RESexp}
  \frac1{\xB} F_a^{\rm heavy}(\xB,Q^2)
  &= \sum_{k=\he} C_{a,k}^{[n_f+1](0)}\(0,\muQ\) \otimes f_{k}^{[n_f+1]}(\mu^2) &&\text{(LO)}\nonumber\\
  &+ \as \sum_{k=g,\he} C_{a,k}^{[n_f+1](1)}\(0,\muQ\) \otimes f_{k}^{[n_f+1]}(\mu^2) &&\text{(NLO)}\nonumber\\
  &+ \as^2 \sum_{k}^{n_f+1} C_{a,k}^{[n_f+1](2)}\(0,\muQ\) \otimes f_{k}^{[n_f+1]}(\mu^2) &&\text{(NNLO)}\nonumber\\
  &+ \as^3 \sum_{k}^{n_f+1} C_{a,k}^{[n_f+1](3)}\(0,\muQ\) \otimes f_{k}^{[n_f+1]}(\mu^2) &&\text{(N$^3$LO)}\nonumber\\
  &+\Ord(\as^4) + \Ord\(\mQ\),
\end{align}
where $C_{a,k}^{[n_f+1](n)}$ represents the $\Ord(\as^n)$ contribution to $C_{a,k}^{[n_f+1]}$
(again, we refer to the ``heavy'' coefficient functions without adding an explicit label).
Note that this expansion also holds after restoring power-suppressed terms as described in section~\ref{sec:VFNS},
by simply replacing $C_{a,k}^{[n_f+1](n)}$ with $\Cb_{a,k}^{[n_f+1](n)} =C_{a,k}^{[n_f+1](n)}+\Delta C_{a,k}^{[n_f+1](n)}$,
Eq.~\eqref{eq:matched2}.
The massless coefficient functions are known up to N$^3$LO~\cite{Moch:2004xu,Vermaseren:2005qc},
and the mass dependent corrections $\Delta C_{a,k}^{[n_f+1]}$ can be constructed
according to the procedure of section~\ref{sec:VFNS}.
Explicit expressions are given in appendix~\ref{app:DeltaC}.

Differently from the fixed-order case, there is here a contribution
at $\Ord(\as^0)$ multiplying the heavy quark PDFs.
Having a LO contribution seems in contradiction with the fact that in the
$n_f$ scheme the process starts at $\Ord(\as)$:
how can resummation of higher order logarithmic contributions lower the overall order of the result?
The answer is that, in the $Q^2\gg m^2$ limit, also the $n_f$-scheme computation
becomes of order 1, because of the logarithmic term $\as\log\mQ\sim1$ in this limit.
However, for scales $Q^2\sim m^2$ this is not the case, and so even the $(n_f+1)$-scheme
result must start at $\Ord(\as)$.

Indeed, the first line of Eq.~\eqref{eq:RESexp} is truly a LO contribution
only if the heavy quark PDFs count as order 1,
i.e.\ they are of the same order as the other non-perturbative PDFs.
This is definitely not the case for scales close to the heavy quark mass,
where the heavy quark PDF is effectively of $\Ord(\as)$.
To see this, we expand the matching functions in the definition of the heavy quark
PDFs in the $(n_f+1)$-scheme, Eq.~\eqref{eq:fres}, to get~\cite{Bonvini:2015pxa,Bonvini:2016fgf}
\begin{align} \label{eq:fhres_exp}
f^{[n_f+1]}_{\h}(\mu^2)
&=
\Bigg[U_{\h g}^{[n_f+1]}(\mu^2,\muh^2)
+ \as U_{\h\h}^{[n_f+1]}(\mu^2,\muh^2)\otimes K_{\h g}^{[n_f+1]\leftarrow [n_f] (1)}\(\mmuh\) \Bigg] f_g^{[n_f]}(\muh^2)
+\, \ldots
\nonumber\\*
&\sim \:\:\:\, \Ord\(\as\log\mumuh\) + \qquad\qquad \qquad\qquad \Ord(\as) \;\:\qquad\qquad\qquad\qquad\qquad\qquad + \Ord(\as^2),
\end{align}
and analogously for the heavy antiquark PDF $f^{[n_f+1]}_{\bar\h}$.
The off-diagonal matching $K_{\h g}^{[n_f+1]\leftarrow [n_f]}$ starts at $\Ord(\as)$
(in fact, it is further suppressed by a $\log\mmuh$ which is small for reasonable choices of $\muh$),
and multiplies a diagonal evolution factor which is of $\Ord(\as^0)$.
The off-diagonal evolution factor $U_{\h g}^{[n_f+1]}(\mu^2,\muh^2)$
also starts at $\Ord(\as)$ (because it contains at least one splitting)
but multiplies (and resums) a $\log\mumuh$.
This log is small (or $\sim1$) for scales close to the heavy quark mass,
but it will eventually become large (such that $\as\log\mumuh\sim1$) for very high scales $\mu\gg m$.
Therefore, it is this off-diagonal evolution term that determines the order of the heavy quark PDF:
$\Ord(\as)$ for small scales, $\Ord(1)$ for large scales.

\begin{figure}
  \includegraphics[width=0.33\textwidth,page=1]{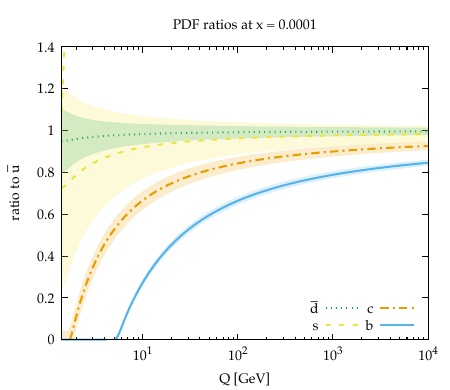}
  \includegraphics[width=0.33\textwidth,page=2]{images/PDF_ratios__PDF4LHC21_mc}
  \includegraphics[width=0.33\textwidth,page=3]{images/PDF_ratios__PDF4LHC21_mc}\\
  \includegraphics[width=0.33\textwidth,page=4]{images/PDF_ratios__PDF4LHC21_mc}
  \includegraphics[width=0.33\textwidth,page=5]{images/PDF_ratios__PDF4LHC21_mc}
  \includegraphics[width=0.33\textwidth,page=6]{images/PDF_ratios__PDF4LHC21_mc}
  \caption{Upper plots: the ratio of the bottom, charm strange and anti-down PDFs over the anti-up PDF,
    as a function of the scale $Q$ and for three values of $x=10^{-4},10^{-2},10^{-1}$,
    using the PDF4LHC21 set and showing the PDF uncertainty as well.
    Lower plots: same as the upper plots, but showing only the charm and bottom PDFs, as a function of $Q/m$,
    where $m$ is the mass of the respective quark.}
  \label{fig:PDFrat}
\end{figure}

To understand how large the scale must be with respect to the heavy quark mass
to truly count the heavy quark PDFs as order 1, we plot in figure~\ref{fig:PDFrat}
the ratio of the bottom, charm, strange and anti-down PDFs over the anti-up PDF,
as a function of the scale $Q$ and for three values of $x=10^{-4},10^{-2},10^{-1}$.
We see that both the anti-down and strange PDFs are of the same order of the anti-up
PDF for the whole range in $Q$ and for all values of $x$,
as expected from them being sea PDFs of non-perturbative origin.
Conversely, the charm and bottom PDFs start from being zero for $Q$ below their mass,
and they rise as $Q$ increases, eventually becoming of $\Ord(1)$ at large $Q$.
We note that the rise is stronger at smaller $x$, while it is very slow at values of $x\gtrsim0.1$.
From the lower plots, that show the charm and bottom PDFs as a function of $Q/m$
being $m$ the respective quark mass, we see that their behaviour is similar,
with the charm PDF rising faster probably due to the larger values of $\as$ involved.
If we arbitrarily require this ratio to be larger than 0.5 to count the PDF as $\Ord(1)$,
we can say that, for the medium value of $x=10^{-2}$,
the heavy quark PDFs need one-two decades in $Q$ above their mass
before being counted as the light PDFs.

We thus conclude that Eq.~\eqref{eq:RESexp} only holds at very high scales
$Q^2\gg m^2$ where $\as\log\mQ\sim1$.
At lower scales, and in particular at scales close to the mass $Q^2\sim m^2$,
the heavy quark PDFs count as $\Ord(\as)$ and the proper expansion is
(omitting arguments for better readability)
\begin{align}\label{eq:RESexp2}
  \frac1{\xB} F_a^{\rm heavy}(\xB,Q^2)
  &= \sum_{k=\he} \Cb_{a,k}^{[n_f+1](0)} \otimes f_{k}^{[n_f+1]}
  + \as \Cb_{a,g}^{[n_f+1](1)} \otimes f_{g}^{[n_f+1]}
  &&\text{(NLO)}\nonumber\\
  &+ \as\sum_{k=\he} \Cb_{a,k}^{[n_f+1](1)} \otimes f_{k}^{[n_f+1]}
  + \as^2 \sum_{k=\li} \Cb_{a,k}^{[n_f+1](2)} \otimes f_{k}^{[n_f+1]}
  &&\text{(NNLO)}\nonumber\\
  &+ \as^2\sum_{k=\he} \Cb_{a,k}^{[n_f+1](2)} \otimes f_{k}^{[n_f+1]}
  + \as^3 \sum_{k=\li} \Cb_{a,k}^{[n_f+1](3)} \otimes f_{k}^{[n_f+1]}
  &&\text{(N$^3$LO)}\nonumber\\
  &+\Ord(\as^4),
\end{align}
where we used the mass-dependent coefficient functions defined in Eq.~\eqref{eq:Cbardef},
which are more appropriate for this kinematic regime where mass effects are relevant.
At this point one could also write explicitly the $(n_f+1)$-scheme PDFs
as in Eq.~\eqref{eq:fres} and expand the evolution factors and the matching functions
similarly to what we did in Eq.~\eqref{eq:FOexp2} for the $n_f$-scheme expression.
As discussed before, the expansion of the evolution factors is not particularly relevant,
while the expansion of the matching functions does matter in this case.
Indeed, for a power counting that ensures
a perfect match between the perturbative ingredients at a given order at the transition point $Q=\muh$ between the fixed-order and resummation regions,
it is important to include in the construction of the PDFs the matching functions at the various orders properly~\cite{Bonvini:2015pxa}.
Details are given in appendix~\ref{app:PDF}.

The counting of Eq.~\eqref{eq:RESexp2} represents the main difference of our VFNS with respect to the other
constructions available in the literature, with the exception of BPT~\cite{Bonvini:2015pxa,Bonvini:2016fgf},
where this construction was suggested for the first time.
As we shall see later in section~\ref{sec:VFNS-N3LO}, this counting naturally leads to a smooth
transition between the two schemes, even at low orders.

We finally note that the scales $Q^2$ at which the heavy quark PDFs become of order 1
are very high, and in particular much higher
than the region where power mass corrections of $\Ord(\mQ)$ are relevant.
Therefore, in the kinematic region where the counting of Eq.~\eqref{eq:RESexp}
should be adopted, it is always legitimate to ignore the power corrections and use just
the massless result.

\subsection{Transitions between kinematic regions}

We thus conclude that there are three different kinematic regimes:
\begin{itemize}
\item $Q^2\lesssim m^2$, where collinear logarithms are not large
  and mass effects are important,
  in which the fixed-order description ($n_f$ scheme) is appropriate, Eq.~\eqref{eq:FOexp};
\item $Q^2\gtrsim m^2$, where resummation of collinear logarithms is convenient
  but mass effects are still important, thus requiring the resummed $n_f+1$ scheme
  in which the perturbative counting accounts for the fact that
  the heavy quark PDFs are of $\Ord(\as)$, Eq.~\eqref{eq:RESexp2};
\item $Q^2\gg m^2$, where resummation of collinear logarithms is mandatory and mass effects are negligible,
  and which is then well described by a massless $(n_f+1)$-scheme result
  with ``standard'' perturbative counting where all PDFs are of order 1, Eq.~\eqref{eq:RESexp}.
\end{itemize}
This classification is summarized in table~\ref{tab:orders},
where we also report the ingredients needed to construct the heavy structure function
at eah order for each kinematic regime.

\begin{table}[t]
  \centering
  \bgroup
  \begin{tabular}{l|ccc}
    & $Q^2\lesssim m^2$ & $Q^2\gtrsim m^2$ & $Q^2\gg m^2$ \\
    \midrule
    active flavours & $n_f$ & $n_f+1$ & $n_f+1$ \\
    heavy quark PDFs & -- & $\Ord(\as)$ & $\Ord(1)$ \\
    power suppressed terms $\Ord\(\mQ\)$ & included & included & negligible \\[1ex]
    LO requires & -- & -- & $C_{a,k=\lh}^{[n_f+1](0)}$ \\[2ex]
    NLO further requires & $C_{a,i=\li}^{[n_f](1)}$ & $\Cb_{a,k=\li}^{[n_f+1](1)}$, $\Cb_{a,k=\he}^{[n_f+1](0)}$ & $C_{a,k=\lh}^{[n_f+1](1)}$ \\[2ex]
    NNLO further requires & $C_{a,i=\li}^{[n_f](2)}$ & $\Cb_{a,k=\li}^{[n_f+1](2)}$, $\Cb_{a,k=\he}^{[n_f+1](1)}$ & $C_{a,k=\lh}^{[n_f+1](2)}$ \\[2ex]
    N$^3$LO further requires & {\color{blue}$C_{a,i=\li}^{[n_f](3)}$} & {\color{blue}$\Cb_{a,k=\li}^{[n_f+1](3)}$}, $\Cb_{a,k=\he}^{[n_f+1](2)}$ & $C_{a,k=\lh}^{[n_f+1](3)}$
  \end{tabular}
  \egroup
  \caption{Details of features and ingredients for the heavy structure functions in each kinematic region.
    The coefficient functions highlighted in blue are not fully known, and will be discussed in the next section.}
  \label{tab:orders}
\end{table}

The transition from the first to the second regime is governed by the matching scale $\muh$
and it corresponds to the actual scheme change.
One can vary the matching scale $\muh$ to study the impact of missing higher orders,
but once the scale is chosen, the transition is sharp.
Conversely, there is no such a clear transition scale between the second and the third regimes.
They are characterised by different perturbative countings,
but within the same scheme, and so there is more arbitrariness in how
and where the transition should be performed.

We thus suggest to perform a smooth transition between these two regions.
We consider a transition function of the form
\beq
\chi\(\mQ\) = \[1-\(\mQ a^2\)^b\]^c \, \theta\(\Qm-a^2\),
\eeq
with $a,b,c$ parameters determining where and how quick the transition between the two regimes is.
This function is zero for $Q^2\leq a^2m^2$ and tends to 1 for $Q^2\gg m^2$.
Our final result in the $n_f+1$ scheme will be thus given by
\begin{align}\label{eq:Fmixcounting}
  F_a^{\rm heavy}(\xB,Q^2)
  &= \left.F_a^{\rm heavy}(\xB,Q^2)\right|_{Q^2\gtrsim m^2\text{ counting}} \\
  &+\chi\(\mQ\)
  \bigg[
  \left.F_a^{\rm heavy}(\xB,Q^2)\right|_{Q^2\gg m^2\text{ counting}}
  -\left.F_a^{\rm heavy}(\xB,Q^2)\right|_{Q^2\gtrsim m^2\text{ counting}}
  \bigg], \nonumber
\end{align}
so that at $Q^2\sim m^2$ the result in the $Q^2\gtrsim m^2$ counting is used,
but going to large $Q^2$ it is smoothly replaced
by the result in the $Q^2\gg m^2$ counting.
Using the information from figure~\ref{fig:PDFrat},
we find that a reasonable choice for the $a,b,c$ parameters is
\begin{align}
a&=4,\qquad
b=\frac12,\qquad
c=4.
\end{align}
The value $a=4$ is chosen such that the asymptotic counting starts contributing sufficiently above $m^2$,
so that in any reasonable variation of the scales $\mu$ and $\muh$ we never enter the asymptotic counting region.
The $b$ parameters govern the speed of the transition:
$b=1/2$ implies that the asymptotic counting starts dominating
for $Q/m\gtrsim6a$, and it is strongly dominating (more than $95\%$) at $Q/m=100a$.
The large value of $c$ finally ensures a slow activation of the asymptotic counting
at scales $Q^2\gtrsim a^2 m^2$.
In principle one could make the $b$ parameter a function of $x$, with larger values of $b$ for smaller $x$
and vice versa, to account for the fact that the behaviour of the heavy quark PDFs depend on $x$ (figure~\ref{fig:PDFrat}).
Alternatively, one could consider fitting these parameters along with the PDFs, to find the values
that best allow to describe the data.
However, since this transition function has no physical meaning and it is just a tool,
we believe that a simpler implementation with fixed parameters is preferred.

We finally note that, for the energy scales probed at past and current DIS experiments,
there is no real need to transition to the $Q^2\gg m^2$ counting,
which is equivalent to say that one can set $\chi=0$.
This would somewhat simplify the practical construction of the resummed result.
While we do not suggest to do so, we take the opportunity to stress that the opposite is not a good option:
using the asymptotic counting also at small scales (i.e., $\chi=1$) would include spurious
higher order terms in the resummed result which were not present in the fixed-order result,
giving rise to a sharp, non-smooth transition between the two regimes.
Unfortunately, this is what was typically done in the construction of a VFNS,
and it led to the introduction of various ad-hoc attempts to mitigate the issues
(see e.g.\ the S-ACOT-$\chi$ variant of S-ACOT~\cite{Tung:2001mv} or the damping function in FONLL~\cite{Forte:2010ta}).

\section{\boldmath Approximate N$^3$LO coefficient functions with mass effects}
\label{sec:approx}

In this section we present a state-of-the-art approximation for the massive coefficient functions at $\ord3$,
namely $C_{a,i}^{[n_f](3)}$ for $a=2,L$ and $i=\li$, appearing in the heavy structure functions Eq.~\eqref{eq:FOexp}.
Our construction is not dissimilar to the analogous one of Ref.~\cite{Kawamura:2012cr},
which represents our main source of inspiration.
In particular, we base the approximation on the same known limits, but we suggest a novel way to combine them.
Moreover, we use the most recent determinations of the various ingredients entring the approximation,
thereby providing a more accurate result.

As shown in section~\ref{sec:VFNS}, the coefficient function is a function of the kinematic variable $z$ and of the ratio $m^2/Q^2$.
(The dependence on the factorization scale $\mu$ is fully determined by lower orders convoluted with DGLAP splitting functions,
so it is known exactly at N$^3$LO.)
The approximation of Ref.~\cite{Kawamura:2012cr} is based on known behaviour in special kinematic limits of these variables.
Specifically:
\begin{itemize}
\item the large-$z$ (threshold) limit is known from threshold resummation;
\item the small-$z$ (high-energy) limit is known from high-energy resummation;
\item the $Q\gg m$ (high-scale) limit is known from collinear factorization, Eq.~\eqref{eq:Csc}.
\end{itemize}
The first two limits describe opposite regimes in the $z$ variable at any value of $m/Q$,
and the third limit further provides knowledge at large scales $Q\gg m$ valid in principle at any $z$.
In fact, this latter limit is incompatible with the threshold limit, as we shall now briefly explain.

At parton level, the variable $s\equiv(p+q)^2=Q^2(1-z)/z$ has a lower limit,\footnote
{As discussed in section~\ref{sec:dis-accuracy}, starting from $\ord3$ there are loop-induced contributions
  to the heavy structure function that do not have heavy quarks in the final state.
  For these terms the lower limit on $s$ is $s\geq0$.
  As these contributions are very small and they appear for the first time at the order for which we are making an approximation,
  we will ignore them.
}
given by the minimum invariant mass squared of the final state: $s\geq 4m^2$.
The threshold limit thus corresponds to small $s\to4m^2$, namely
\beq\label{eq:zmaxdef}
z\to z_{\rm max}\equiv\frac1{1+\frac{4m^2}{Q^2}},
\eeq
having used the relation $z=Q^2/(Q^2+s)$.
This dependence on $m/Q$ of the threshold limit
is not associated with mass logarithms.
This means that in the right-hand side of Eq.~\eqref{eq:Csc} this dependence is hidden in the neglected power corrections,
while the first term, which represents the known $Q\gg m$ limit used in the approximation,
is defined for all $z\leq 1$ and thus has the wrong behaviour in the $z\to z_{\rm max}$ limit.
In other words, an approximation based on the high-scale limit will never be accurate at large $z$.
Instead, the small-$z$ ($s\gg Q^2$) limit is perfectly compatible with
the high-scale ($Q^2\gg m^2$) limit Eq.~\eqref{eq:Csc},
as the two regions overlap in the region $s\gg Q^2\gg m^2$ where both approximations are valid.

These considerations already suggest that it is convenient to combine first the high-energy and the high-scale limits
in a proper way to obtain an approximation which is valid at small/medium $z$ and which gets better and better
as $Q^2$ increases, and then match it with the threshold limit which provides complementary information.
This is the main difference in our construction with respect to Ref.~\cite{Kawamura:2012cr},
where instead the three limits were treated as independent and matched on the same level.

In the rest of this section we will briefly review the various ingredients of the approximation,
referring to Ref.~\cite{Kawamura:2012cr} and to our appendices for all the details.
Here, we rather focus on the differences between our construction and the original one of Ref.~\cite{Kawamura:2012cr}.
To validate our construction, we apply it to the NNLO coefficient functions,
which are fully known~\cite{Laenen:1992zk,Laenen:1992cc,Riemersma:1994hv,Harris:1995tu,Hekhorn:2018ywm,Klann:2026svr},
also providing an uncertainty to account for the approximate nature of our result.
We then present the new results for the N$^3$LO coefficient functions, and compare them with those of Ref.~\cite{Kawamura:2012cr}.
For simplicity, here we focus on $C_{2,g}^{[n_f]}$, which gives the leading contribution to the DIS cross section,
while the results for $C_{2, q}^{[n_f]}$, $C_{L, g}^{[n_f]}$ and $C_{L, q}^{[n_f]}$ are reported in appendix~\ref{sec:app:coeff-func}.

\subsection{High scale plus high energy: the asymptotic limit}
\label{sec:asy}

Let us start from the high-scale limit.
As already stated, this approximation is obtained from the collinear factorization formula Eq.~\eqref{eq:Csc}
after neglecting the power-suppressed terms,
\beq\label{eq:Chighscale}
C_{a,i}^{[n_f]\,\rm h.s.}\(\mQ,\muQ\) \equiv \sum_k^{n_f+1} C_{a,k}^{[n_f+1]}\(0,\muQ\) \otimes K_{ki}^{[n_f+1]\leftarrow[n_f]}\(\mmu\).
\eeq
The right-hand side is given in terms of massless coefficient functions, fully known up to N$^3$LO~\cite{Moch:2004xu,Vermaseren:2005qc},
and the matching functions, now also completely known at the same order~\cite
{Ablinger:2010ty,Ablinger:2014vwa,Ablinger:2014nga,Behring:2014eya,Ablinger:2023ahe,Ablinger:2024xtt},
enabling the complete construction of the high-scale approximation at $\ord3$.
The perturbative expansion of Eq.~\eqref{eq:Chighscale} is explicitly given in appendix~\ref{app:DeltaC}.

Note that at the time of Ref.~\cite{Kawamura:2012cr} the matching coefficients $K_{hq}^{[n_f+1]\leftarrow[n_f]}$
and $K_{hg}^{[n_f+1]\leftarrow[n_f]}$ were not fully known at $\ord3$, thus introducing an additional source of uncertainty
in their results affecting $C_{2,q}^{[n_f]}$ and $C_{2,g}^{[n_f]}$ respectively.\footnote
{The N$^3$LO longitudinal structure functions are not sensitive to these terms because
  the massless coefficient $C_{a,h}^{[n_f+1]}$ is zero at $\ord0$, so they start contributing at one order higher.}
The results of Ref.~\cite{Kawamura:2012cr} were based on an approximation of these contributions
based on known Mellin moments~\cite{Bierenbaum:2009mv},
improved in Ref.~\cite{Alekhin:2017kpj} thanks to the exact computation of $K_{hq}^{[n_f+1]\leftarrow[n_f]}$
at three loops~\cite{Ablinger:2014nga}.
The recent exact computation of $K_{hg}^{[n_f+1]\leftarrow[n_f]}$~\cite{Ablinger:2023ahe,Ablinger:2024xtt}
allows us to provide state-of-the-art predictions for the high-scale limit
(see also Refs.~\cite{Ablinger:2025awb,Ablinger:2025joi}).

Let us now move to the high-energy limit.
The leading behaviour of the coefficient function in the $z\to0$ limit has the form
\beq
C_{a,i}^{[n_f]}\(z,\mQ,\muQ\) = C_{a,i}^{[n_f]\,\rm h.e.}\(z,\mQ,\muQ\) + \Ord(z^0)
\eeq
with
\beq\label{eq:Chighenergy}
C_{a,i}^{[n_f]\,\rm h.e.}\(z,\mQ,\muQ\) \equiv \sum_{n=2}^\infty \as^n \sum_{p=0}^{n-2} c_{a,i}^{(n,p)}\(\mQ,\muQ\) \frac{\log^{n-2-p}z}z,
\eeq
where we have restored the explicit $z$ dependence in the coefficient, which is now at the core of our discussion.
The leading high-energy terms are those diverging at $z\to0$ as $1/z$, possibly times a power of $\log z$.
These contributions appear for the first time at $\ord2$.
The largest power of the log, corresponding to the terms with $p=0$ in the inner sum in Eq.~\eqref{eq:Chighenergy},
are the leading logarithms (LL), the terms with $p=1$ are the next-to-leading logarithms (NLL), and so on.
At NNLO there is only the LL term, while at N$^3$LO also the NLL contributes.

The technology to predict (and resum) these logarithmic terms has been developed in the 90s
and it is based on the so-called $k_t$ factorization~\cite{Catani:1990eg,Catani:1994sq}.
This approach is currently able to predict only the LL contributions.
This means that, while at NNLO we have full control of the small-$z$ limit,
at N$^3$LO we do not know the exact NLL terms $c_{a,i}^{(3,1)}$.

Differently from Ref.~\cite{Kawamura:2012cr}, where these NLL contributions were guessed from the known high-scale limit,
we prefer to construct approximate expressions for them using knowledge from resummation.
Indeed, modern approaches to resumation do produce some (but not all) subleading contributions,
which can provide a better way of including them in our prediction.
Since in any case these predictions are incomplete, we associate an uncertainty to them,
by varying these NLL terms in a controlled manner (see section~\ref{sec:fullApprox}).
Technical details as well as explicit formulae are collected in appendix~\ref{sec:app:nll-approx}.

As we already mentioned at the beginning of section~\ref{sec:approx}, the high-scale and high-energy limits
are compatible among themselves (in other words, they commute).
This means that the high-energy limit of Eq.~\eqref{eq:Chighscale} coincides with the high-scale limit of Eq.~\eqref{eq:Chighenergy}.
This enables us to combine them in a very simple way, i.e.\ using an additive matching.
We call this the \emph{asymptotic limit},
\beq\label{eq:Casy}
C_{a,i}^{[n_f]\,\rm asy} \equiv C_{a,i}^{[n_f]\,\rm h.s.} + C_{a,i}^{[n_f]\,\rm h.e.} - C_{a,i}^{[n_f]\,\rm h.s.\ h.e.},
\eeq
where the last term contains the double counting between the first two terms.
This expression reproduces by construction the same correct result in both the high-scale or the high-energy limits,
and differs only far from them, where it is indeed an approximation.
One obvious advantage of this construction is that it does not depend on any arbitrary matching parameters,
in contrast with the result of Ref.~\cite{Kawamura:2012cr} where the high-energy contribution is artificially
switched off at high scales.
Another advantage is its accuracy, as we shall see in a moment.

We observe that there is another natural way of combining
the high-energy and high-scale limits, namely through a multiplicative matching,
where the high-scale coefficient would be multiplied by the ratio
$C_{a,i}^{[n_f]\,\rm h.e.} / C_{a,i}^{[n_f]\,\rm h.s.\ h.e.}$
between the high-energy term and its high-scale limit.
While legitimate, this approach is potentially problematic if the denominator approaches zero,
or if the ratio becomes negative.
In order to avoid these problems, we consider a modified version of the multiplicative matching,
which is in fact a mixture of multiplicative and additive matching.
Specifically, we write
\beq\label{eq:Casy'}
C_{a,i}^{[n_f]\,\rm asy'} \equiv
\[C_{a,i}^{[n_f]\,\rm h.s.} + C_{a,i}^{[n_f]\,\rm h.e.} - C_{a,i}^{[n_f]\,\rm h.s.\ h.e.} + \(\frac1{R_a}-1\) C_{a,i}^{[n_f]\,\rm h.e.\ LL}\] R_a
\eeq
where $C_{a,i}^{[n_f]\,\rm h.e.\ LL}$ is just the LL term of the high-energy contribution.
This expression, that we generically denote asy$'$, clearly coincides with the additive matching
Eq.~\eqref{eq:Casy} for $R_a=1$.
It is also easy to see that the LL behaviour at high energy is correctly reproduced for any value of $R_a$.
Indeed, the goal of this expression is to be accurate at high energy at LL level,
because that is currently the only exactly known term.
Consistently, to reproduce a fully multiplicative matching at LL (and treat NLL terms additively)
we should choose $R_a = C_{a,i}^{[n_f]\,\rm h.e.\ LL} / C_{a,i}^{[n_f]\,\rm h.s.\ h.e.\ LL}$.
This form of $R_a$ is independent of $z$, and thus this choice is advantageous with respect to
the option of putting in the ratio the whole high-energy terms because that would be $z$ dependent,
possibly leading to $R_a<0$ or divergent during the $z$ integration.
However, it is still possible that far from the high-scale region this ratio becomes negative or divergent independently of $z$.
In order to avoid this, we suggest to use a modified version of $R_a$, constructed such that it equals
$C_{a,i}^{[n_f]\,\rm h.e.\ LL} / C_{a,i}^{[n_f]\,\rm h.s.\ h.e.\ LL}$ at high scales
and such that it remains positive at any scale.
In practice, we use different expressions for $F_2$ and $F_L$, that we present and motivate in appendix~\ref{sec:app:nll-approx}.

We now focus on the NNLO, which serves as a validation of our procedure.
At this order the asymptotic approximations read
\begin{align}
  \label{eq:Casy2}
  C_{a,i}^{[n_f](2)\,\rm asy}\(z,\mQ,\muQ\)
  &= C_{a,i}^{[n_f](2)\,\rm h.s.}\(z,\mQ,\muQ\) + \frac{c_{a,i}^{(2,0)}\(\mQ,\muQ\)-c_{a,i}^{(2,0)\,\rm h.s.}\(\mQ,\muQ\)}z,
  \\
  \label{eq:Casy2'}
  C_{a,i}^{[n_f](2)\,\rm asy'}\(z,\mQ,\muQ\)
  &= \[C_{a,i}^{[n_f](2)\,\rm h.s.}\(z,\mQ,\muQ\)  + \frac{c_{a,i}^{(2,0)}\(\mQ,\muQ\)}{z\, R_a^{(2)}\(\mQ,\muQ\)}-\frac{c_{a,i}^{(2,0)\,\rm h.s.}\(\mQ,\muQ\)}z \]\nonumber\\
  &\quad\times R_a^{(2)}\(\mQ,\muQ\),
\end{align}
where $c_{a,i}^{(2,0)\,\rm h.s.}$ is the high-scale limit of $c_{a,i}^{(2,0)}$ presented in appendix~\ref{sec:app:nll-approx}
and $R_a^{(2)}$ is the ratio given by Eqs.~\eqref{eq:R2}--\eqref{eq:RL}.
In figure~\ref{fig:asy:n2lo} we plot the exact result for $a=2$ and $i=g$
(other functions are presented in appendix~\ref{sec:app:coeff-func})
along with the high-scale limit, the high-energy limit and their overlapping limit in the left plots,
and with the asymptotic approximation in the right plots.
Different rows correspond to different values of $Q^2/m^2$, from low values (where power-suppressed mass dependence is important)
to high values (where the high-scale approximation becomes more accurate).
Each plot shows the $z$ dependence, but the curves are shown as a function of another variable,
\beq
\eta = \frac{s}{4 m^2} -1 = \frac{Q^2}{4 m^2}\frac{1-z}z -1,
\eeq
to make more direct contact with previous works~\cite{Kawamura:2012cr, Laenen:1992zk,Laenen:1992cc,Riemersma:1994hv,Harris:1995tu}.
This variable tends to zero in the threshold limit, and to infinity in the high-energy limit.
Finally, we set the factorization (and renormalization) scale to $\mu^2=m^2$,
for a direct comparison with Ref.~\cite{Kawamura:2012cr}.
In this and all the other parton-level plots we set the electric charge of the heavy quark to one;
to obtain the actual coefficient function the plots must be multiplied by the electric charge squared.

\begin{figure}[tp]
  \centering
  \includegraphics[width=0.49\textwidth]{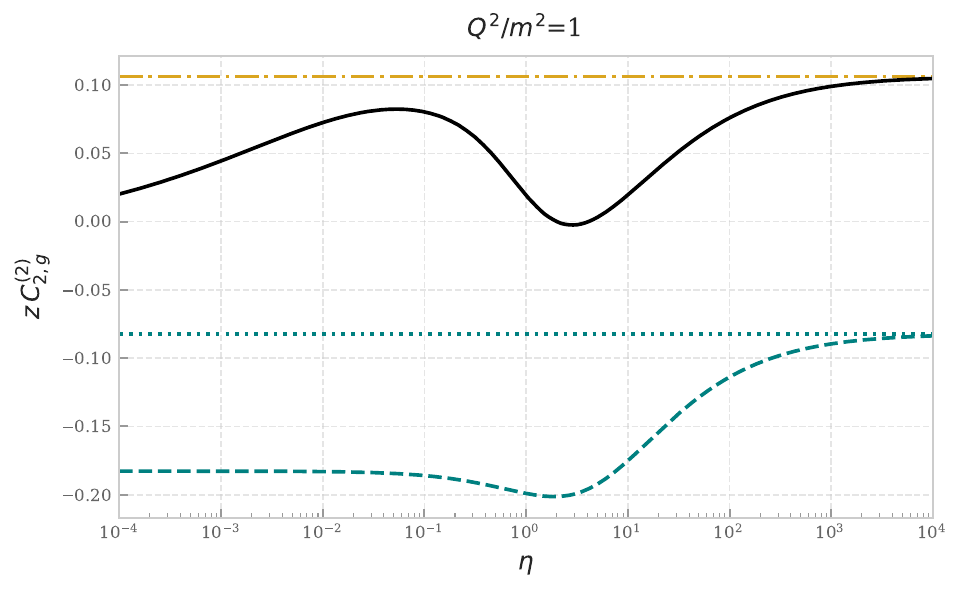}
  \includegraphics[width=0.49\textwidth]{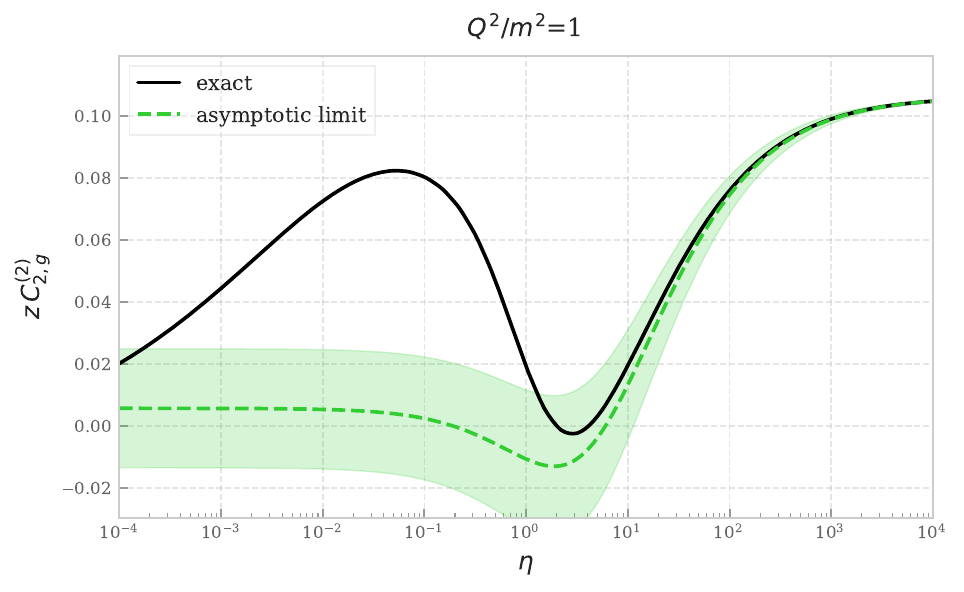}
  \includegraphics[width=0.49\textwidth]{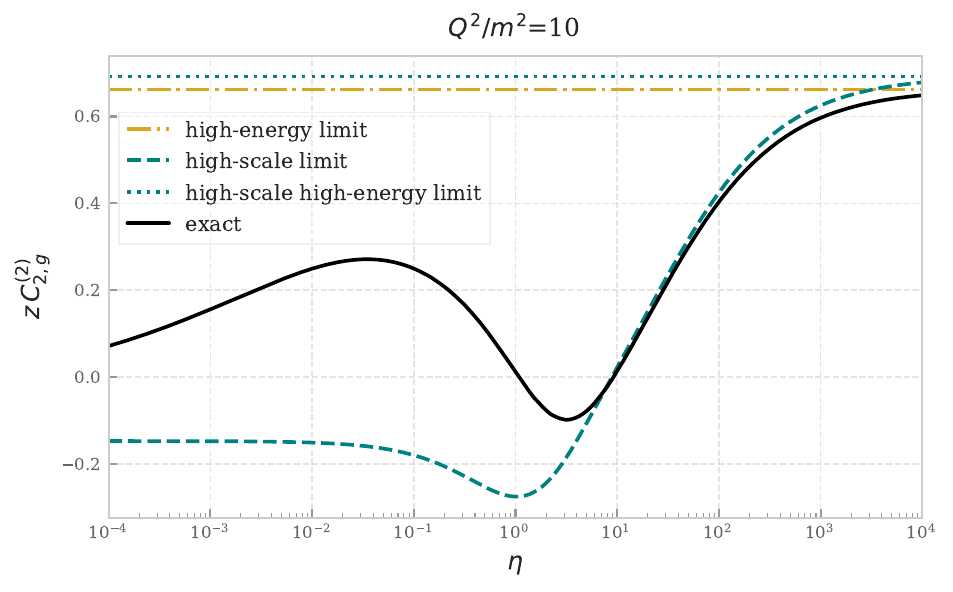}
  \includegraphics[width=0.49\textwidth]{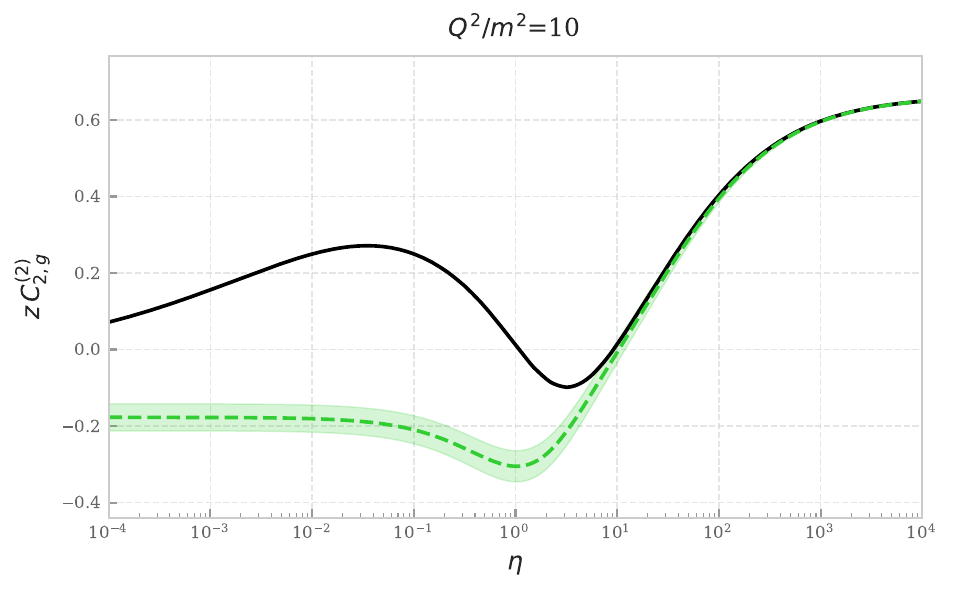}
  \includegraphics[width=0.49\textwidth]{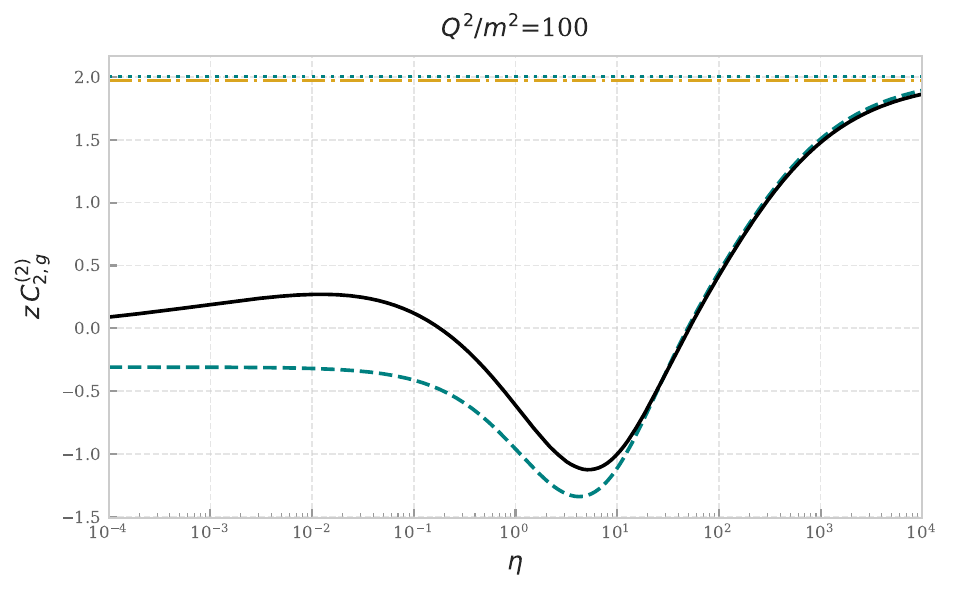}
  \includegraphics[width=0.49\textwidth]{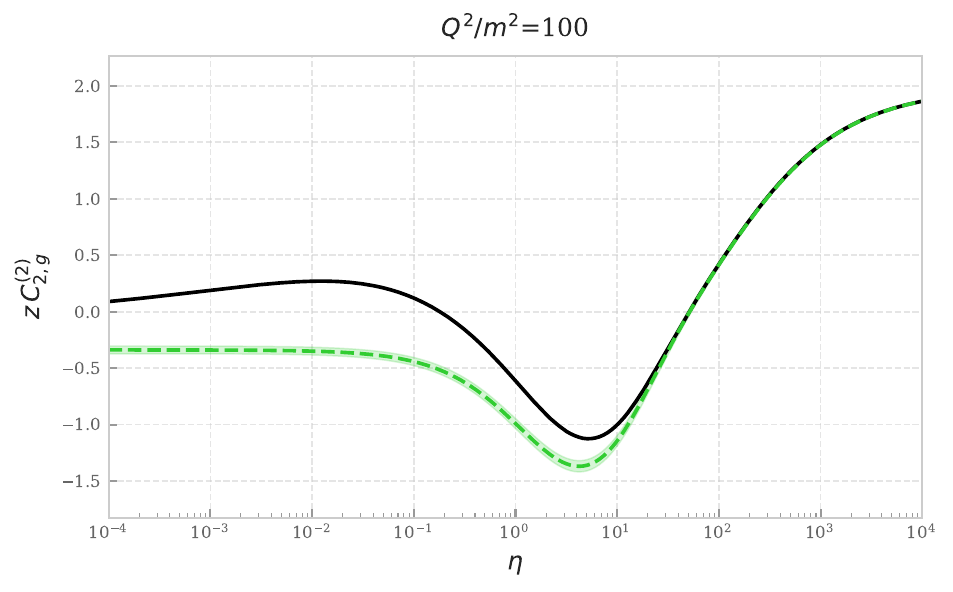}
  \includegraphics[width=0.49\textwidth]{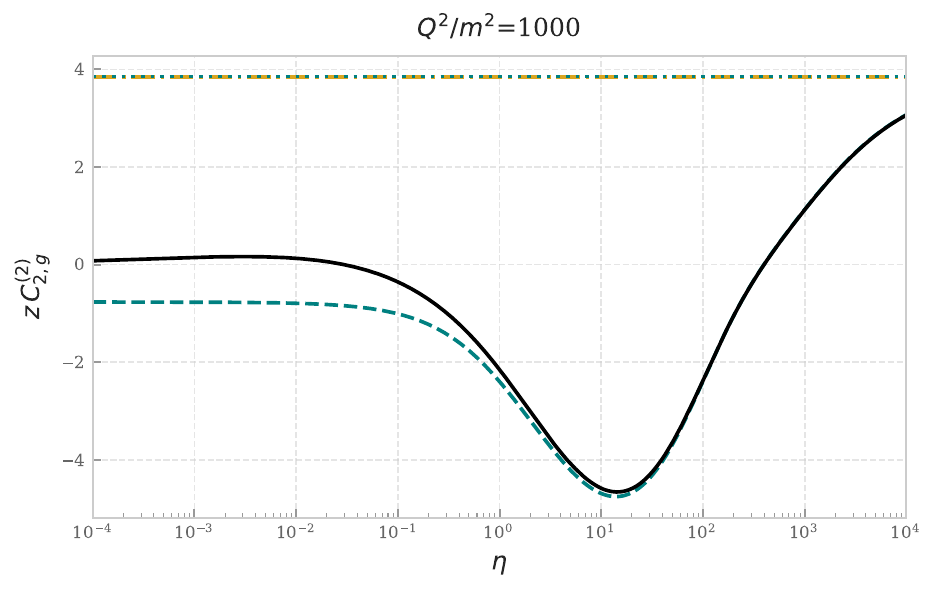}
  \includegraphics[width=0.49\textwidth]{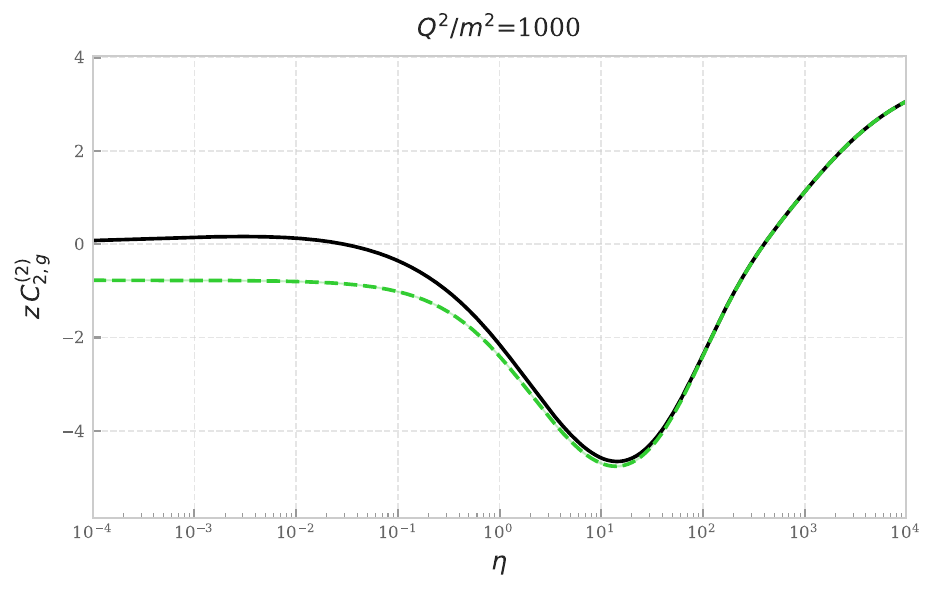}
  \caption{Left: comparison between the exact coefficient function $C_{2,g}^{[n_f]}$ and its high-energy limit,
    high-scale limit and simultaneous high-energy high-scale limit.
    Right: comparison between the exact coefficient function $C_{2,g}^{[n_f]}$ and its asymptotic limit,
    the latter including an uncertainty band as described in the text.
    All plots show the $\ord{2}$ coefficient function computed at $\mu=m$ as a function of $\eta$ for different values of $Q^2/m^2$.}
  \label{fig:asy:n2lo}
\end{figure}

We observe that the high-scale approximation is indeed never good at threshold (small $\eta$),
while it provides a good approximation over a wide range of $\eta\gtrsim1$ when $Q^2\gg m^2$.
However, at smaller values of $Q^2/m^2$ the high-scale approximation becomes inaccurate even in this range.
More precisely, we observe that the shape is similar to the exact, but there is an offset.
Because we plot the coefficient function multiplied by $z$, this offset precisely corresponds to a high-energy LL term.
Indeed we see that the high-energy limit of the exact and of the high-scale approximation are straight lines,
which become identical at large $Q^2/m^2$ but can be rather different at small $Q^2/m^2$.
Our asymptotic approximation Eq.~\eqref{eq:Casy2}, based on the additive matching Eq.~\eqref{eq:Casy},
has thus the effect of shifting the high-scale approximation to match the correct high-energy limit.
As a result, we see that the asymptotic approximation is rather good for a wide range of $\eta\gtrsim10$
at any scale $Q^2$, range that extends to smaller values as long as $Q^2$ increases.

The right plots also show an uncertainty band on the asymptotic approximation,
which estimates the uncertainty due to the missing power corrections $\Ord(m^2/Q^2)$ in the high-scale term
as well as the missing $\Ord(z^0)$ corrections in the high-energy term.
To obtain it we simply considered the asy$'$ result Eq.~\eqref{eq:Casy2'},
obtained with the modified multiplicative matching Eq.~\eqref{eq:Casy'}.
The band is obtained by symmetrising  the difference between this variant and our central curve Eq.~\eqref{eq:Casy2}.
We note that, by construction, the band squeezes at large $\eta$ where the asymptotic result
is correct and both matching (additive and multiplicative) must give the same result.
The band covers well the exact result for $\eta\gtrsim10$, where the asymptotic approximation is supposed to be valid.
Below this value, the band is no longer able (in general) to cover the exact result,
because the asymptotic approximation is no longer valid.

We conclude that our asymptotic-limit result provides an excellent approximation at high $\eta$ (small $z$)
which extends down to values of $\eta=\Ord(10)$ with a reliable uncertainty estimate in this region.
Matching this result with the information from the opposite regime, $\eta\to0$,
has thus the chance to provide a good approximation over the whole range.

\subsection{The threshold limit}
\label{sec:thr}

We now move to the threshold approximation, describing the limit $z\to z_{\rm max}$, i.e.\ $\eta\to0$.
In the threshold limit, where the center-of-mass energy $s$ is just above the production threshold of the pair $4m^2$,
the initial state gluon leg can only emit soft radiation, that factorizes in the coefficient function. 
This radiation produces logarithmically enhanced terms in the limit $z\to z_{\rm max}$,
that can be conveniently written in terms of the heavy quark velocity
\beq\label{eq:beta}
\beta
\equiv \sqrt{1- \frac{4m^2}{s}}
= \sqrt{1-\frac{4m^2}{Q^2}\frac{z}{1-z}}
= \sqrt{\frac\eta{1+\eta}},
\eeq
which tends to zero in the threshold limit.
The coefficient function in the dominant gluon channel behaves in the threshold ($\beta\to0$) limit as
\beq\label{eq:Cthreshold}
\frac{C_{a,g}^{[n_f]\,\rm thr}\(z,\mQ,\muQ\)}{\as C_{a,g}^{[n_f](1)}\(z,\mQ,\muQ\)}
\equiv 1 + \sum_{n=1}^\infty \as^n \sum_{r=0}^n \frac1{\beta^r} \sum_{p=0}^{2(n-r)} b_{a}^{(n,r,p)}\(\mQ,\muQ\) \log^p\beta,
\eeq
where the terms with $r=0$ correspond to the usual threshold logarithms which can be described
by threshold resummation~\cite{Sterman:1986aj, Catani:1989ne, Contopanagos:1996nh, Kidonakis:1997gm, Bonciani:1998vc}
while terms with $r>0$, which are more singular in the threshold limit, are due to Coulomb (or Glauber) gluon exchanges
between the initial-state gluon and the final-state heavy quarks.
They can also be resummed to all orders~\cite{Hoang:2000yr}.

Note that the quark channel is strongly suppressed in the threshold limit,
and it vanishes much faster than the gluon channel as $\beta\to0$.
Therefore we cannot provide any approximation for it in this limit, which means that our approximation
for the quark channels (presented in appendix~\ref{sec:app:coeff-func})
is entirely given by the asymptotic limit described in section~\ref{sec:asy}.

For the dominant structure function $F_2$ all the ingredients needed to approximate at threshold
the NNLO and the N$^3$LO coefficients are presented in Ref.~\cite{Kawamura:2012cr}
(we report some detail in appendix~\ref{app:threshold}).
We have to stress that the constant term at N$^3$LO, corresponding to $n=2$, $r=0$, $p=0$,
is not known yet.
Ref.~\cite{Kawamura:2012cr} uses a Pad\'e approximant for it, assigning a 100\% uncertainty to this value
(which means that it is varied between 0 and twice its value).
We also follow the same strategy.
Moreover, on top of this variation, we further provide an additional uncertainty due to subleading power terms,
which contributes also at NNLO.
Specifically, we consider a variation of the approximation obtained by keeping the same right-hand side of Eq.~\eqref{eq:Cthreshold}
but replacing the NLO coefficient in the denominator of the left-hand side with its leading threshold limit,
\beq\label{eq:CthresholdNLO}
C_{a,g}^{[n_f](1)}\(z,\mQ,\muQ\) = C_{a,g}^{[n_f](1)\,\rm thr}\(z,\mQ,\muQ\) \[1+\Ord(\beta^2)\].
\eeq
Note that because of the presence of negative powers of $\beta$ in the right-hand side of Eq.~\eqref{eq:Cthreshold}
this replacement should be compensated by changing the values of some of the $b_{a}^{(n,r,p)}$ coefficients.
However, as the full NLO coefficient differs from its leading threshold limit by $\Ord(\beta^2)$ terms,\footnote
{This is true both for $F_2$ and the longitudinal structure function $F_L$, so the same reasoning applies to boh cases.}
the effect is beyond the accuracy at NNLO and it only affects the currently unknown constant term $b_{a}^{(2,0,0)}$ at N$^3$LO.
Therefore, this variation probes, on top of non-threshold terms, exactly the unknown constant terms at N$^3$LO,
and thus provides an additional, controlled way of assessing the uncertainty due to the lack of its knowledge.\footnote
{Numerically, the $\beta^0$ contribution produced by the interference of subleading NLO terms with the Coulomb term
  is very small, and in particular much smaller than the effect of including or not the Pad\'e approximant for the constant N$^3$LO term.}
Finally, at N$^3$LO we take the sum in quadrature of this variation and the one of the Pad\'e approximant.

Most of the ingredients for the threshold approximation of the longitudinal structure function $F_L$
are the same as for $F_2$, as they are due to the universal structure of soft radiation,
the only exception being the ``hard'' terms, which at NNLO 
can be found in Ref.~\cite{Hekhorn:2018ywm}.
However, the way the purely threshold terms and the Coulomb terms are combined
and appear in the coefficients of Eq.~\eqref{eq:Cthreshold} is different
and technically nontrivial.
We present the construction of the threshold approximation for $F_L$ in appendix~\ref{app:threshold}.

\begin{figure}[tp]
  \centering
  \includegraphics[width=0.49\textwidth]{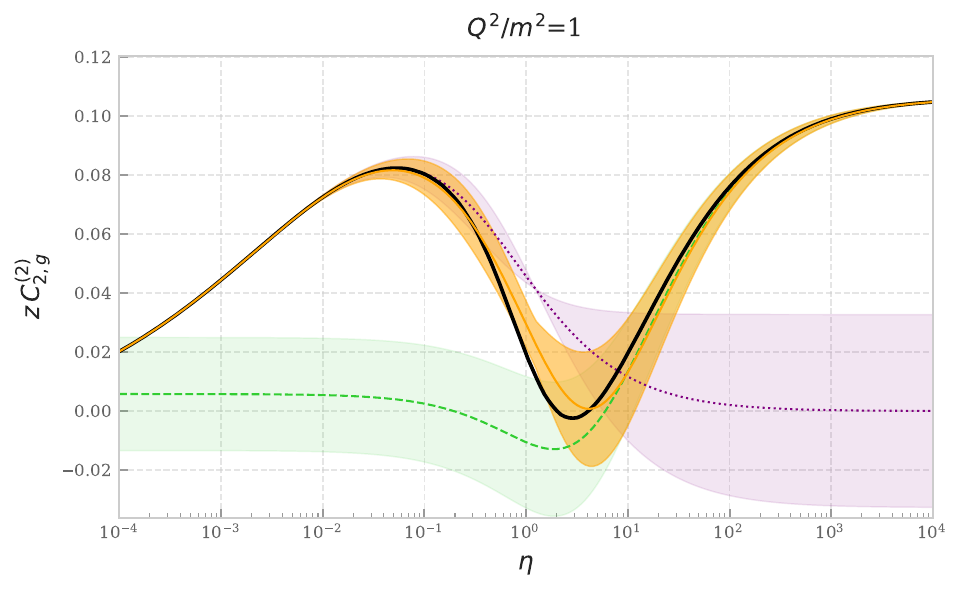}
  \includegraphics[width=0.49\textwidth]{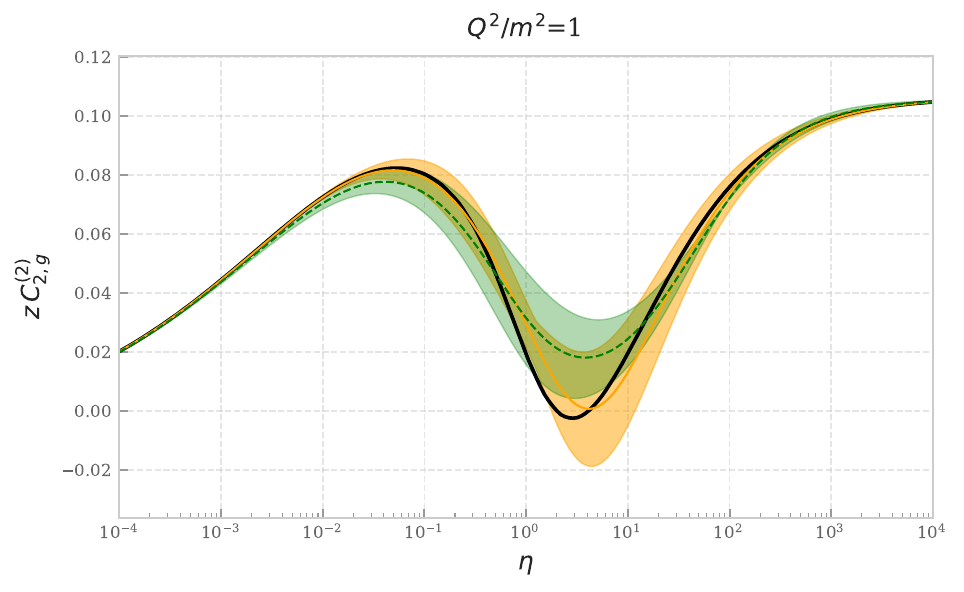}
  \includegraphics[width=0.49\textwidth]{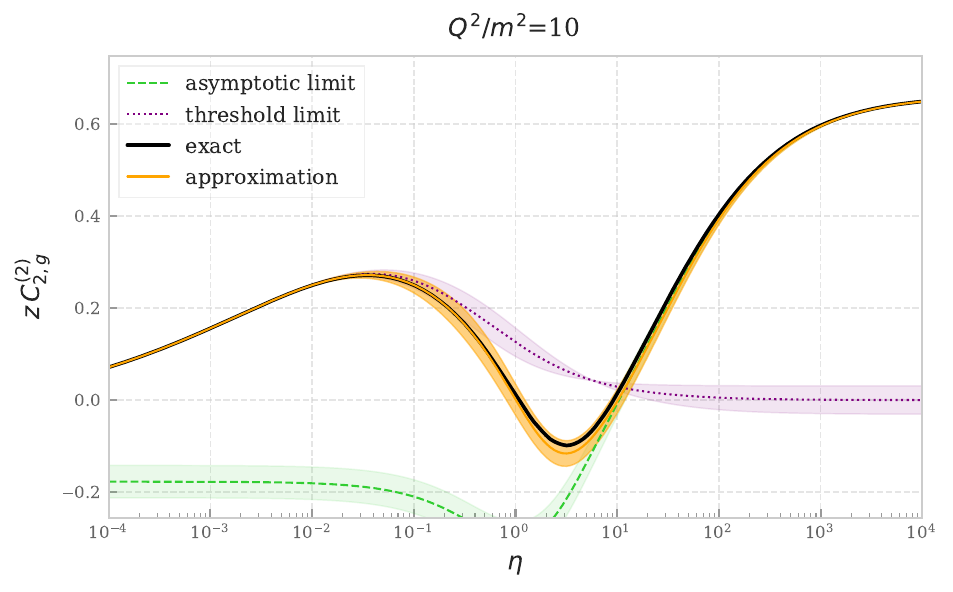}
  \includegraphics[width=0.49\textwidth]{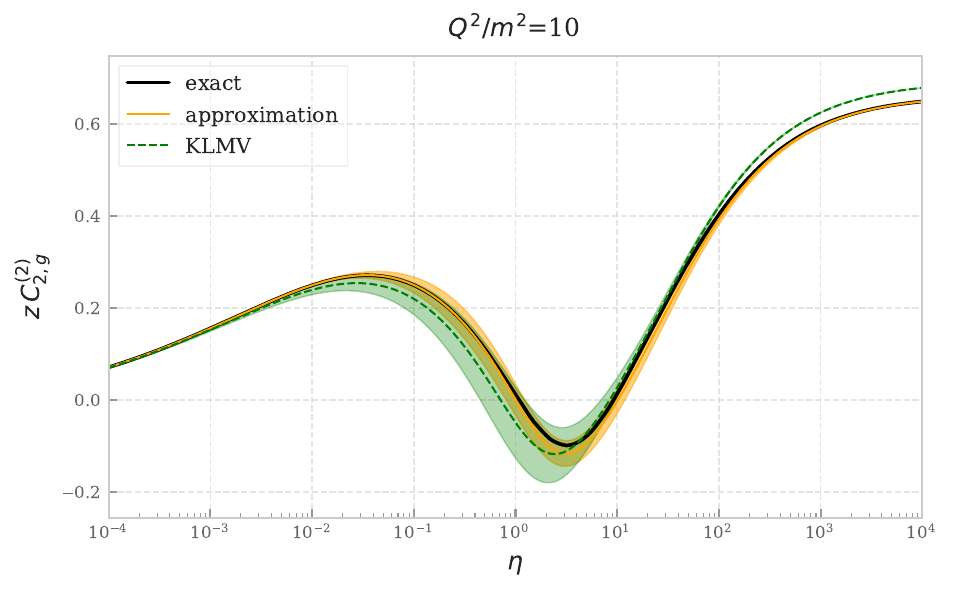}
  \includegraphics[width=0.49\textwidth]{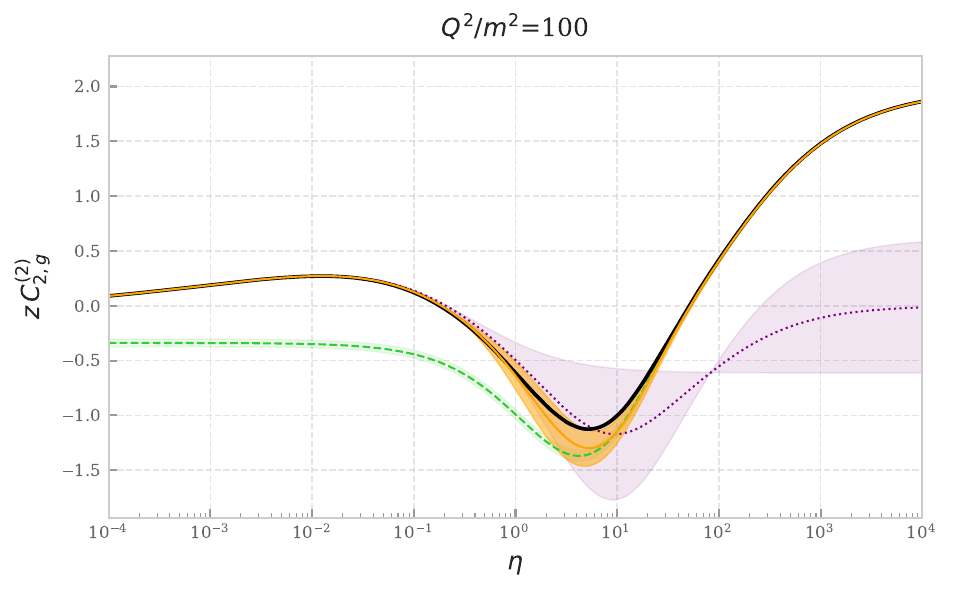}
  \includegraphics[width=0.49\textwidth]{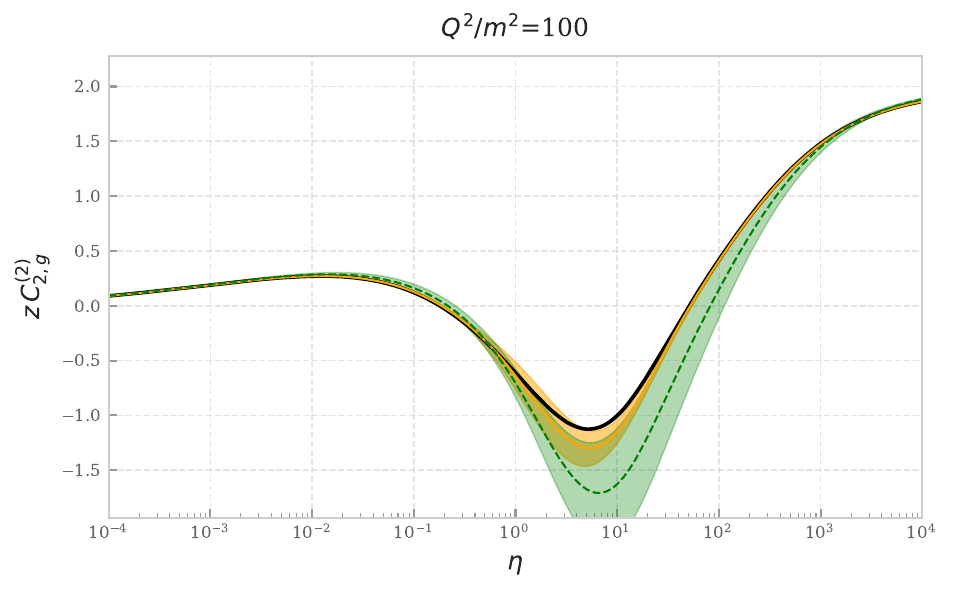}
  \includegraphics[width=0.49\textwidth]{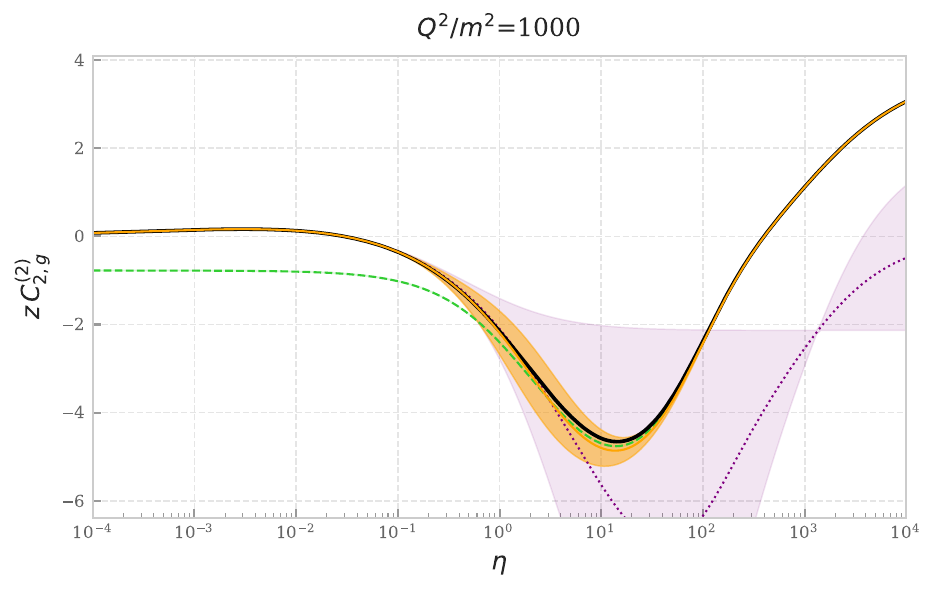}
  \includegraphics[width=0.49\textwidth]{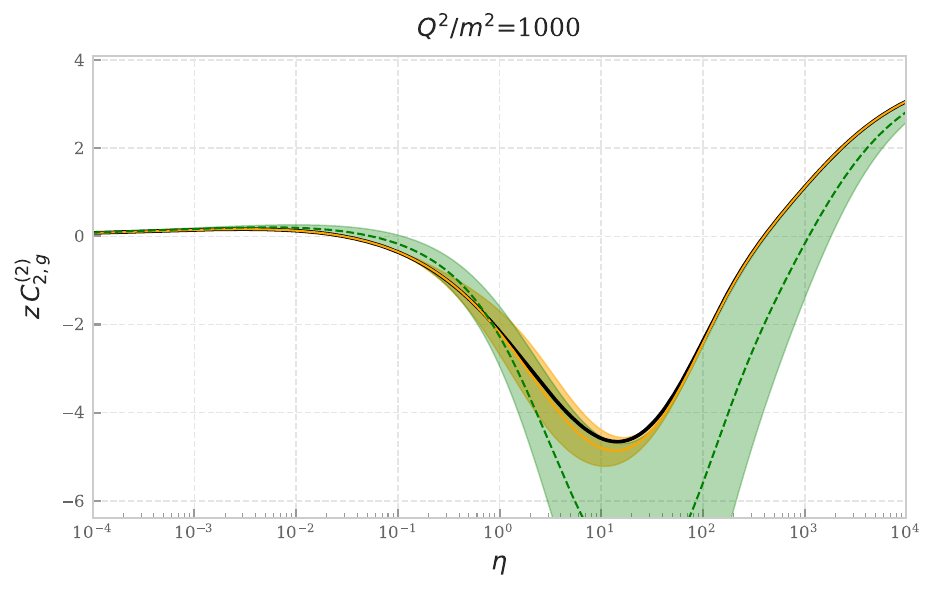}
  \caption{Left: Comparison between the exact coefficient function $C_{2,g}^{[n_f]}$ and its threshold and asymptotic limits,
    as well as the full approximation (with uncertainties).
    Right: Comparison between the exact coefficient function $C_{2,g}^{[n_f]}$ and our full approximation
    with the approximation of Ref.~\cite{Kawamura:2012cr}, dubbed KLMV.
    All plots show the $\ord{2}$ coefficient function computed at $\mu=m$ as a function of $\eta$ for different values of $Q^2/m^2$.}
  \label{fig:appr:n2lo}
\end{figure}

In figure~\ref{fig:appr:n2lo} (left) we show the threshold approximation compared to the exact result,
in the same fashion of figure~\ref{fig:asy:n2lo}.
The uncertainty band is constructed by symmetrising the variation mentioned above with respect to the central value
given by Eq.~\eqref{eq:Cthreshold}.
We observe that the threshold approximation is excellent for $\eta\lesssim 0.1$ at all scales,
becoming more accurate even at higher values of $\eta$ as $Q^2$ gets larger.

\subsection{Full approximation}
\label{sec:fullApprox}

We are now ready to combine the threshold and asymptotic approximations to obtain our
complete approximation.
We do this by simply summing the two approximations, each with a damping function,
\begin{equation}
  \label{eq:approx}
  C_{a,i}^{[n_f]\, \rm approx}\(z,\mQ,\muQ\)
  = D_{\rm thr}\(z,\mQ\) C_{a,i}^{[n_f]\, \rm thr}\(z,\mQ,\muQ\) + D_{\rm asy}\(z,\mQ\) C_{a,i}^{[n_f]\, \rm asy}\(z,\mQ,\muQ\) . 
\end{equation}
These damping functions must satisfy the obvious limits
\begin{align}
  \label{eq:damp:func}
  D_{\rm thr}\(z,\mQ\) &\xrightarrow{z \rightarrow 0} 0 ,&
  D_{\rm thr}\(z,\mQ\) &\xrightarrow{z \rightarrow \zmax} 1, \nonumber\\
  D_{\rm asy}\(z,\mQ\) &\xrightarrow{z \rightarrow 0} 1 ,&
  D_{\rm asy}\(z,\mQ\) &\xrightarrow{z \rightarrow \zmax} 0,
\end{align}
ensuring that each contribution is fully active in its limit of validity and fully switched off in the opposite limit.
The specific form that we use for the damping functions is
\beq\label{eq:damping}
1-D_{\rm asy}\(z,\mQ\) = D_{\rm thr}\(z,\mQ\) = \frac1{1+\(\frac{\eta}{\eta_0(m^2/Q^2)}\)^{\rho(m^2/Q^2)}},
\eeq
where $\eta_0$ and $\rho$ are parameters governing the form of the damping function.
In principle, they could depend on the scale $Q^2$ in order to accomodate different level of accuracy of the approximation at different scales.
However, in practice, for the sake of simplicity and to avoid a pointless fine tuning we fix them to
a scale independent value: $\eta_0=2$, $\rho=1.3$.

Figure~\ref{fig:appr:n2lo} (left) shows the comparison between the exact coefficient function at $\ord{2}$
with the asymptotic and threshold limits and our full approximation Eq.~\eqref{eq:approx}.
It is clear that at any scale $Q^2$ the small and large $\eta$ limits are very well described,
and the only region which is delicate is the ``central'' one, i.e.\ the one around $\eta\sim1$,
where the full approximation must transition from one limit to the other.
This transition region is not controlled by the approximation, and it is thus affected by larger uncertainties.
As the scale $Q^2$ increases, the approximations in the two limits become more and more accurate,
and so does the full approximation, whose uncertainty reduces accordingly.
In all regions, the approximate prediction is always compatible within uncertainties with the exact result.

The right plots of figure~\ref{fig:appr:n2lo} show the comparison of our approximation with the one of Ref.~\cite{Kawamura:2012cr},
with the exact result as a reference.
We see that the central value of our approximation is generally closer to the exact result than the one of Ref.~\cite{Kawamura:2012cr},
especially at medium and high $\eta$.
Moreover, our uncertainty band is typically smaller at medium/high scales $Q^2$, where the approximation is supposed
to be more accurate, and still well compatible with the exact result, even more than the larger one of Ref.~\cite{Kawamura:2012cr}
(see e.g.\ the region of mid $\eta$ at $Q^2/m^2=100$).
The relative sizes of the bands are inverted at low $Q^2$, where indeed the missing power corrections of $\Ord(m^2/Q^2)$
are large and make the approximations less accurate. In fact, our larger band covers well the exact result at $Q^2/m^2=1$,
while the smaller one of Ref.~\cite{Kawamura:2012cr} does not in some regions of $\eta$.
Finally, we observe that the large-$\eta$ limit of the result of Ref.~\cite{Kawamura:2012cr} is less accurate than ours,
in particular at $Q^2/m^2=10$ where their asymptotic value does not match the exact result within uncertainties,
confirming our expectation that the construction based on additive/multiplicative matching provides a superior approximation.

\begin{figure}[tp]
  \centering
  \includegraphics[width=0.49\textwidth]{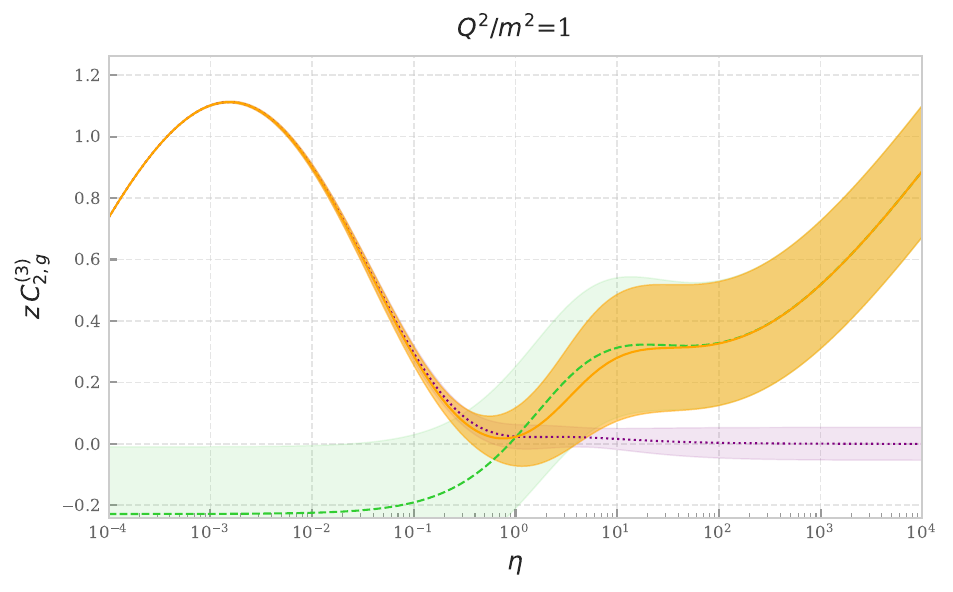}
  \includegraphics[width=0.49\textwidth]{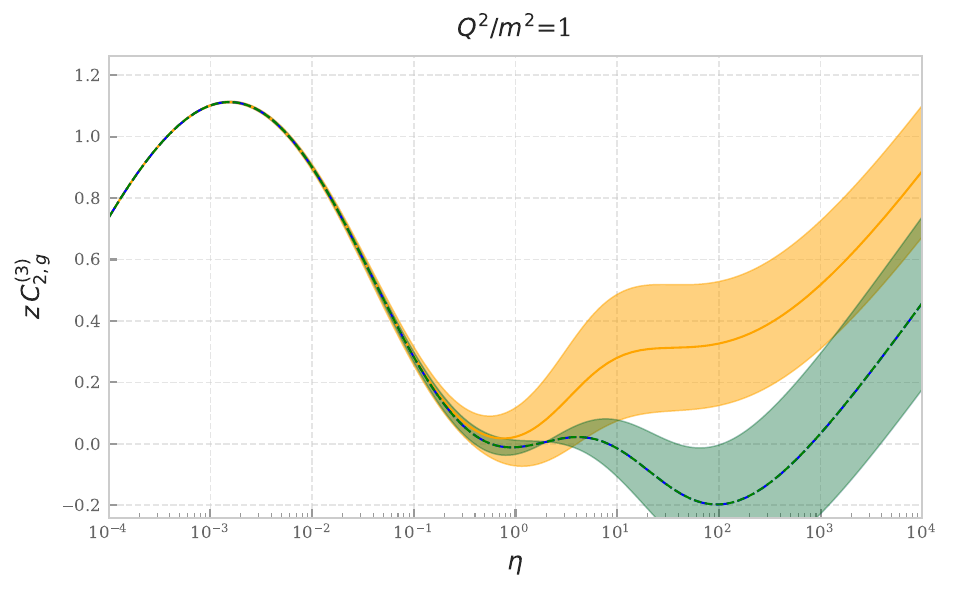}
  \includegraphics[width=0.49\textwidth]{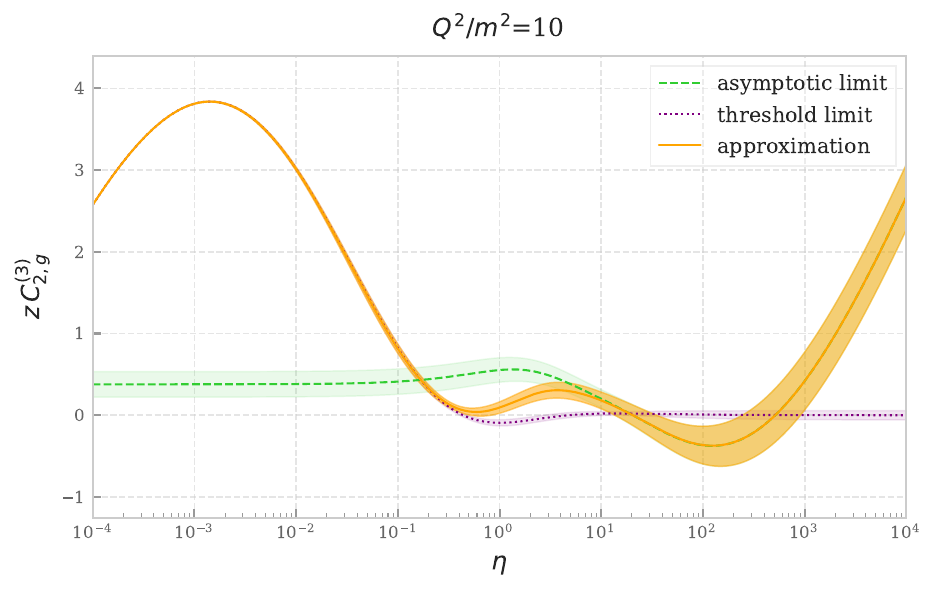}
  \includegraphics[width=0.49\textwidth]{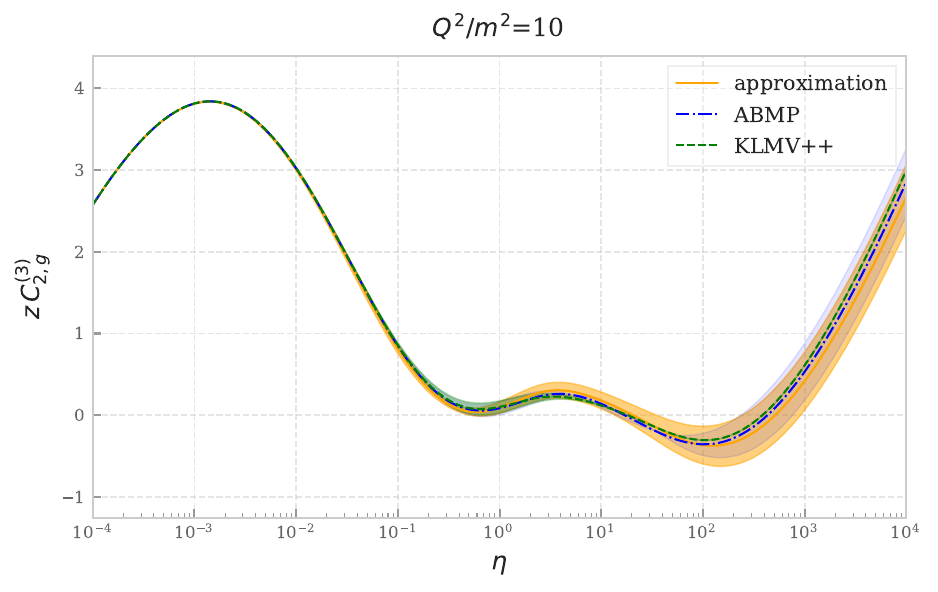}
  \includegraphics[width=0.49\textwidth]{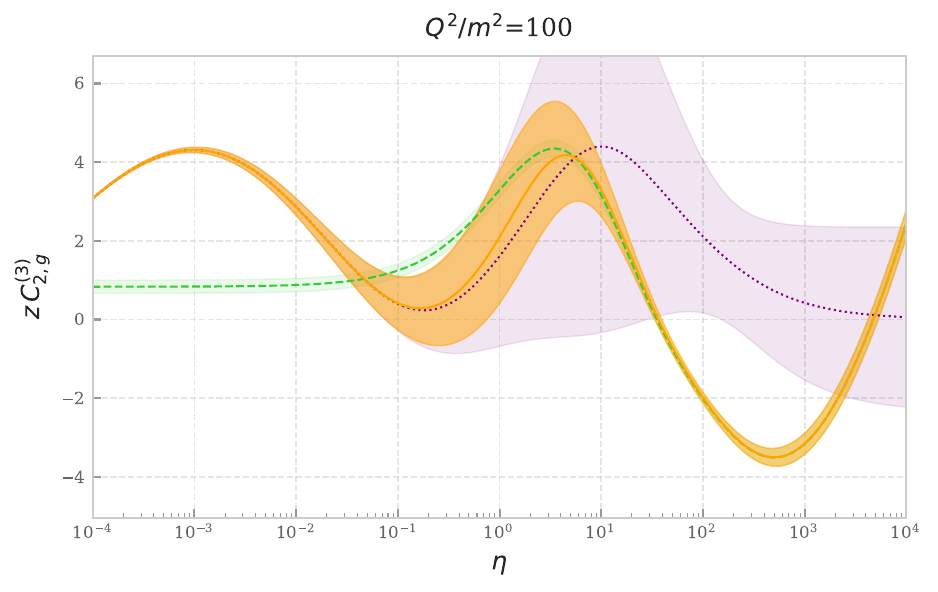}
  \includegraphics[width=0.49\textwidth]{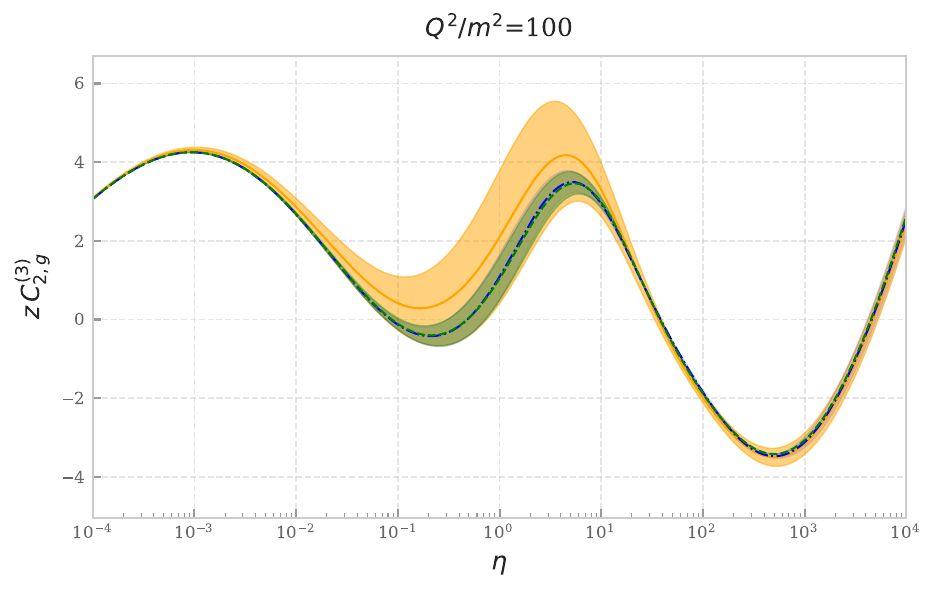}
  \includegraphics[width=0.49\textwidth]{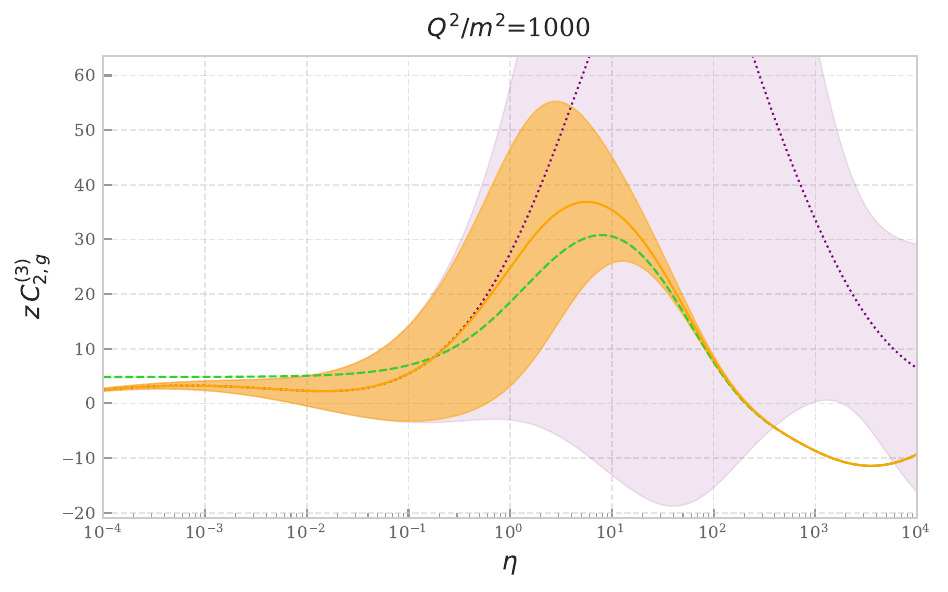}
  \includegraphics[width=0.49\textwidth]{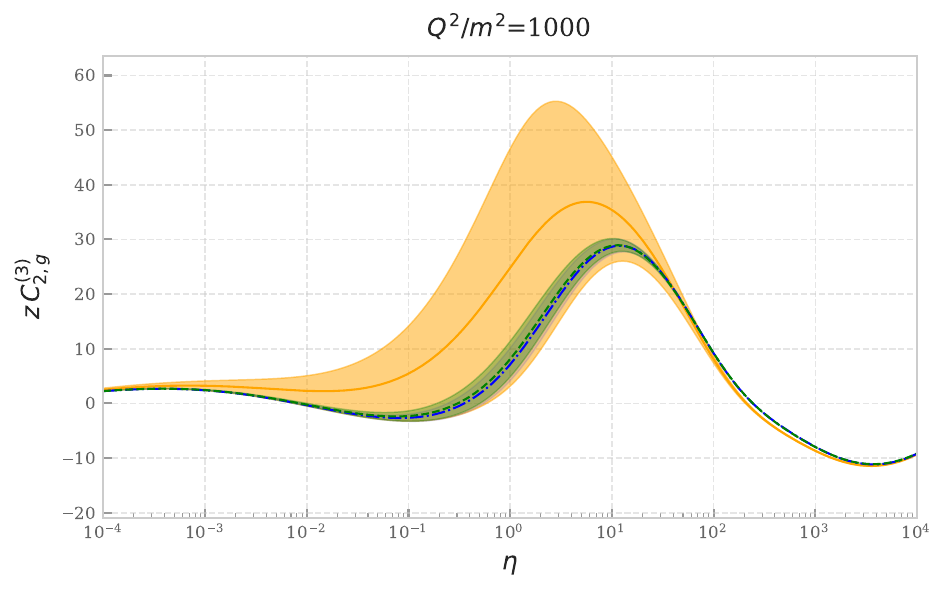}
  \caption{Same as figure~\ref{fig:appr:n2lo}, but for the $\ord3$ coefficient,
    and without showing the exact result which is unknown at this order.
    In the right plots we show both the approximation of Ref.~\cite{Kawamura:2012cr} in the updated version of Ref.~\cite{Alekhin:2017kpj} (ABMP curve)
    and what it would look like with the current knowledge of the heavy-quark matching coefficients (KLMV++ curve).}
  \label{fig:appr:n3lo}
\end{figure}

We are now in the position of moving to N$^3$LO.
In figure~\ref{fig:appr:n3lo} we show the same approximate curves as in figure~\ref{fig:appr:n2lo}, but at the unknown $\ord3$.
Remember that at this order both the threshold and the asymptotic approximation are not fully known:
the former lacks of the knowledge of the constant ($\beta^0$) term, the latter of the NLL coefficient at high energy.
Our way of estimating the uncertainty on the threshold approximation at this order includes a variation of the $\beta^0$ term,
as discussed in section~\ref{sec:thr}.
For the high-energy part, we construct our asymptotic limit in a way that we also cover the uncertainty due to the missing NLL coefficient.
Specifically, on top of the additive matching Eq.~\eqref{eq:Casy} that at this order reads
\begin{align}\label{eq:Casy3}
  C_{a,i}^{[n_f](3)\,\rm asy}\(z,\mQ,\muQ\)
  &= C_{a,i}^{[n_f](3)\,\rm h.s.}\(z,\mQ,\muQ\) + \frac{c_{a,i}^{(3,1)}\(\mQ,\muQ\)-c_{a,i}^{(3,1)\,\rm h.s.}\(\mQ,\muQ\)}z \nonumber\\
  &\quad+ \frac{c_{a,i}^{(3,0)}\(\mQ,\muQ\)-c_{a,i}^{(3,0)\,\rm h.s.}\(\mQ,\muQ\)}z \log z,
\end{align}
we consider two variants of the modified multiplicative matching Eq.~\eqref{eq:Casy'}, given by\footnote
{Note that the coefficient $c_{a,i}^{(3,1)\,\rm h.s.}$ is the high-scale limit of the not-fully-known NLL coefficient,
  for which we provide an approximation in appendix~\ref{sec:app:nll-approx}.
  This guarantees that in the high-scale limit the correct NLL term from the high-scale coefficient is correctly reproduced.}
\begin{align}\label{eq:Casy3'}
  C_{a,i}^{[n_f](3)\,\rm asy'}\(z,\mQ,\muQ\)
  &= \vast[ C_{a,i}^{[n_f](3)\,\rm h.s.}\(z,\mQ,\muQ\) + \frac{c_{a,i}^{(3,1)}\(\mQ,\muQ\)-c_{a,i}^{(3,1)\,\rm h.s.}\(\mQ,\muQ\)}z \nonumber\\
  &\quad\quad + \(\frac{c_{a,i}^{(3,0)}\(\mQ,\muQ\)}{z\, R_a^{(3)}\(\mQ,\muQ\)}-\frac{c_{a,i}^{(3,0)\,\rm h.s.}\(\mQ,\muQ\)}z\) \log z\vast] R_a^{(3)}\(\mQ,\muQ\),
  \\
  \label{eq:Casy3''}
  C_{a,i}^{[n_f](3)\,\rm asy''}\(z,\mQ,\muQ\)
  &= \vast[ C_{a,i}^{[n_f](3)\,\rm h.s.}\(z,\mQ,\muQ\) \nonumber\\
  &\quad\quad + \(\frac{c_{a,i}^{(3,0)}\(\mQ,\muQ\)}{z\, R_a^{(3)}\(\mQ,\muQ\)}-\frac{c_{a,i}^{(3,0)\,\rm h.s.}\(\mQ,\muQ\)}z\) \log z\vast] R_a^{(3)}\(\mQ,\muQ\).
\end{align}
All these constructions have the correct LL behaviour at small $z$, but they differ at NLL:
the first one behaves according to the approximate NLL coefficient $c_{a,i}^{(3,1)}$,\footnote
{Plus the exact NLL term of the high-scale coefficient function minus the high-scale limit of $c_{a,i}^{(3,1)}$,
  which are not the same in our construction.}
the second one rescales it by the factor $R_a^{(3)}$, Eqs.~\eqref{eq:R2}--\eqref{eq:RL},
and the last one simply sets it to zero.
Therefore, the use of these three results allows us to probe different values of the yet unknown NLL coefficient.
Specifically, we construct the uncertainty on the asymptotic contribution at N$^3$LO by summing in quadrature the difference between
each of the modified multiplicative matching expressions Eq.~\eqref{eq:Casy3'} and Eq.~\eqref{eq:Casy3''}
and our central value given by the additive matching result Eq.~\eqref{eq:Casy3}.
Finally, the uncertainty band of the full approximation is given by the envelope
between the threshold and asymptotic bands.

The comparison with the result of Ref.~\cite{Kawamura:2012cr} is shown in the right plots.
We actually do not plot the original result of Ref.~\cite{Kawamura:2012cr}, because
one of the ingredients (the matching function $K_{hg}^{[n_f+1]\leftarrow[n_f]}$) was not fully known at the time and it is fully known now.
So, we decided to plot both the last published result, which is the updated prediction of Ref.~\cite{Alekhin:2017kpj}
(ABMP curve) as well as what the curve would look like today with the complete knowledge of the coefficient
(KLMV++ curve).
We observe that the difference between these two curves is very small, while there is a difference in the uncertainty band,
which is strongly reduced at mid $Q^2$ scales in the large-$\eta$ region.
Similarly to what happend at NNLO, there are sizable differences between our approximation and the construction of Ref.~\cite{Kawamura:2012cr}.
At low $Q^2$ the difference is mosty in the region of mid/high $\eta$, where the uncertainties of both results
are large, though only partially overlapping.
At higher scales the large-$\eta$ region becomes more stable upon different construction, and with smaller uncertainty,
due to the dominance of the exact high-scale result,
while at low $\eta$ the results agree within uncertainties, but our uncertainty is much larger.

Owing to the validation at NNLO given above, we believe that our construction is reliable and our uncertainty realistic.
The discrepancies with the result of Ref.~\cite{Kawamura:2012cr} at low $Q^2$ and high $\eta$ are mostly due
to a different NLL term and the different combination of high-energy and high-scale limits,
and we have already discussed why we think that our construction is more robust.
At high $Q^2$ and mid/low $\eta$ the discrepancies in the uncertainty bands (and central values) are more subtle.
We have verified that most of our uncertainty in this region, which descend from the threshold part of our result,
comes from the variation of the Pad\'e approximant for the N$^3$LO $\beta^0$ term of Eq.~\eqref{eq:Cthreshold}.
In principle Ref.~\cite{Kawamura:2012cr} has a similar uncertainty and should thus produce a comparable band.
However, in the practical implementation of their N$^3$LO approximation, Eqs.~(4.17) and (4.18) of Ref.~\cite{Kawamura:2012cr},
the constant term is multiplied by a damping function that suppresses its contribution at large scales $Q^2$.
The effect of this choice is to shrink the band towards our lower variation, which is indeed what we see from the plot.
While it is true that the uncertainty on the Pad\'e coefficient leads to a large band,
we believe that our band represents in a more realistic way the actual uncertainty on the result.

\section{Results for N$^3$LO structure functions}
\label{sec:VFNS-N3LO}

Now that we have presented our novel approximate expression for the $\ord3$ coefficient functions in the $n_f$ scheme,
we have all the ingredients to construct the DIS structure functions at N$^3$LO
according to the procedure discussed in section~\ref{sec:dis-accuracy}.

On the technical side, we recall that the $\ord2$ massive coefficient functions are known only in numerical form
from Refs.~\cite{Laenen:1992zk,Laenen:1992cc,Riemersma:1994hv,Harris:1995tu,Hekhorn:2018ywm,Klann:2026svr},
as tables in the variables $\eta$ and $m^2/Q^2$ to be interpolated
and matched to the small- and large-$\eta$ asimptotic limits, which are known in analytic form.
Everything else is known analytically.
As far as the approximation of the N$^3$LO massive coefficients is concerned,
we adopt the strategy of Ref.~\cite{Kawamura:2012cr} of approximating the coefficient
at the factorization and renormalization scale $\mu^2=m^2$ (as presented in section~\ref{sec:approx}),
and then evolve to any other value of the scales exactly.
In our codes we have implemented all the necessary ingredients,
using the latest exact results available from the literature.
In particular, we have collected in the public code \adani all the coefficient functions needed
to construct the massive and massless results, as well as the various matching functions,
so that the construction of the structure functions at both fixed-order and resummed level
is just a matter of computing convolutions with the appropriate PDFs.

To begin with, we present the $F_2^{\rm heavy}$ and $F_L^{\rm heavy}$ structure functions for bottom quark production computed in the $n_f$ scheme
at any scale considered in the plots (from $Q=1$~GeV to $Q=1$~TeV), including our approximate results both at NNLO and N$^3$LO.
These plots, representing the fixed-order part of the VFNS result,
serve as a validation at NNLO, complementary to the parton-level one performed in section~\ref{sec:approx},
and allow us to present our new N$^3$LO approximate prediction without mixing it
with the effects of the resummation of collinear mass logarithms.
In these and all subsequent plots we consider only the photon-mediated DIS,
despite the fact that we reach values of $Q^2$ where the $Z$ contribution cannot be neglected.
Therefore, in the region of high scales the results should not be considered as representative of the actual DIS cross section,
but merely as an ingredient of it.
Note that, differently from the parton-level plots of section~\ref{sec:approx} and appendix~\ref{sec:app:coeff-func},
we now include the electric charge of the bottom quark in the structure functions,
corresponding to an overall factor $e_h^2=1/9$.

\begin{figure}[tp]
  \centering
  \includegraphics[width=0.328\textwidth, page=1]{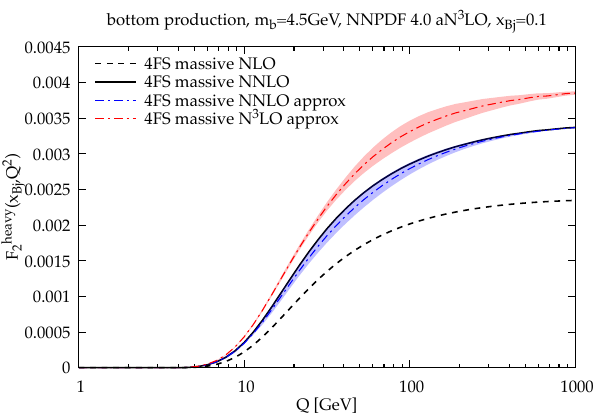}
  \includegraphics[width=0.328\textwidth, page=2]{images/DIS_F2_paper.pdf}
  \includegraphics[width=0.328\textwidth, page=3]{images/DIS_F2_paper.pdf}\\
  \includegraphics[width=0.328\textwidth, page=1]{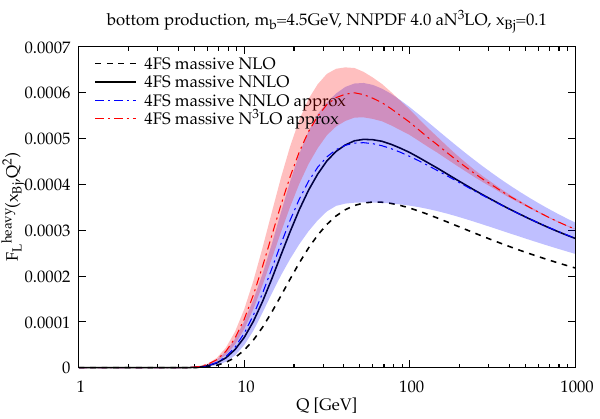}
  \includegraphics[width=0.328\textwidth, page=2]{images/DIS_FL_paper.pdf}
  \includegraphics[width=0.328\textwidth, page=3]{images/DIS_FL_paper.pdf}
  \caption{The structure functions $F_2^{\rm heavy}$ (upper plots) and $F_L^{\rm heavy}$ (lower plots)
    for three different values of $\xB=10^{-1}, 10^{-2}, 10^{-4}$ (from left to right)
    for bottom quark production in the $n_f=4$ scheme (fixed-order massive result) at $\mu=Q$.
    In addition to the exact results at NLO and NNLO (dashed and solid black curves),
    we plot our approximation at NNLO and N$^3$LO (blue and red dot-dashed curves)
    with their own uncertainty bands.}
  \label{fig:F2FLfixedorder}
\end{figure}

Figure~\ref{fig:F2FLfixedorder} shows the heavy structure functions for bottom production,
with $m=4.92$~GeV, using the NNPDF 4.0 aN$^3$LO PDF set~\cite{NNPDF:2024nan},\footnote
{We use the same N$^3$LO set for all orders as we want to focus on the perturbative
  progression of the perturbative ingredient, namely the partonic coefficient function.}
that we have consistently evolved in the $n_f=4$ scheme starting from the fit scale of $1.65$~GeV using the \texttt{EKO} code~\cite{Candido:2022tld}.
The factorization and renormalization scales are set equal to the hard scale, $\mu=Q$,
and we take the value of the strong coupling from the PDF set for consistency.\footnote
{The original PDF set uses $\as(m_Z^2)=0.118$ in the $n_f=5$ scheme, corresponding to $\as(1.65\text{GeV})=0.330$ at the initial scale,
  that we take as our reference value for constructing the $n_f=4$ set.
  As a consequence, after evolving from this scale in the $n_f=4$ scheme, the value of the strong coupling at the $Z$ mass is modified.}
The upper plots show $F_2^{\rm heavy}$ and the lower plots $F_L^{\rm heavy}$,
and in each row there are three plots corresponding to three values of $\xB=10^{-1}, 10^{-2}, 10^{-4}$ (large, mid and low).
We can see the exact curves at NLO (dashed black) and NNLO (solid black)
as well as our approximations at NNLO (dot-dashed blue) and N$^3$LO (dot-dashed red)\footnote
{Obviously, the NNLO and N$^3$LO approximate curves are constructed using the exact result
up the previous order and using the approximate expression only for the last order.}
along with their uncertainty bands constructed as describes in section~\ref{sec:approx}.
We immediately observe that the approximate NNLO results are perfectly compatible within uncertainties
with the exact ones, providing a further confirmation of the goodness of our construction.
We also note that the bands for the longitudinal structure function are much bigger than
the analogous for $F_2$, in particular at NNLO.
In fact, the large NNLO band even at high scales is due to the way the uncertainty
on the asymptotic part is constructed (see appendix~\ref{sec:app:coeff-func}
and the discussion at the end of appendix~\ref{sec:app:nll-approx}).\footnote
{In particular, in order to be compatible with exact result at low $Q$,
  the NNLO band has been constructed with a rather large
  variation of the parameter $A_L$ introduced in Eq.~\eqref{eq:RL},
  leading to a large uncertainty also at higher $Q$.}
As the construction of the N$^3$LO uncertainty is different (and more similar to the one of $F_2$),
the resulting band is more robust and reliable, and indeed it gets small at high scales as expected.
Finally, we note that the perturbative progression of these results is very reasonable,
even though the $\ord3$ corrections are generally rather large, especially for $F_L$.
This is possibly due to the presence of the unresummed collinear mass logarithms in these plots.

\begin{figure}[tp]
  \centering
  \includegraphics[width=0.5\textwidth, page=4]{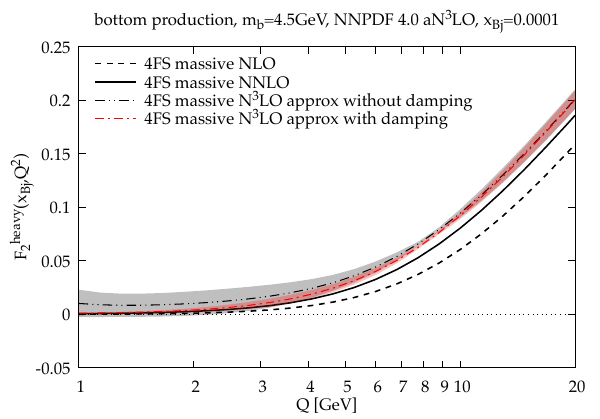}
  \caption{A zoom at low $Q$ of the structure functions $F_2^{\rm heavy}$ at $\xB=10^{-4}$,
    showing the approximate N$^3$LO result with (dot-dashed red) and without (dot-dot-dashed gray)
    the damping function Eq.~\eqref{eq:dampingF2}.}
  \label{fig:F2lowQ}
\end{figure}

In the plots of figure~\ref{fig:F2FLfixedorder} we have also included in the N$^3$LO prediction
a damping factor acting at low $Q$, of the form
\beq\label{eq:dampingF2}
d\(\mQ\) = \frac1{1+m^2/Q^2}.
\eeq
This function multiplies the approximate coefficient functions at $\ord3$ defined in section~\ref{sec:approx}.
The reason for introducing this function is that we have noticed that, without it, the N$^3$LO structure function
$F_2^{\rm heavy}$ at low $\xB$ did not tend to zero at low $Q$, as previous orders do.
We show in figure~\ref{fig:F2lowQ} a zoom of the behaviour of this structure function at low $Q$ for $\xB=10^{-4}$
(which is where the effect is larger), with and without the damping Eq.~\eqref{eq:dampingF2}.
As we can see, the N$^3$LO structure function is compatible with zero within the uncertainty,
but the central value is higher, and the uncertainty band is large in comparison with the size of the previous orders.
This is probably due to the large uncertainty induced by the unkonwn NLL contributions at small $z$,
that indeed give rise to large uncertainties at high $\eta$ and low $Q$
(see figures~\ref{fig:appr:n3lo}, \ref{fig:appr:n3lo:2q}, \ref{fig:appr:Lg}, \ref{fig:appr:Lq}),
and it reflects the fact that our knowledge in this region of very low $Q$ is very limited.
While the small-$z$ NLL term remains unknown, we believe that the damping function Eq.~\eqref{eq:dampingF2}
provides a reasonable solution, thanks to the physical constraint that the structure function must go to zero in
the low $Q$ limit, also implying that the cross section there is very small, and thus an imperfect prediction is of little relevance.

\begin{figure}[tp]
  \centering
  \includegraphics[width=0.328\textwidth, page= 4]{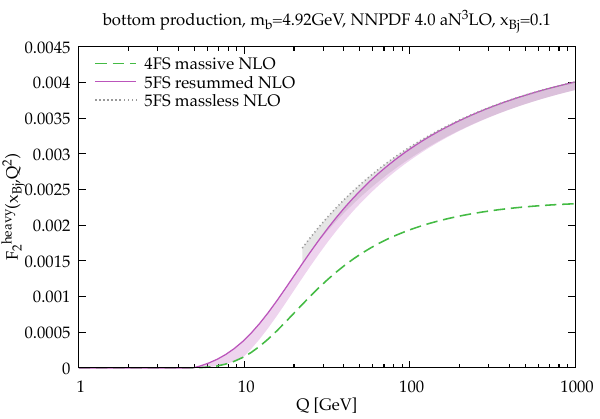}
  \includegraphics[width=0.328\textwidth, page= 5]{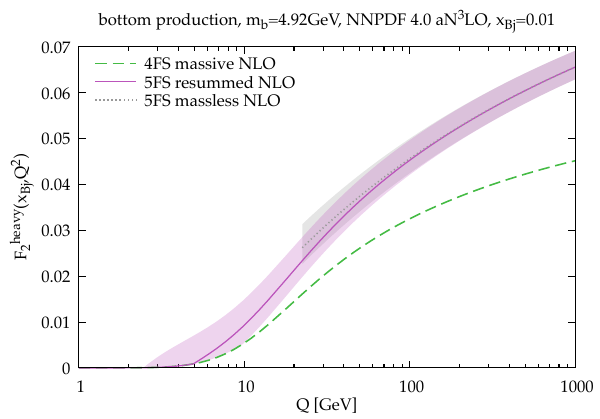}
  \includegraphics[width=0.328\textwidth, page= 6]{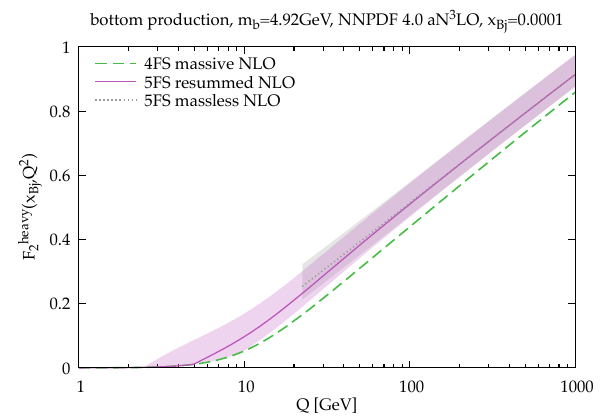}\\
  \includegraphics[width=0.328\textwidth, page= 7]{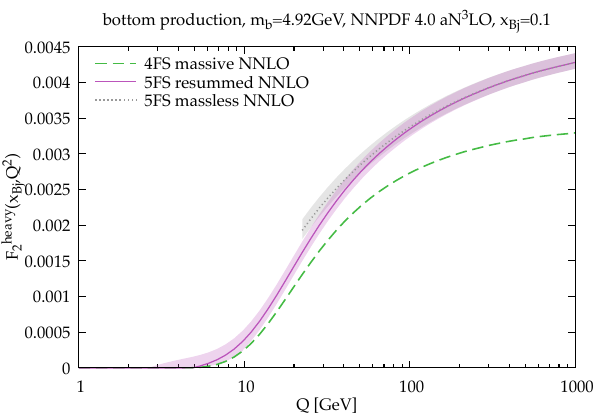}
  \includegraphics[width=0.328\textwidth, page= 8]{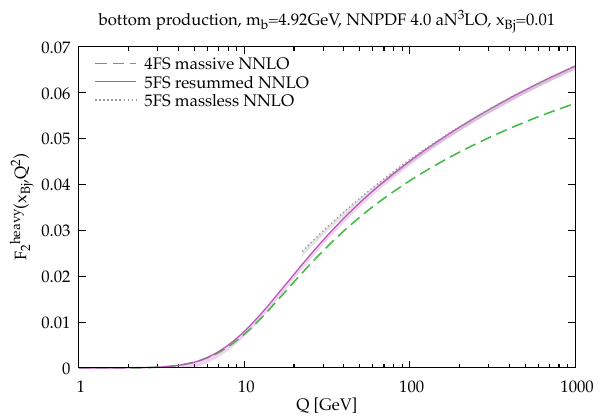}
  \includegraphics[width=0.328\textwidth, page= 9]{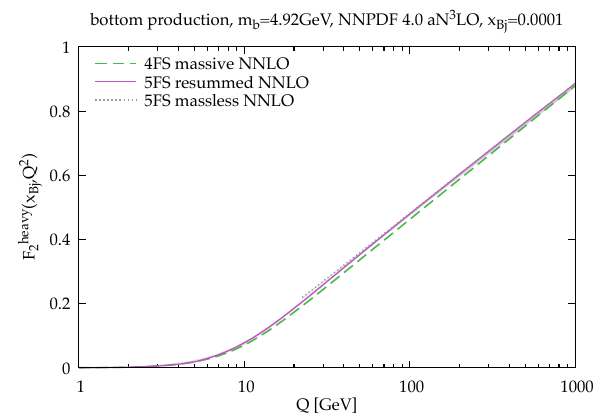}\\
  \includegraphics[width=0.328\textwidth, page=10]{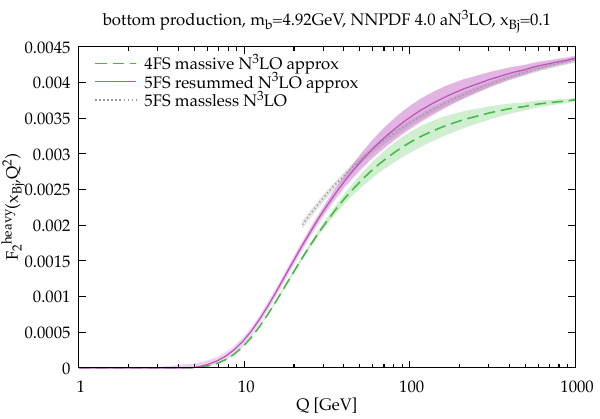}
  \includegraphics[width=0.328\textwidth, page=11]{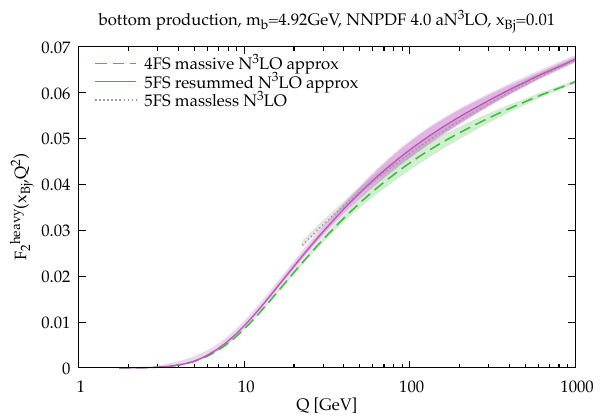}
  \includegraphics[width=0.328\textwidth, page=12]{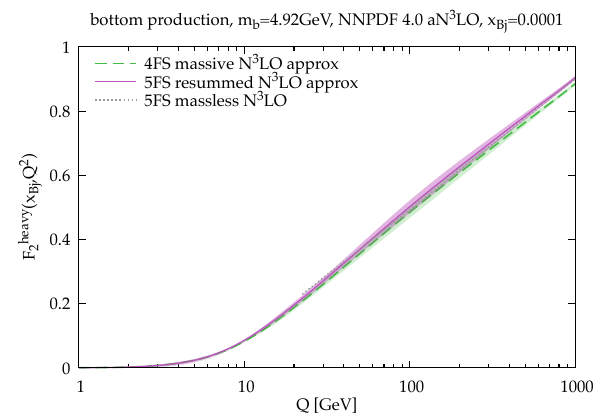}
  \caption{The structure function $F_2^{\rm heavy}$ at NLO (top), NNLO (middle) and N$^3$LO (bottom)
    computed in the fixed-order $n_f=4$ scheme (dashed green),
    the resummed $n_f=5$ scheme (solid purple) and, at high scales, its massless limit (dotted gray),
    at the representative values $\xB=10^{-1}, 10^{-2}, 10^{-4}$ (from left to right).
    For the $n_f=5$ results we also include an uncertainty band due to the variation of the matching scale $\muh$
    by a factor of 2 about its central value $\muh=m$.
    At N$^3$LO this uncertainty is summed in quadrature (lighter band) with the uncertainty of our approximation (darker band).}
  \label{fig:F2}
\end{figure}

\begin{figure}[tp]
  \centering
  \includegraphics[width=0.328\textwidth, page= 4]{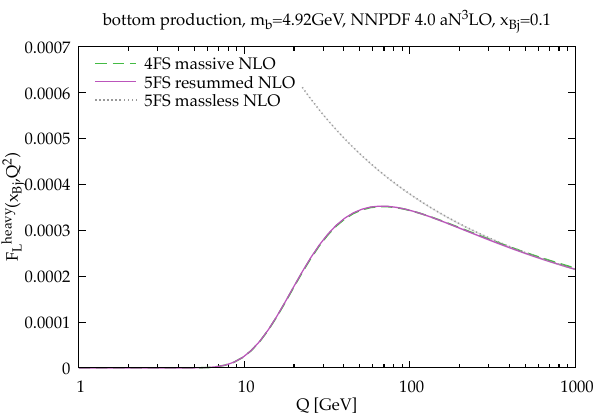}
  \includegraphics[width=0.328\textwidth, page= 5]{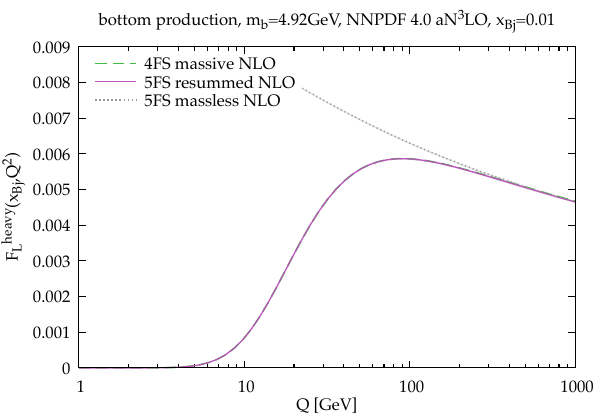}
  \includegraphics[width=0.328\textwidth, page= 6]{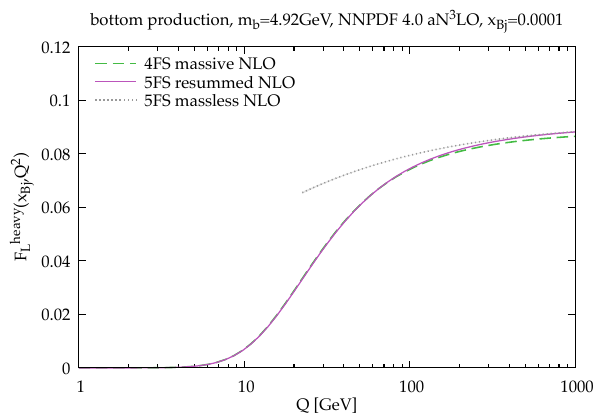}\\
  \includegraphics[width=0.328\textwidth, page= 7]{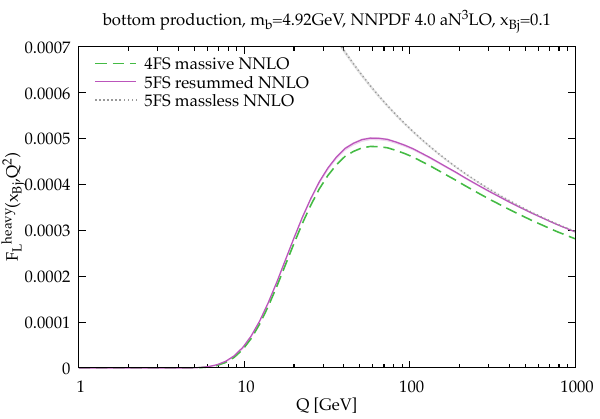}
  \includegraphics[width=0.328\textwidth, page= 8]{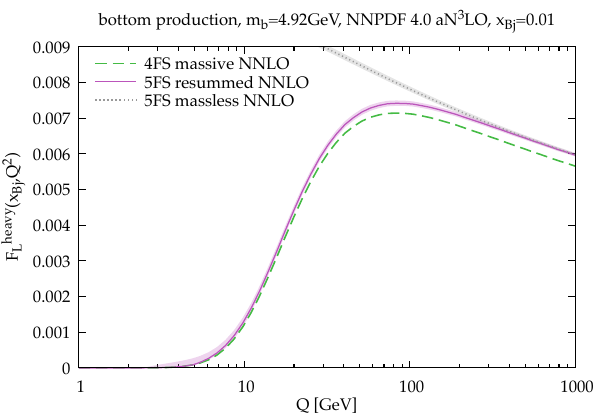}
  \includegraphics[width=0.328\textwidth, page= 9]{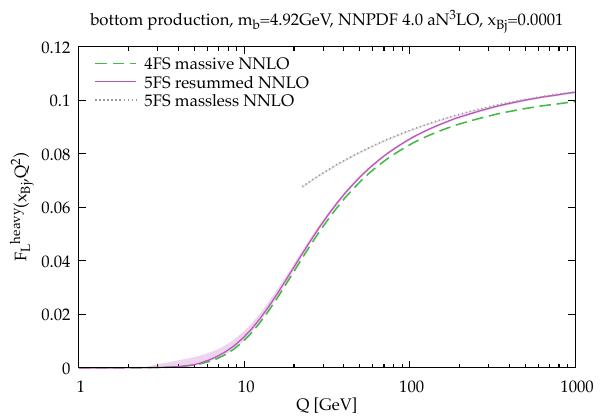}\\
  \includegraphics[width=0.328\textwidth, page=10]{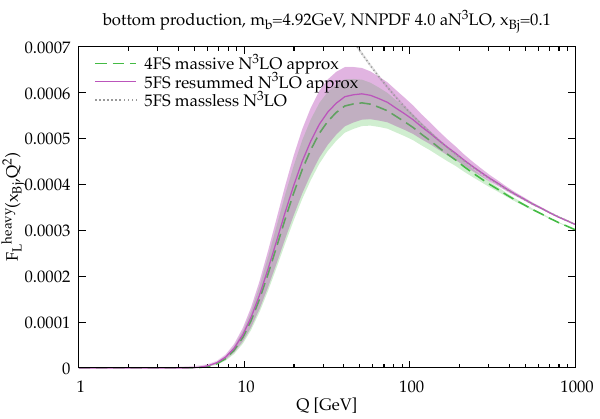}
  \includegraphics[width=0.328\textwidth, page=11]{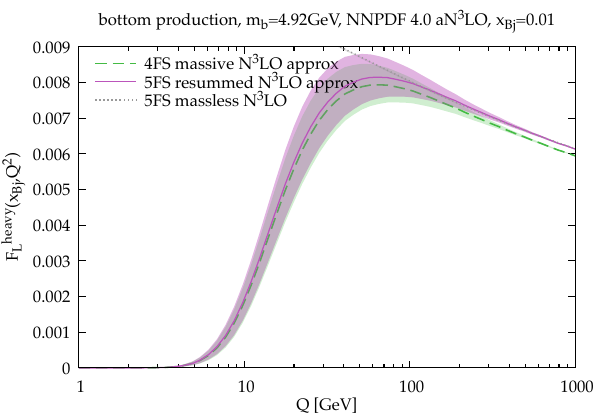}
  \includegraphics[width=0.328\textwidth, page=12]{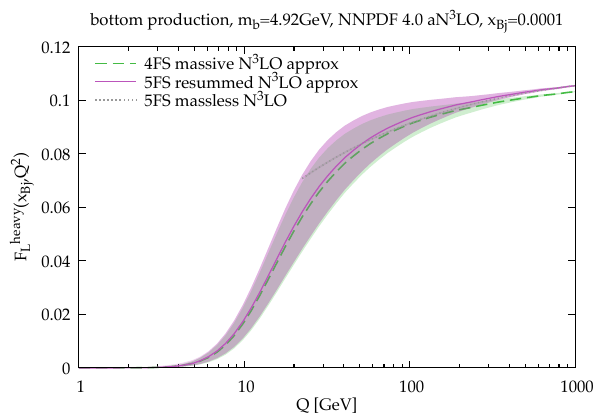}
  \caption{Same as figure~\ref{fig:F2}, but showing the results for the longitudinal structure function.}
  \label{fig:FL}
\end{figure}

We now move to the resummed result.
We first show separately the various orders, for each of which we report the fixed-order, resummed and massless contributions.
These are presented for $F_2^{\rm heavy}$ in figure~\ref{fig:F2} and for $F_L^{\rm heavy}$ in figure~\ref{fig:FL},
in a grid in which each row corresponds to the order (from NLO to N$^3$LO) and each column to the value of $\xB$
(as in figure~\ref{fig:F2FLfixedorder}).
For the sake of simplicity, we adopt the $Q^2\gtrsim m^2$ counting for the resummed result throughout the whole
scale range, namely we set the function $\chi(m^2/Q^2)$ to zero in Eq.~\eqref{eq:Fmixcounting}.
Moreover, for the best description of the transition region between fixed-order and resummed regimes,
we adopt the construction of the resummed result according to Eqs.~\eqref{eq:RESexp2NLO}, \eqref{eq:RESexp2NNLO}
and \eqref{eq:RESexp2N3LO}, as discussed in appendix~\ref{app:PDF}, which makes use of $n_f=5$ PDFs
constructed ad hoc (by re-evolving the NNPDF 4.0 aN3LO PDF set~\cite{NNPDF:2024nan} as we did for the construction of the $n_f=4$ set)
with different orders for the matching functions.
This construction guarantees a smooth transition at the matching scale $\muh$, without discontinuities.

\begin{figure}[tp]
  \centering
  \includegraphics[width=0.328\textwidth, page=1]{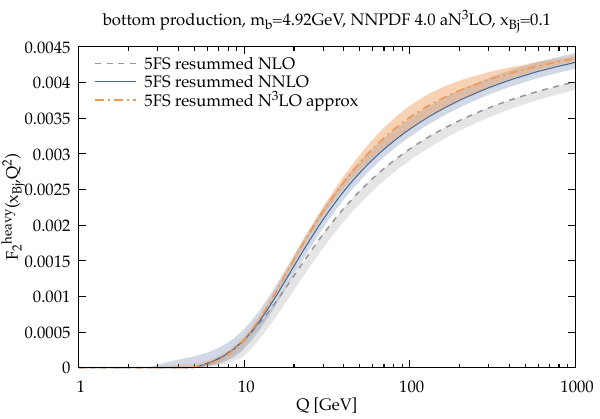}
  \includegraphics[width=0.328\textwidth, page=2]{images/DIS_F2_Res_paper.pdf}
  \includegraphics[width=0.328\textwidth, page=3]{images/DIS_F2_Res_paper.pdf}\\
  \includegraphics[width=0.328\textwidth, page=1]{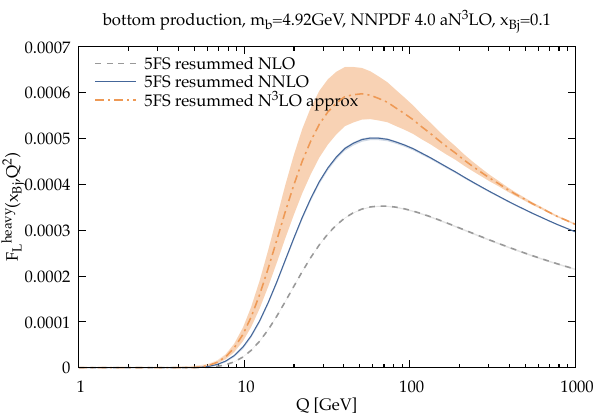}
  \includegraphics[width=0.328\textwidth, page=2]{images/DIS_FL_Res_paper.pdf}
  \includegraphics[width=0.328\textwidth, page=3]{images/DIS_FL_Res_paper.pdf}
  \caption{Same as figure~\ref{fig:F2FLfixedorder}
    but showing the results in the $n_f=5$ scheme (resummed result)
    at NLO (dashed gray), NNLO (solid blue) and N$^3$LO (dot-dashed orange).
    The uncertainty bands due to $\muh$ variations are also shown in the same fashion as in figures~\ref{fig:F2},~\ref{fig:FL}.}
  \label{fig:F2FLres}
\end{figure}

We immediately observe that in all plots the resummed result tends to its massless limit at high scales.
While this is obvious, it does provide an important and non-trivial cross check of the implementation.
At low scale the massless result is inadequate, so we plot it only for $Q>20$~GeV.
The comparison between the resummed $n_f=5$ result and the fixed-order $n_f=4$ result is more interesting.
We note that, for $F_2$, the difference between the two is large at NLO, especially at high $\xB$, and grows with increasing the scale.
Moving to NNLO and eventually to N$^3$LO, the difference is reduced, especially at low $x$.
This is an indication that both results are converging towards the same value, as they should, but the resummed result
does it faster (this will also be clear later from figure~\ref{fig:F2FLres}).
The transition from fixed order to resummation, happening at $Q=\muh$, is continuous, thanks to our construction,
but with a discontinuous derivative which is particularly noticeable at NLO.
Increasing the order, such discontinuity is reduced, becoming totally negligible at N$^3$LO, especially at low $x$.
The situation for $F_L$ is slightly different, due to the fact that in this case the effect of the resummation is very small.
At NLO, the fixed-order and resummed results are basically identical (except at very high scales),
and at the next orders there is some effect taking on as $Q$ increases but it is very small.
As a consequence, the transition from fixed order to resummation is very smooth in all configurations.

In each plot we also show an uncertainty on the resummed result obtained through a variation of the heavy-quark matching scale $\muh$,
as originally suggested in Refs.~\cite{Bonvini:2015pxa,Bonvini:2016fgf}.
In particular, the band is constructed by varying the scale by a factor of two up and down about its central value $\muh=m$.
This band is a measure of the scheme uncertainty, related in particular to the transition
from the $n_f$ scheme to the $n_f+1$ scheme.
As we can see this uncertainty is much larger for $F_2$ than for $F_L$ --- in the latter the band is almost invisible.
This is possibly due to a more effective cancellation of this uncertainty
in the longitudinal structure function, in turn due to the dominance of the gluon channel:
once the resummation is turned on, part of the gluon is converted into the heavy quark, but the gluon PDF itself does not change much,
so the net effect on the result is very small, as confirmed by the similarity of the resummed result with the fixed order.
For $F_2$ instead, the resummation plays a more important role, due to the dominance of the quark channel and thus
to the enhanced sensitivity to the heavy quark PDF, and consequently the result is more sensitive to
the value of the matching scale, leading to a larger band.
As the band represents a perturbative uncertainty, its size reduces by increasing the order, becoming very small at N$^3$LO for all values of $\xB$ considered.
At N$^3$LO, the fixed-order curve also shows the uncertainty associated to our approximation (darker inner band),
while the lighter outer band is the sum in quadrature of this uncertainty and the $\muh$ uncertainty.
We note that the uncertainty on our N$^3$LO approximation is by far the dominant one,
except at low scales, where the result (and its uncertainty band) is small anyway.

Finally, in figure~\ref{fig:F2FLres} we show the resummed result at the various orders,
in the same grid of structure functions and $\xB$ values used for figure~\ref{fig:F2FLfixedorder}.
Each curve is supplemented by the $\muh$ variation band, and the N$^3$LO result
also shows its sum in quadrature with the uncertainty due to our approximation.
By comparing these plots with those of figure~\ref{fig:F2FLfixedorder} we observe that
the perturbative convergence is very much improved for $F_2$, thanks to the large impact of the
mass logarithms to this observable, showing the importance of the resummation procedure and the reliability
of the results obtained.
For $F_L$, the resummation of mass logarithms is less substantial, and therefore the benefits obtained are less significant.
Yet, the availability of the resummed results for all the structure functions allows a consistent treatment of heavy quark
contributions to DIS, enabling a precise determination of PDFs at N$^3$LO accuracy.

\section{Conclusions}
\label{sec:conclusions}

In this work we have presented a careful construction of the DIS cross section that allows us to reach N$^3$LO accuracy,
as needed in particular for state-of-the-art PDF fits.
The focus of our work is on the treatment of heavy-quark mass dependence over a wide range of scales,
which spans from regions where power-behaving mass effects are important and collinear mass logarithms are small
to regions where these logarithms need to be resummed to all orders in $\as$ but power corrections are negligible.

One of the key aspects of our work is the construction of a variable flavour number scheme
with a proper $\as$ counting of the perturbative ingredients at the different energy scales.
Our approach reflects the one introduced in Ref.~\cite{Bonvini:2015pxa,Bonvini:2016fgf},
in which the main observation is that, close to the heavy-quark mass scale, the perturbative heavy-quark PDF
counts as an $\Ord(\as)$ object, impacting the way the various channels are combined together to
achieve a given perturbative accuracy.
This has the advantage of leading to smooth transitions across the heavy-quark matching scale,
as the results in the two sides,
corresponding to different factorization schemes with a different number of active flavours,
are compatible order by order in perturbation theory.
We also propose a way to smoothly transition from such a counting, which is appropriate close to the matching scale,
to a more standard counting (where the heavy-quark PDF is considered of order one) that is more suitable at higher scales.

The other main achievement of this work is a novel approximation for the yet unknown
$\ord3$ contribution to neutral-current DIS structure functions with full mass dependence.
Following the seminal work of Ref.~\cite{Kawamura:2012cr}, where such an approximation
has been constructed for $F_2$,
we have revisited the way the various ingredients are combined together and proposed
an improved construction which better reproduces the known $\ord2$ result that we used as a benchmark.
In addition to this, we have also used the recently completed determination of the matching functions
that were only partially known at the time of Ref.~\cite{Kawamura:2012cr},
thus providing a state-of-the-art prediction.
Moreover, we have extended this construction to the longitudinal structure function $F_L$,
for which the N$^3$LO massive result was not available in any form so far.

We have then presented representative results for the structure functions at N$^3$LO
obtained according to our construction. 
These allowed us to show that our construction of the resummation is reliable throughout the wide
hard-scale range considered and that our approximate N$^3$LO prediction is very reasonable.
In particular, for $F_2$ our prediction is rather accurate, with sufficiently small uncertainties,
while $F_L$ is affected by larger uncertainties.
The inclusion of the resummation of collinear mass logarithms leads to a fast converging perturbative expansion for $F_2$,
much faster than at fixed order,
while $F_L$ is affected by larger perturbative corrections, only mildly reduced by the inclusion of resummation.

The results of this work have already been used (in a preliminary form~\cite{Laurenti:2021akt})
for approximate N$^3$LO PDF fits by the NNPDF group~\cite{NNPDF:2024nan,Barontini:2024dyb,Ball:2025xgq},
and they are being used in forthcoming PDF fit studies by xFitter~\cite{Alekhin:2014irh,xFitter:2022zjb}
thanks to their implementation in \texttt{APFEL++}~\cite{Bertone:2013vaa, Bertone:2017gds}.
The approximate N$^3$LO coefficient functions for $F_2$ and $F_L$ are publicly available
through the \adani code at \href{https://github.com/niclaurenti/adani}{\texttt{https://github.com/niclaurenti/adani}}.

\acknowledgments
{
  We thank
  Valerio Bertone,
  Stefano Camarda,
  Stefano Forte,
  Francesco Giuli,
  Marco Guzzi,
  Felix Hekhorn,
  Giacomo Magni,
  Sven Moch,
  Tanjona Rabemananjara,
  Luca Rottoli
  for discussions and correspondence.
  The work of MB was supported by the Italian Ministry of University and Research (MUR) grant PRIN 2022SNA23K
  funded by the European Union -- Next Generation EU, Mission 4, Component 2, CUP I53D23001410006.
}

\FloatBarrier
\appendix

\section{\boldmath Explicit expressions for VFNS coefficient functions}
\label{app:DeltaC}

In this appendix we collect explicit order-by-order expressions for the VFNS coefficient functions.
We start considering Eq.~\eqref{eq:DeltaCdef}, which defines the power-suppressed massive contribution to the $(n_f+1)$-scheme
coefficient functions.
As explained in section~\ref{sec:VFNS}, the explicit form for these coefficients $\Delta C_{a,k}^{[n_f+1]}$
is obtained by inverting the matching functions $K_{ki}^{[n_f+1]\leftarrow [n_f]}$ on the subspace of light flavours.
In other words, calling $\tilde K_{ki}^{[n_f+1]\leftarrow [n_f]}$ its restriction to this subspace~\cite{Ball:2015dpa},
we have
\begin{align}\label{eq:DeltaC}
\Delta C_{a,k}^{[n_f+1]}\(\mQ,\muQ\)
  &= \sum_i^{n_f} \[C_{a,i}^{[n_f]}\(\mQ,\muQ\) - \sum_p^{n_f+1} C_{a,p}^{[n_f+1]}\(0,\muQ\) \otimes K_{pi}^{[n_f+1]\leftarrow [n_f]}\(\mmu\) \]
    \nonumber\\
  &\quad \otimes \tilde K_{ik}^{-1\,[n_f+1]\leftarrow [n_f]}\(\mmu\) , \qquad k=g,q,\bar q.
\end{align}
Note that, by definition, in the light subspace
\beq
\sum_i^{n_f} K_{pi}^{[n_f+1]\leftarrow [n_f]}\(\mmu\) \otimes \tilde K_{ik}^{-1\,[n_f+1]\leftarrow [n_f]}\(\mmu\) = \delta_{pk},
\qquad p,k=g,q,\bar q,
\eeq
times obviously an implicit $\delta(1-z)$ in the longitudinal variable.
This means that we can also write
\begin{align}\label{eq:DeltaCv2}
\Delta C_{a,k}^{[n_f+1]}\(\mQ,\muQ\)
  &= \sum_i^{n_f} \[C_{a,i}^{[n_f]}\(\mQ,\muQ\) - \sum_{p=h,\bar h} C_{a,p}^{[n_f+1]}\(0,\muQ\) \otimes K_{pi}^{[n_f+1]\leftarrow [n_f]}\(\mmu\) \]
    \nonumber\\
  &\quad \otimes \tilde K_{ik}^{-1\,[n_f+1]\leftarrow [n_f]}\(\mmu\) - C_{a,k}^{[n_f+1]}\(0,\muQ\), \qquad k=g,q,\bar q,
\end{align}
or equivalently, from Eq.~\eqref{eq:Cbardef},
\begin{align}\label{eq:Cbar}
\bar C_{a,k}^{[n_f+1]}\(\mQ,\muQ\)
  &= \sum_i^{n_f} \[C_{a,i}^{[n_f]}\(\mQ,\muQ\) - \sum_{p=h,\bar h} C_{a,p}^{[n_f+1]}\(0,\muQ\) \otimes K_{pi}^{[n_f+1]\leftarrow [n_f]}\(\mmu\) \]
    \nonumber\\
  &\quad \otimes \tilde K_{ik}^{-1\,[n_f+1]\leftarrow [n_f]}\(\mmu\), \qquad k=g,q,\bar q.
\end{align}
To limit the proliferation of terms, we shall now consider Eq.~\eqref{eq:Cbar} and present its expansion order by order.
The analogous expansion of $\Delta C_{a,k}^{[n_f+1]}$ is simply obtained from Eq.~\eqref{eq:DeltaCv2}
by subtracting the corresponding order of the massless coefficient $C_{a,k}^{[n_f+1]}$.

Before proceeding, we recall that it is natural to expect that coefficients in the $n_f$ scheme be computed
with $n_f$-scheme $\as$, and coefficients in the $n_f+1$ scheme be computed with $(n_f+1)$-scheme $\as$.\footnote
{The scheme-change matching functions are more ambiguous and there is no
  natural choice for the scheme to be used for $\as$, since these functions live at the border between the two schemes.
  Therefore, any choice is reasonable, but of course the coefficients of the expansion must be consistent with that choice.
}
This implies that a \emph{renormalization} scheme change to write all the expansions in terms of the same $\as$ is needed
before combining and comparing the expansions.
In the following we do not specify the actual choice of renormalization scheme for $\as$, but we give now the transformation rules
needed to apply the scheme change.
The relation between the strong coupling in the two schemes is~\cite{Chetyrkin:1997sg}
\begin{align}\label{eq:alphasmatching}
  \frac{\as^{[n_f+1]}(\mu^2)}{\as^{[n_f]}(\mu^2)}
  &= 1 + \frac{\as^{[n_f]}(\mu^2)}{6\pi} \log\frac{\mu^2}{m^2}
    + \frac{\as^2(\mu^2)}{\pi^2} \[\frac7{24} + \frac{19}{24}\log\frac{\mu^2}{m^2} + \frac1{36}\log^2\frac{\mu^2}{m^2}\] + \ord3
+\ldots
\end{align}
where $m$ is the heavy quark pole mass.
In the $\ord2$ term we did not specify the scheme of $\as$ as the difference affects higher orders only.
Starting from a perturbative expansion in $\as^{[n_f+1]}$, we can use Eq.~\eqref{eq:alphasmatching} to convert it
to an expansion in $\as^{[n_f]}$ as
\begin{align}
  A(\as)
  &= A_0 + A_1 \as^{[n_f+1]} + A_2 \(\as^{[n_f+1]}\)^2  + A_3 \(\as^{[n_f+1]}\)^3 + ... \nonumber\\
  &= A_0 + A_1 \as^{[n_f]} + \[A_2+\frac{A_1}{6\pi}\log\frac{\mu^2}{m^2}\] \(\as^{[n_f]}\)^2 \nonumber\\
  &\quad + \[A_3+\frac{A_2}{3\pi}\log\frac{\mu^2}{m^2}
    +\frac{A_1}{\pi^2}\(\frac7{24} + \frac{19}{24}\log\frac{\mu^2}{m^2} + \frac1{36}\log^2\frac{\mu^2}{m^2}\)\] \(\as^{[n_f]}\)^3 + ...
\end{align}
for a generic quantity $A$.
Similarly, if we start from an expansion in powers of $\as^{[n_f]}$ we can convert it to powers of $\as^{[n_f+1]}$
using the inverse of Eq.~\eqref{eq:alphasmatching}. As it is straightforward to obtain analogous expressions,
we do not write them down explicitly.

Assuming now that all the coefficients have been converted to the same renormalization scheme
and thus use the same $\as$, we can write down the expansions.
To improve readability, we omit the arguments and the scheme labels, using the identifications
\begin{align}
  \bar C_{k} &\to \bar C_{a,k}^{[n_f+1]}\(z,\mQ,\muQ\), \\
  C_{k} &\to
            \begin{cases}
              C_{a,k}^{[n_f+1]}\(z,0,\muQ\) & k=h,\bar h, \\
              C_{a,k}^{[n_f]}\(z,\mQ,\muQ\) & k=g,q,\bar q,
            \end{cases}
                                               \\
  K_{ij} &\to K_{ij}^{[n_f+1]\leftarrow [n_f]}\(z,\mmu\).
\end{align}
We find
\begin{subequations}\label{eq:CbarofCExp}
\begin{align}
  \bar C_{g}^{(1)}
  &= C_{g}^{(1)} -2C_{h}^{(0)} \otimes K_{hg}^{(1)},
  \\
  \bar C_{g}^{(2)}
  &= C_{g}^{(2)} - C_{g}^{(1)} \otimes K_{gg}^{(1)}
    -2C_{h}^{(0)}\otimes \[K_{hg}^{(2)}-K_{hg}^{(1)}\otimes K_{gg}^{(1)}\]
    -2C_{h}^{(1)}\otimes K_{hg}^{(1)},
    \\
  \bar C_{q}^{(2)}
  &= C_{q}^{(2)} - 2C_{h}^{(0)}\otimes K_{hq}^{(2)},
    \\
  \bar C_{g}^{(3)}
  &= C_{g}^{(3)} - C_{g}^{(2)}\otimes K_{gg}^{(1)}
    - C_{g}^{(1)}\otimes \[K_{gg}^{(2)}-K_{gg}^{(1)}\otimes K_{gg}^{(1)}\]
    \nonumber\\
  &\quad -2C_{h}^{(0)}\otimes \[K_{hg}^{(3)}-K_{hg}^{(2)}\otimes K_{gg}^{(1)}
    -K_{hg}^{(1)}\otimes \(K_{gg}^{(2)}- K_{gg}^{(1)} \otimes K_{gg}^{(1)}\)\]
    \nonumber\\
  &\quad -2C_{h}^{(1)}\otimes \[K_{hg}^{(2)}-K_{hg}^{(1)}\otimes K_{gg}^{(1)}\]
    -2C_{h}^{(2)}\otimes K_{hg}^{(1)},
    \\
  \bar C_{q}^{(3)}
  &= C_{q}^{(3)} 
    -C_{g}^{(1)}\otimes K_{gq}^{(2)}
    -2C_{h}^{(0)}\otimes \[K_{hq}^{(3)}-K_{hg}^{(1)}\otimes K_{gq}^{(2)}\]
    -2C_{h}^{(1)}\otimes K_{hq}^{(2)}
    ,
\end{align}
\end{subequations}
which are valid both for $a=2,L$. In fact, for the longitudinal structure function $a=L$,
these results further simplify because $C_{h}^{(0)}=0$.
Note that in the equation above, rather than summing over $h$ and $\bar h$, we have doubled
each $h$ contribution, owing to the fact that in our case $C_{h}=C_{\bar h}$ and $K_{hk}=K_{\bar hk}$ for $k=g,q,\bar q$.
We have also written expressions for $C_q$ only, but those for $C_{\bar q}$ are identical up the change $q\to\bar q$.

The equations above can be inverted to obtain the (massive) coefficient function in the $n_f$ scheme
in terms of those in the $n_f+1$ scheme,
\begin{subequations}\label{eq:CofCbarExp}
\begin{align}
  C_{g}^{(1)}
  &= \bar C_{g}^{(1)} + 2C_{h}^{(0)} \otimes K_{hg}^{(1)},
  \\
  C_{g}^{(2)}
  &= \bar C_{g}^{(2)} + \bar C_{g}^{(1)} \otimes K_{gg}^{(1)}
    +2C_{h}^{(0)}\otimes K_{hg}^{(2)}
    +2C_{h}^{(1)}\otimes K_{hg}^{(1)},
    \\
  C_{q}^{(2)}
  &= \bar C_{q}^{(2)} + 2C_{h}^{(0)}\otimes K_{hq}^{(2)},
    \\
  C_{g}^{(3)}
  &= \bar C_{g}^{(3)} + \bar C_{g}^{(2)} \otimes K_{gg}^{(1)}
    + \bar C_{g}^{(1)} \otimes K_{gg}^{(2)}
    +2C_{h}^{(0)}\otimes K_{hg}^{(3)}
    +2C_{h}^{(1)}\otimes K_{hg}^{(2)}
    +2C_{h}^{(2)}\otimes K_{hg}^{(1)},
  \\
  C_{q}^{(3)}
  &= \bar C_{q}^{(3)} 
    +\bar C_{g}^{(1)}\otimes K_{gq}^{(2)}
    +2C_{h}^{(0)}\otimes K_{hq}^{(3)}
    +2C_{h}^{(1)}\otimes K_{hq}^{(2)}
    .
\end{align}
\end{subequations}
Note that in the right-hand side of this equation all the coefficients could be written as $\bar C$,
since $C_{a,k}^{[n_f+1]}=\bar C_{a,k}^{[n_f+1]}$ for $k=h,\bar h$.
These equations are the direct expansion of Eq.~\eqref{eq:Csc} once the power corrections are included
in the $n_f+1$ coefficients, namely
\beq\label{eq:Cbarsc}
C_{a,i}^{[n_f]}\(\mQ,\muQ\) = \sum_k^{n_f+1} \bar C_{a,k}^{[n_f+1]}\(\mQ,\muQ\) \otimes K_{ki}^{[n_f+1]\leftarrow[n_f]}\(\mmu\),
\eeq
which can be derived from the direct comparison of Eq.~\eqref{eq:matched2} with Eq.~\eqref{eq:main},
or by inverting Eq.~\eqref{eq:Cbar}.
Eqs.~\eqref{eq:CofCbarExp} also provide the high-scale limit of the massive $n_f$-scheme coefficients,
Eq.~\eqref{eq:Chighscale}, simply obtained by replacing $\bar C_{a,k}^{[n_f+1]}$ with their massless limits $C_{a,k}^{[n_f+1]}$.

We finally notice that using Eq.~\eqref{eq:Chighscale} it is possible to rewrite Eq.~\eqref{eq:DeltaC}
in the form
\begin{align}\label{eq:DeltaCv3}
\Delta C_{a,k}^{[n_f+1]}\(\mQ,\muQ\)
  &= \sum_i^{n_f} \[C_{a,i}^{[n_f]}\(\mQ,\muQ\) - C_{a,i}^{[n_f]\rm h.s.}\(0,\muQ\) \]
    \otimes \tilde K_{ik}^{-1\,[n_f+1]\leftarrow [n_f]}\(\mmu\) ,
\end{align}
out of which we can easily write the $\Delta C$ coefficients (or similarly the $\bar C$ coefficients)
in a form that is simpler than Eq.~\eqref{eq:CbarofCExp}, namely
\begin{subequations}\label{eq:DeltaCofCExp}
\begin{align}
  \Delta C_{g}^{(1)}
  &= C_{g}^{(1)} -C_{g}^{(1)\rm h.s.},
  \\
  \Delta C_{g}^{(2)}
  &= C_{g}^{(2)}-C_{g}^{(2)\rm h.s.} - \[C_{g}^{(1)}-C_{g}^{(1)\rm h.s.}\] \otimes K_{gg}^{(1)},
    \\
  \Delta C_{q}^{(2)}
  &= C_{q}^{(2)}-C_{q}^{(2)\rm h.s.},
    \\
  \Delta C_{g}^{(3)}
  &= C_{g}^{(3)}-C_{g}^{(3)\rm h.s.} - \[C_{g}^{(2)}-C_{g}^{(2)\rm h.s.}\]\otimes K_{gg}^{(1)}
    - \[C_{g}^{(1)}-C_{g}^{(1)\rm h.s.}\]\otimes \[K_{gg}^{(2)}-K_{gg}^{(1)}\otimes K_{gg}^{(1)}\],
    \\
  \Delta C_{q}^{(3)}
  &= C_{q}^{(3)} -C_{q}^{(3)\rm h.s.}
    -\[C_{g}^{(1)}-C_{g}^{(1)\rm h.s.}\]\otimes K_{gq}^{(2)}
    .
\end{align}
\end{subequations}

\section{Construction of PDFs with proper power counting}
\label{app:PDF}

We have seen in section~\ref{sec:PCres} that for scales above the
heavy-quark scale but not too high, the heavy-quark PDF should count
as an $\Ord(\as)$ object.
This observation leads to Eq.~\eqref{eq:RESexp2}, that we report here for convenience:
\begin{align}\label{eq:RESexp2xx}
  \frac1{\xB} F_a^{\rm heavy}(\xB,Q^2)
  &= \sum_{k=\he} \Cb_{a,k}^{[n_f+1](0)} \otimes f_{k}^{[n_f+1]}
  + \as \Cb_{a,g}^{[n_f+1](1)} \otimes f_{g}^{[n_f+1]}
  &&\text{(NLO)}\nonumber\\
  &+ \as\sum_{k=\he} \Cb_{a,k}^{[n_f+1](1)} \otimes f_{k}^{[n_f+1]}
  + \as^2 \sum_{k=\li} \Cb_{a,k}^{[n_f+1](2)} \otimes f_{k}^{[n_f+1]}
  &&\text{(NNLO)}\nonumber\\
  &+ \as^2\sum_{k=\he} \Cb_{a,k}^{[n_f+1](2)} \otimes f_{k}^{[n_f+1]}
  + \as^3 \sum_{k=\li} \Cb_{a,k}^{[n_f+1](3)} \otimes f_{k}^{[n_f+1]}
  &&\text{(N$^3$LO)}\nonumber\\
  &+\Ord(\as^4).
\end{align}
This result shall be used for scales $Q>\muh$.
For $Q<\muh$, the fixed-order counting applies, Eq.~\eqref{eq:FOexp}, that we also report with a simplified notation:
\begin{align}\label{eq:FOexpxx}
  \frac1{\xB} F_a^{\rm heavy}(\xB,Q^2)
  &= \as C_{a,g}^{[n_f](1)} \otimes f_{g}^{[n_f]} &&\text{(NLO)}\nonumber\\
  &+ \as^2 \sum_{i}^{n_f} C_{a,i}^{[n_f](2)} \otimes f_{i}^{[n_f]} &&\text{(NNLO)}\nonumber\\
  &+ \as^3 \sum_{i}^{n_f} C_{a,i}^{[n_f](3)} \otimes f_{i}^{[n_f]} &&\text{(N$^3$LO)}\nonumber\\
  &+\Ord(\as^4).
\end{align}
Ideally, we would like to two results to be identical at the matching scale $Q=\muh$,
so that the structure functions are continuous curves across the whole $Q$ range.

We already know that formally the two results are equivalent, by construction,
thanks to Eq.~\eqref{eq:fsc} and Eq.~\eqref{eq:Cbarsc}.
However, this all-order equivalence does not guarantee that at a given finite order the two expressions
give identical results.
In particular, once writing both expressions in terms of a common set of PDFs, there may be higher order interference terms
that contaminate the finite-order result giving discontinuous contribution at the matching scale.
We thus now want to focus on how one can guarantee continuity at the matching scale $\muh$ at each finite order.

Let us then consider both the $n_f$-scheme result Eq.~\eqref{eq:FOexpxx} and the $(n_f+1)$-scheme result Eq.~\eqref{eq:RESexp2xx}
at the scale $Q=\muh$.\footnote
{What we actually want is to have the factorization scale $\mu=\muh$.
  In order to avoid any complication, we just assume that $\mu=Q$, and focus on $Q=\muh$.}
For a direct comparison, we shall convert the $(n_f+1)$-scheme PDF into the $n_f$-scheme ones at the scale $\muh$,
using Eq.~\eqref{eq:fsc} that order by order reads
\begin{subequations}\label{eq:PDFexp}
\begin{align}
  f_g^{[n_f+1]}\label{eq:PDFexp-g}
  &= f_{g}^{[n_f]}
  + \as K_{gg}^{(1)}\otimes f_{g}^{[n_f]}
  + \as^2 \sum_{k=g,q,\bar q} K_{gk}^{(2)} \otimes f_{k}^{[n_f]}
    +...\\
  f_q^{[n_f+1]}\label{eq:PDFexp-q}
  &= f_{q}^{[n_f]}
  + \as^2 \sum_{k=g,q,\bar q} K_{qk}^{(2)} \otimes f_{k}^{[n_f]}
    +...\\
  f_h^{[n_f+1]}\label{eq:PDFexp-h}
  &= \as K_{hg}^{(1)} \otimes f_{g}^{[n_f]}
  + \as^2 \sum_{k=g,q,\bar q} K_{hk}^{(2)} \otimes f_{k}^{[n_f]}
  + \as^3 \sum_{k=g,q,\bar q} K_{hk}^{(3)} \otimes f_{k}^{[n_f]}
    +...
\end{align}
\end{subequations}
having omitted the arguments and the scheme labels in the matching functions for better readability,
and showing only the PDFs for light quark $q$ and heavy quark $h$, as those for the anti-quarks are identical.
The goal here is to understand which terms must be included in the PDFs appearing in Eq.~\eqref{eq:RESexp2xx}
to match exactly Eq.~\eqref{eq:FOexpxx} at a given order.

To do so, we note that in terms of the $n_f$-scheme PDFs the result of Eq.~\eqref{eq:FOexpxx}
contains at N$^n$LO all and only the perturbative contributions up to $\as^n$.
In principle we could write them in terms of the $(n_f+1)$-scheme coefficients $\bar C$
for a direct comparison with Eq.~\eqref{eq:RESexp2xx}, using Eqs.~\eqref{eq:CofCbarExp},
but we already know that by construction the results must be the same,
and this operation does not change the powers of $\as$ appearing in each row of Eq.~\eqref{eq:FOexpxx},
which is our focus now.
Therefore, to guarantee continuity at $Q=\muh$, we just need to make sure that the
result Eq.~\eqref{eq:RESexp2xx}, when written in terms of the $n_f$-scheme PDFs,
produces at N$^n$LO all and \emph{only} terms up to $\as^n$.

Before drawing the general conclusion, let us look explicitly at the first couple of orders.
At NLO, we need the first line of Eq.~\eqref{eq:RESexp2xx} to produce all terms at order $\as$,
and nothing else, when rewriting it in terms of $n_f$-scheme PDFs.
This is achieved by converting the gluon at $\Ord(\as^0)$ and the heavy quark at $\Ord(\as)$,
corresponding to the first term in each line of Eqs.~\eqref{eq:PDFexp} (the light-quark PDFs do not play any role here).
Note that this is not equivalent to including all $\Ord(\as)$ matching, as this would also
introduce the second term of Eq.~\eqref{eq:PDFexp-g} which would generate a term of $\Ord(\as^2)$
from the first line of Eq.~\eqref{eq:RESexp2xx}.

This observation can be easily generalised to higher orders. Indeed, we see that in each line of
Eq.~\eqref{eq:RESexp2xx} the terms multiplying a light-quark PDF have always an extra explicit power of $\as$
with respect to the terms multiplying a heavy-quark PDF.
This immediately implies that, to produce the same powers of $\as$ from each line,
the matching functions $K_{jk}$ must be included at different orders depending whether $j$ is light or heavy.
Specifically, the light-to-light matching ($j=g,q,\bar q$) must be included at one order less
than the light-to-heavy matching ($j=h,\bar h$).

This is not the end of the story though. Indeed, if we move to NNLO, we see that it is not sufficient
to construct the $(n_f+1)$-scheme PDFs including $K_{jk}$ at $\Ord(\as^2)$ for $j=h,\bar h$ and at $\Ord(\as)$ for $j=g,q,\bar q$.
Indeed, this choice does produce the correct $\Ord(\as^2)$ terms from the first line of Eq.~\eqref{eq:RESexp2xx},
but it also produces spurious $\Ord(\as^3)$ terms from the second line of the same equation.
So, the correct way to proceed is to use $(n_f+1)$-scheme PDFs constructed with
matching functions at this order only for the first line, while for the second line
the order of each matching function must be lower by a unity (and then be equal to the PDFs used at NLO).
In this way, all and only terms up to $\Ord(\as^2)$ are generated by the first two lines of Eq.~\eqref{eq:RESexp2xx},
thus matching exactly the first two lines of Eq.~\eqref{eq:FOexpxx}.

\begin{table}[t]
  \centering
  \begin{tabular}{lcc}
    & \multicolumn{2}{c}{max order in $K_{jk}$ with} \\
    PDF matching order & $j=h,\bar h$ & $j=g,q,\bar q$ \\
    \midrule
    NLO$_{\rm m}$ & $\as$ & $\as^0$ \\
    NNLO$_{\rm m}$ & $\as^2$ & $\as$ \\
    N$^3$LO$_{\rm m}$ & $\as^3$ & $\as^2$
  \end{tabular}
  \caption{Proper orders in the matching functions $K_{jk}$ to be included in the construction of the PDFs at a given matching order,
    denoted N$^n$LO$_{\rm m}$.}
  \label{tab:PDFmatching}
\end{table}

We are now ready to generalise our findings.
First of all, for a proper power counting that guarantees continuity it is necessary to include
the light-to-light matching functions at one order less than the light-to-heavy matching functions.
We shall thus construct PDFs at a given ``matching order'' according to table~\ref{tab:PDFmatching}.
Secondly, such PDFs shall be used at different matching orders in the different lines of Eq.~\eqref{eq:RESexp2xx},
depending on the desired order for the structure function.
Specifically, at NLO we have
\begin{align}\label{eq:RESexp2NLO}
  \frac{F_a^{\rm heavy,\,NLO}(\xB,Q^2)}{\xB}
  &= \sum_{k=\he} \Cb_{a,k}^{[n_f+1](0)} \otimes f_{k}^{[n_f+1],\rm NLO_m}
  + \as \Cb_{a,g}^{[n_f+1](1)} \otimes f_{g}^{[n_f+1],\rm NLO_m},
\end{align}
at NNLO we have
\begin{align}\label{eq:RESexp2NNLO}
  \frac{F_a^{\rm heavy,\,NNLO}(\xB,Q^2)}{\xB}
  &= \sum_{k=\he} \Cb_{a,k}^{[n_f+1](0)} \otimes f_{k}^{[n_f+1],\rm NNLO_m}
    + \as \Cb_{a,g}^{[n_f+1](1)} \otimes f_{g}^{[n_f+1],\rm NNLO_m}
    \\
  &+ \as\sum_{k=\he} \Cb_{a,k}^{[n_f+1](1)} \otimes f_{k}^{[n_f+1],\rm NLO_m}
  + \as^2 \sum_{k=\li} \Cb_{a,k}^{[n_f+1](2)} \otimes f_{k}^{[n_f+1],\rm NLO_m},\nonumber
\end{align}
and finally at N$^3$LO we have
\begin{align}\label{eq:RESexp2N3LO}
  \frac{F_a^{\rm heavy,\,N^3LO}(\xB,Q^2)}{\xB}
  &= \sum_{k=\he} \Cb_{a,k}^{[n_f+1](0)} \otimes f_{k}^{[n_f+1],\rm N^3LO_m}
    + \as \Cb_{a,g}^{[n_f+1](1)} \otimes f_{g}^{[n_f+1],\rm N^3LO_m}
    \\
  &+ \as\sum_{k=\he} \Cb_{a,k}^{[n_f+1](1)} \otimes f_{k}^{[n_f+1],\rm NNLO_m}
    + \as^2 \sum_{k=\li} \Cb_{a,k}^{[n_f+1](2)} \otimes f_{k}^{[n_f+1],\rm NNLO_m}\nonumber
    \\
  &+ \as^2\sum_{k=\he} \Cb_{a,k}^{[n_f+1](2)} \otimes f_{k}^{[n_f+1],\rm NLO_m}
  + \as^3 \sum_{k=\li} \Cb_{a,k}^{[n_f+1](3)} \otimes f_{k}^{[n_f+1],\rm NLO_m}, \nonumber
\end{align}
where the label N$^n$LO$_{\rm m}$ refers to the order of the matching functions according to table~\ref{tab:PDFmatching}.
This construction has been already proposed and used in Ref.~\cite{Bonvini:2015pxa},
where indeed it leads to the sought continuity at the matching scale.
To our knowledge, no other applications of this construction exist.

Note that in the whole discussion so far we have not mentioned the order of the evolution.
From the point of view of continuity, the evolution does not matter, so any order can be used without affecting our conclusions.
Of course, it is natural to use evolution at an order that matches the order of the computation,
even though in principle one could use a higher order evolution without changing the overall accuracy.
What is important to stress is that in the construction of the result at a given order,
all PDFs are evolved at the same perturbative accuracy, even if the matching order changes between different contributions.
So, for instance, in Eq.~\eqref{eq:RESexp2N3LO}, the evolution must be performed at N$^3$LL
for all PDFs at NLO$_{\rm m}$, NNLO$_{\rm m}$ and N$^3$LO$_{\rm m}$.

Practically, this implies that different PDF sets are needed for computing a result at a given order.
This might seem surprising and inconvenient, especially for users used to access PDFs from LHAPDF~\cite{Buckley:2014ana}.
However, we need to stress that what we suggest to modify is a perturbative part of the PDFs,
while we leave untouched the actual nonperturbative input, which is the PDFs at the initial (low) scale, where they are fitted.
For practical reasons, LHAPDF grids contain also the perturbative part, such that one can quickly access PDFs at any scale.
Therefore, for convenience, different sets evolved from the same nonperturbative initial condition
with the same order for the evolution but different matching oders can be constructed and provided
in LHAPDF format, and then used in the construction of theoretical predictions according to Eqs.~\eqref{eq:RESexp2NLO}--\eqref{eq:RESexp2NLO},
which is what we did for producing the resummed plots of section~\ref{sec:VFNS-N3LO}.
For the goal of fitting PDFs, the evolution and matching is treated as a part of the theoretical prediction,
so in this case implementing the construction that we suggest should not be difficult.

As a final remark, we stress that a more conventional construction in which the same matching order is used
for all terms (for instance N$^3$LO$_{\rm m}$ for all PDFs in the N$^3$LO result) remains a reasonable
option, that has no formal problems.
The only limit is that such a result would present a discontinuity at $Q=\muh$,
which however should become smaller and smaller increasing the order.

\section{More on the approximate coefficients}
\label{sec:app:coeff-func}

In this appendix we collect results for the approximate coefficient functions at NNLO and N$^3$LO
for the quark-channel $F_2$ structure function and for both gluon- and quark-channel $F_L$ structure function.

Let us start from $F_2$. The quark-channel coefficient is constructed similarly to the gluon-channel one,
with the difference that there is no threshold contribution because the quark channel is strongly suppressed at threshold.
Therefore the approximation is based on the asymptotic part only,
constructed in the very same way that we have discussed in section~\ref{sec:approx},
supplemented by its own damping function according to Eq.~\eqref{eq:approx}.

\begin{figure}[tp]
  \centering
  \includegraphics[width=0.49\textwidth]{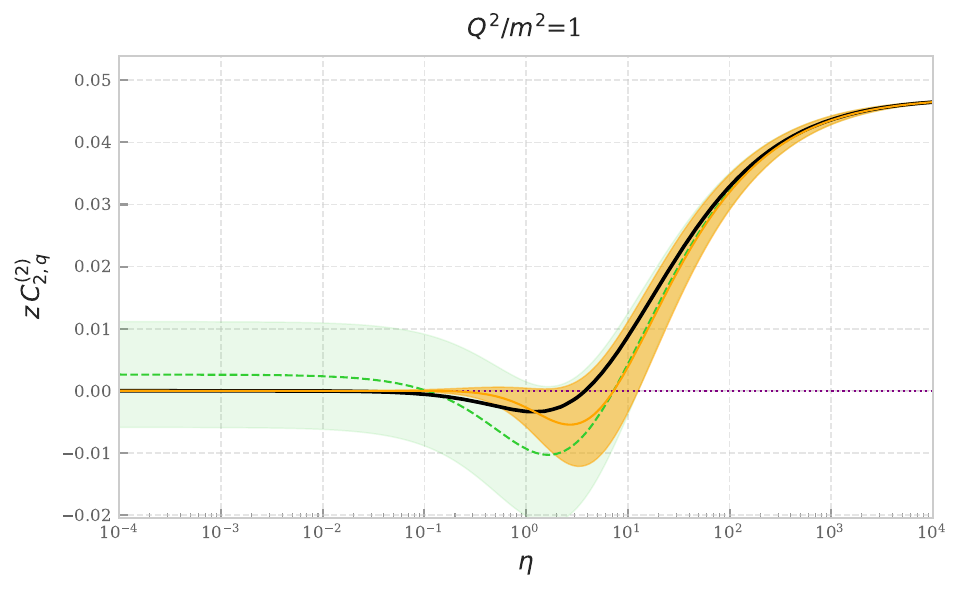}
  \includegraphics[width=0.49\textwidth]{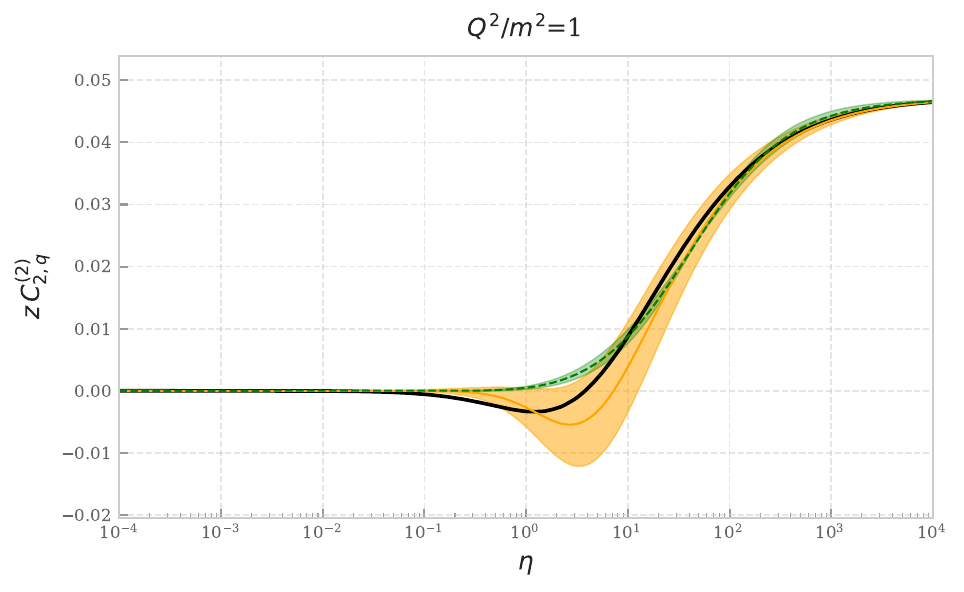}
  \includegraphics[width=0.49\textwidth]{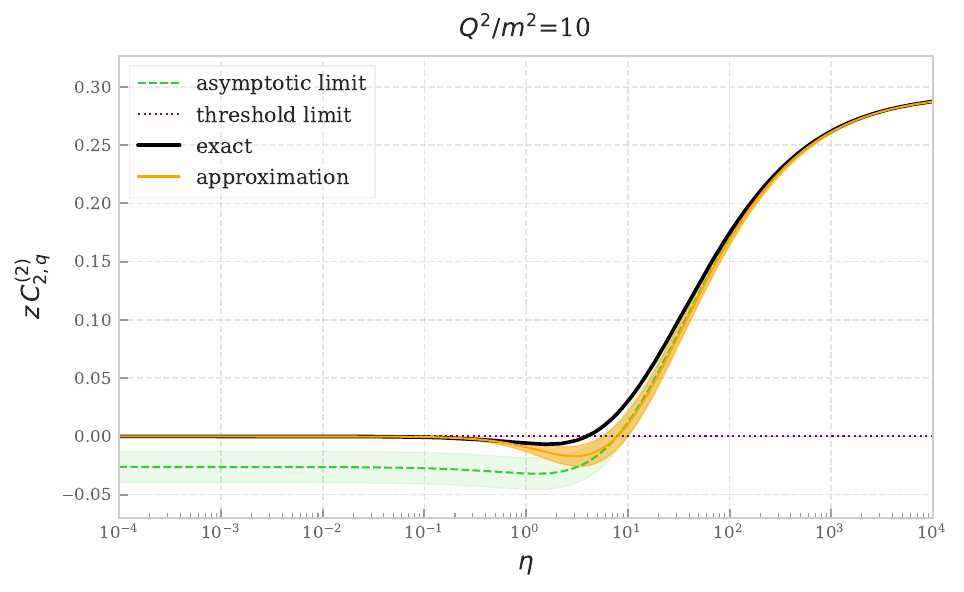}
  \includegraphics[width=0.49\textwidth]{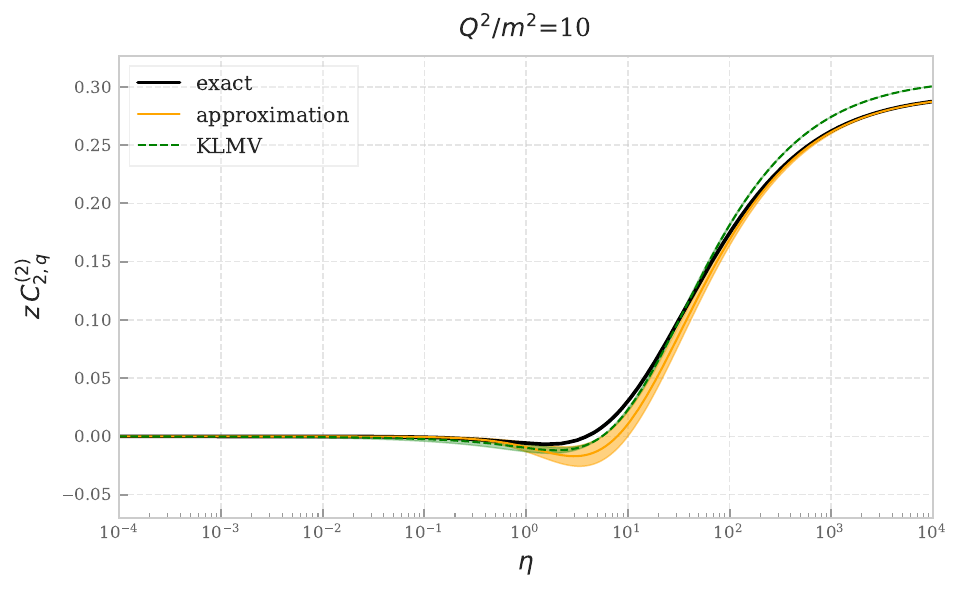}
  \includegraphics[width=0.49\textwidth]{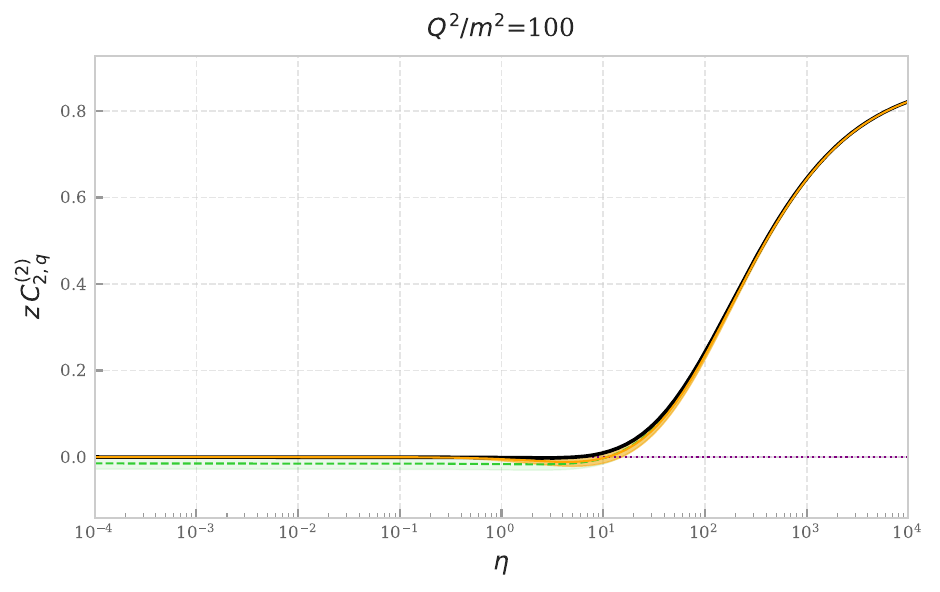}
  \includegraphics[width=0.49\textwidth]{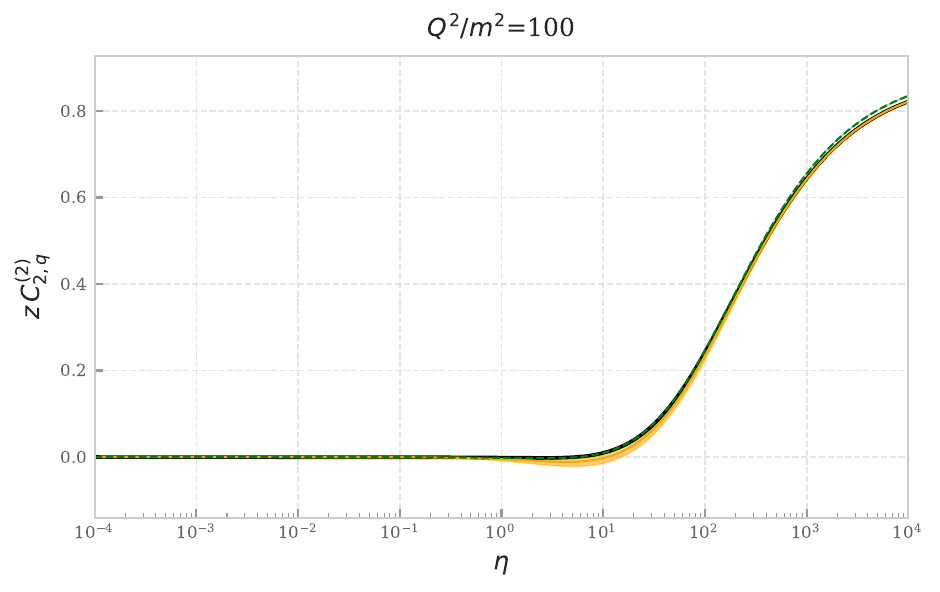}
  \includegraphics[width=0.49\textwidth]{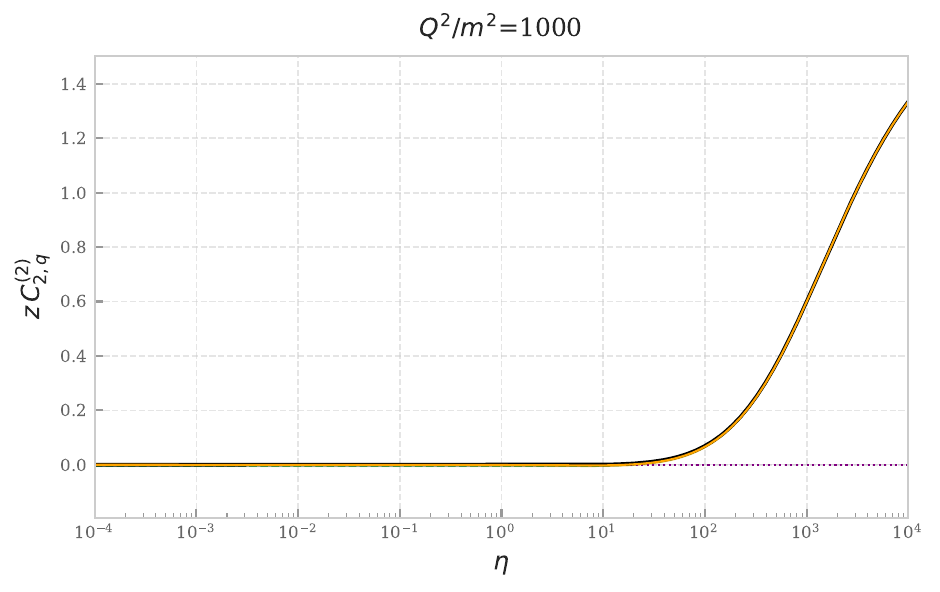}
  \includegraphics[width=0.49\textwidth]{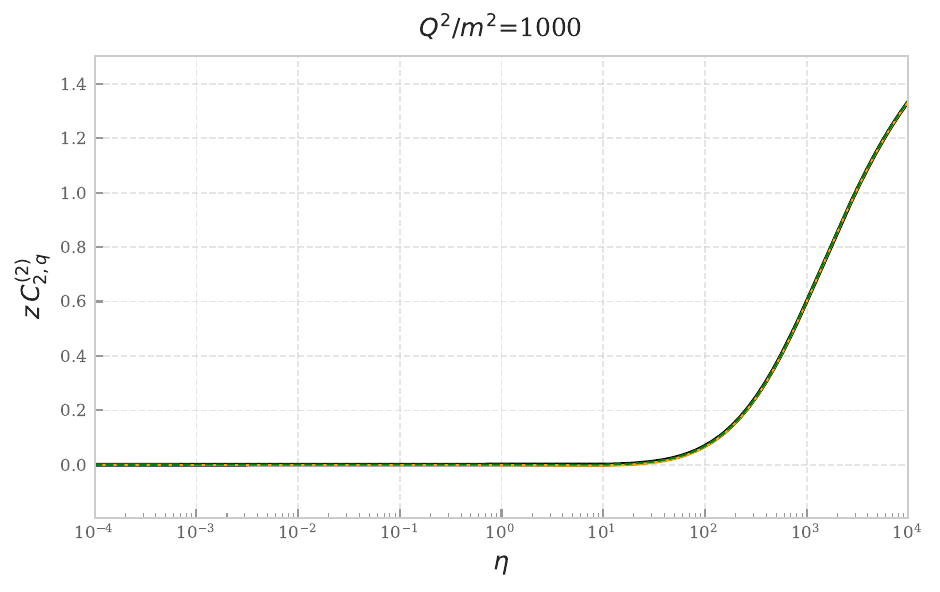}
  \caption{Same as Fig.~\ref{fig:appr:n2lo} but for the quark coefficient $C_{2,q}$.}
  \label{fig:appr:n2lo:2q}
\end{figure}
\begin{figure}[tp]
  \centering
  \includegraphics[width=0.49\textwidth]{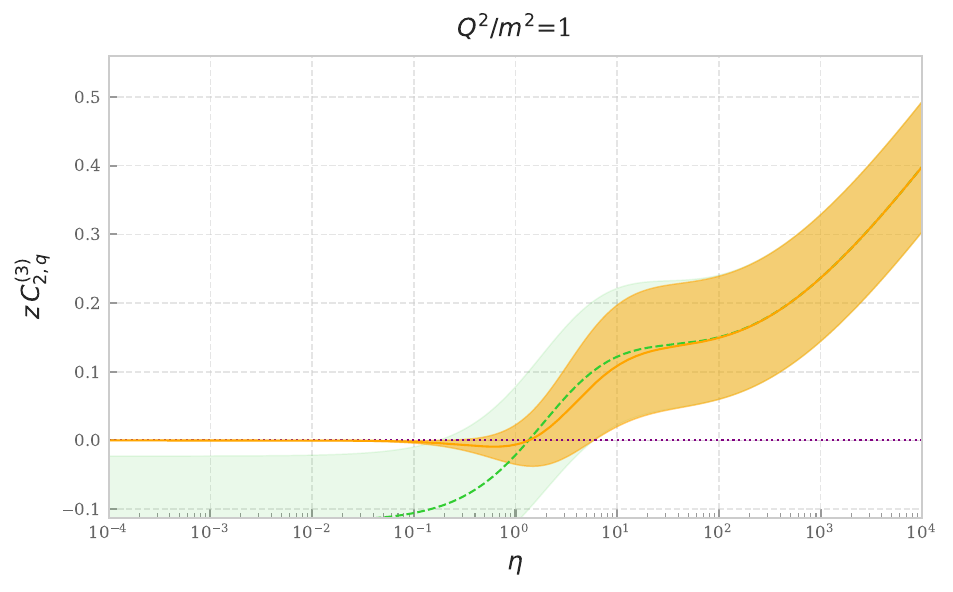}
  \includegraphics[width=0.49\textwidth]{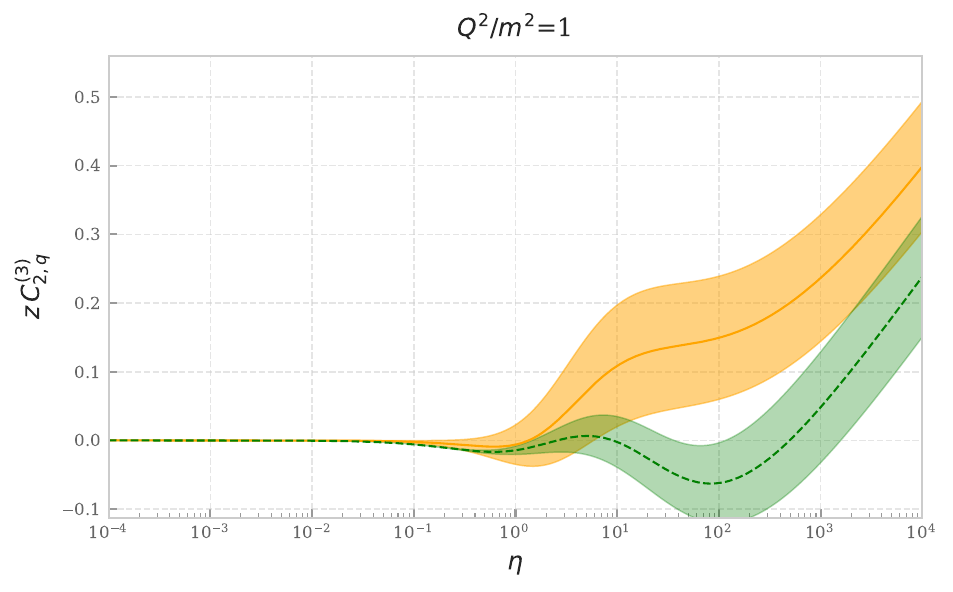}
  \includegraphics[width=0.49\textwidth]{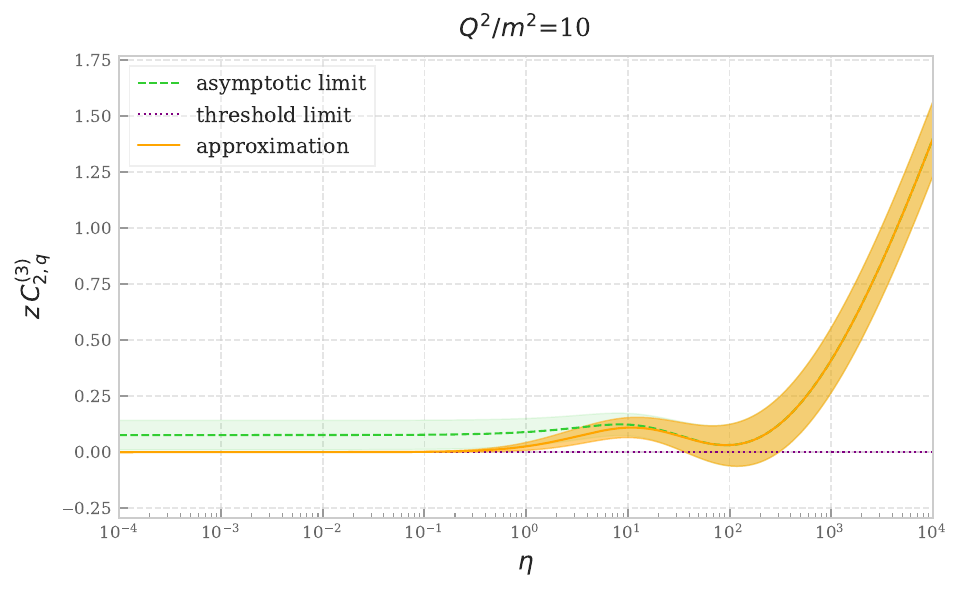}
  \includegraphics[width=0.49\textwidth]{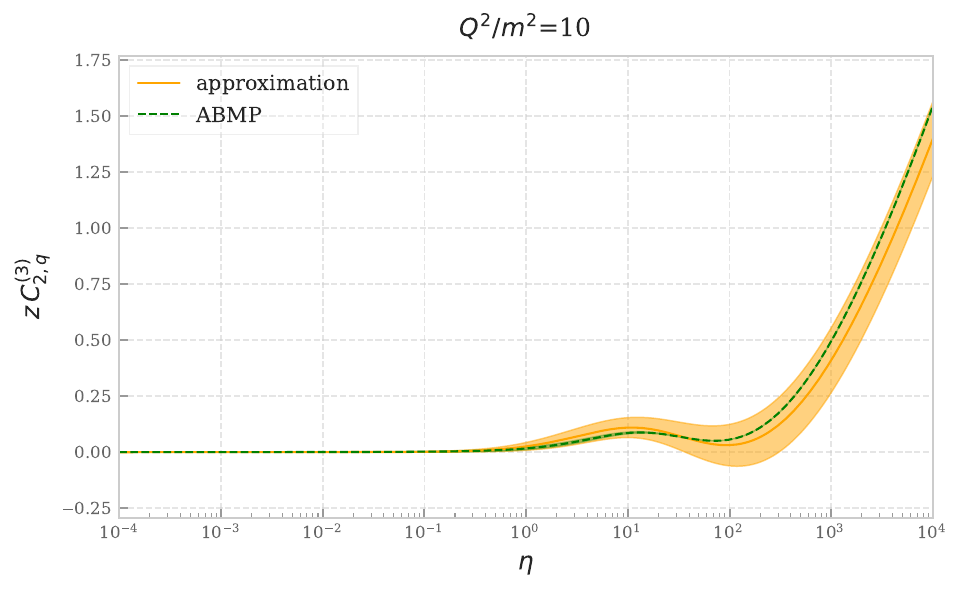}
  \includegraphics[width=0.49\textwidth]{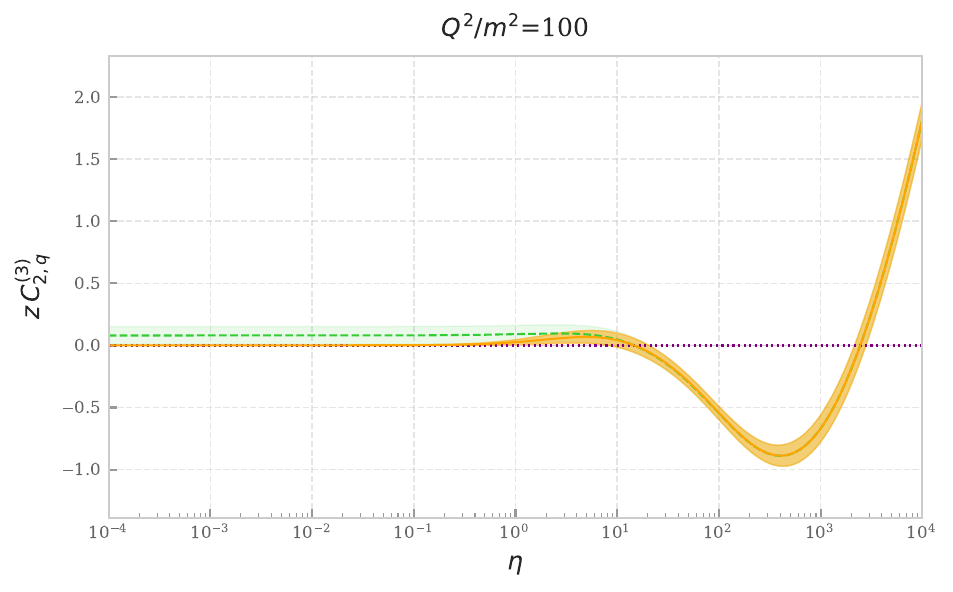}
  \includegraphics[width=0.49\textwidth]{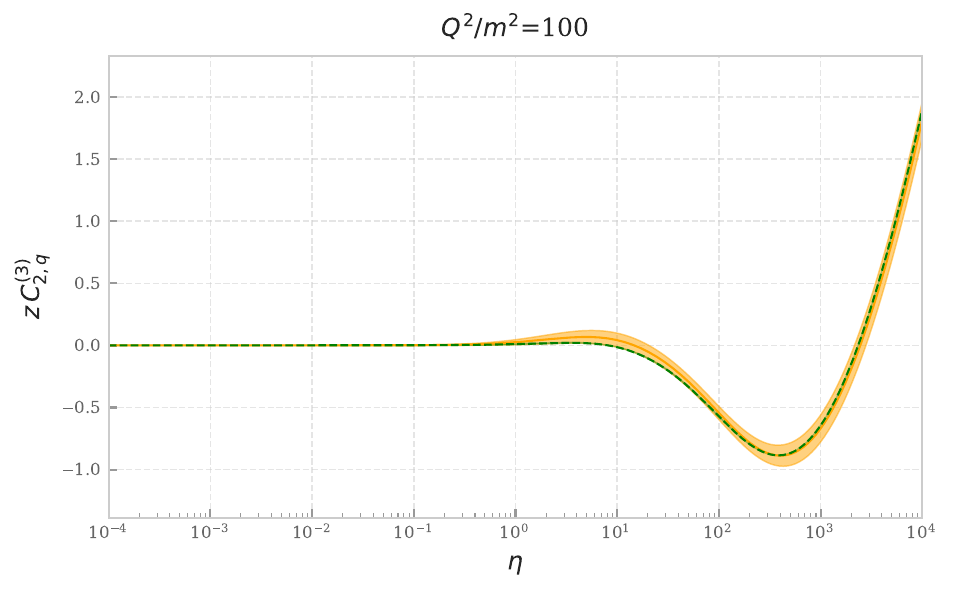}
  \includegraphics[width=0.49\textwidth]{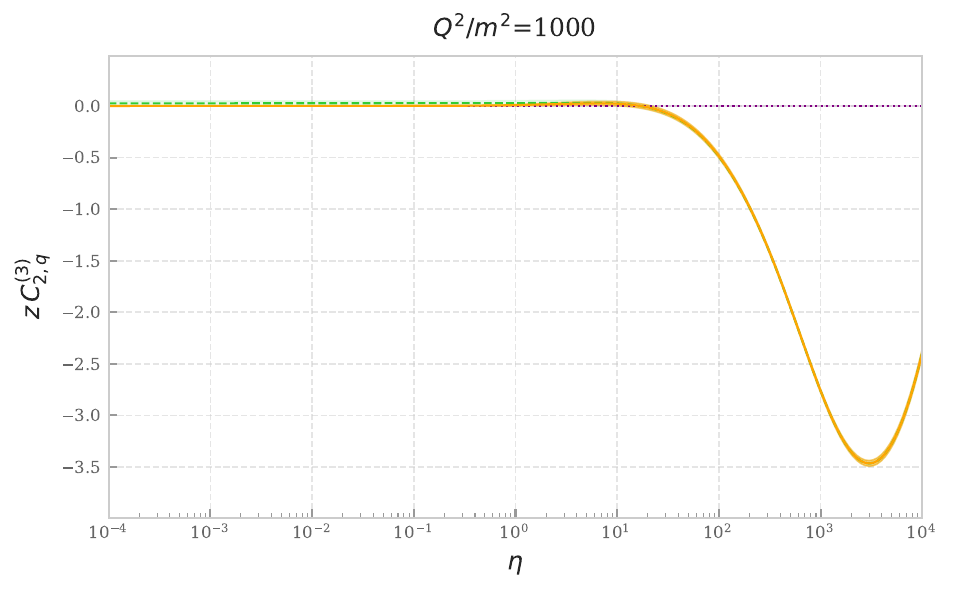}
  \includegraphics[width=0.49\textwidth]{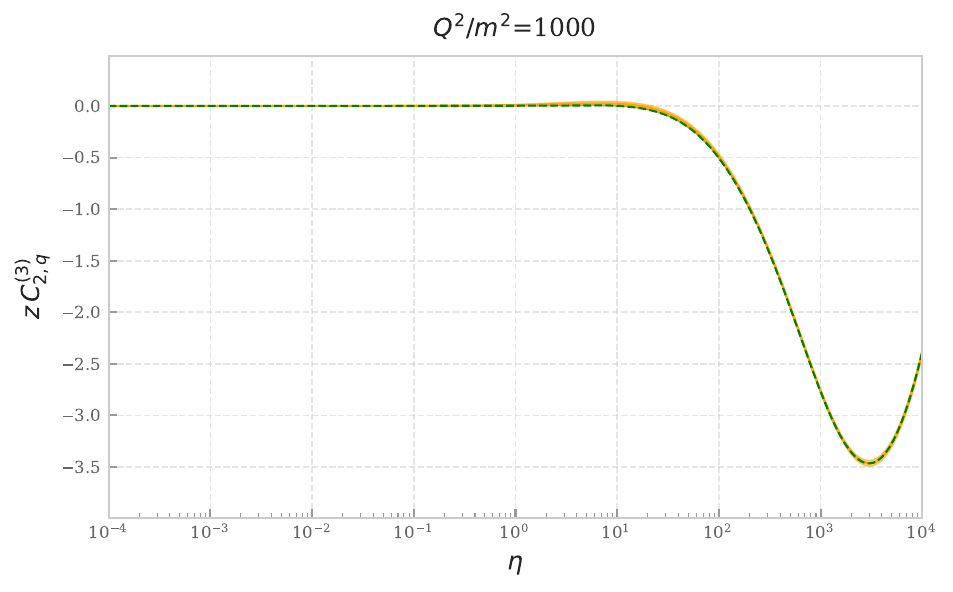}
  \caption{Same as Fig.~\ref{fig:appr:n3lo} but for the quark coefficient $C_{2,q}$.
    Only the ABMP curve is shown in the right plots, as the matching functions needed
    for the approximation of the quark channel were all known exactly at that time.}
  \label{fig:appr:n3lo:2q}
\end{figure}

The results at NNLO and N$^3$LO are presented in figures~\ref{fig:appr:n2lo:2q} and \ref{fig:appr:n3lo:2q}, respectively.
At NNLO we observe that our approximation covers well the exact result,
with the execption of the $\eta\sim10$ region in the $Q^2/m^2=10$ plot, where the exact curve lies slightly outside our band.
This is probably due to an imprecise description of the transition from the asymptotic (large $\eta$) region to the threshold region,
which in our construction is governed by the damping function that simply switches off the asymptotic contribution
at low $\eta$.
The absence of a better description of the opposite limit ($\eta\to0$)
makes it impossible to merge the two regions in a physically motivated way,
giving this kind of artefacts. Overall, we believe that the approximation is rather good anyway.
In the same figures we also compare our result with the one of Ref.~\cite{Kawamura:2012cr} (KLMV curve) at NNLO
and of Ref.~\cite{Alekhin:2017kpj} (ABMP curve) at N$^3$LO.\footnote
{At N$^3$LO the approximation of Ref.~\cite{Kawamura:2012cr} was affected in the quark channel by the incomplete
  knowledge of the matching function $K_{hq}^{[n_f+1]\leftarrow[n_f]}$.
  The update of Ref.~\cite{Alekhin:2017kpj} used instead the exact function computed in Ref.~\cite{Ablinger:2014nga},
  and it is therefore based on the same ingredients that we also use.}
As for the gluon channel, we observe that our description of the large-$\eta$ limit is superior,
and our uncertainty estimate is more reliable.
Finally, at N$^3$LO we observe similar differences as in the gluon channel at low $Q^2$,
while the two results behave the same at higher $Q^2$,
where indeed the difference in the gluon channel was mostly due to the threshold term, which is absent here.

\begin{figure}[tp]
  \centering
  \includegraphics[width=0.49\textwidth]{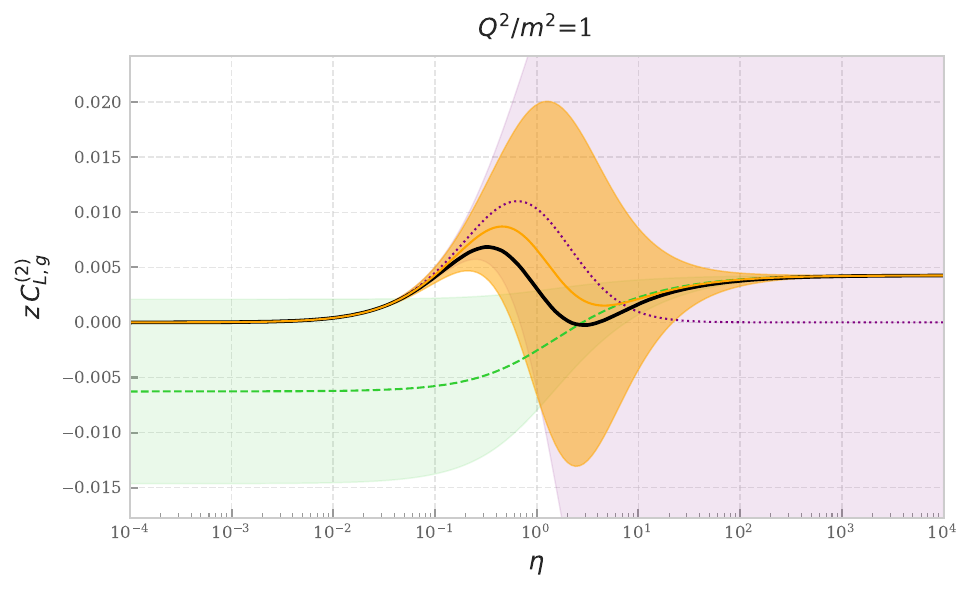}
  \includegraphics[width=0.49\textwidth]{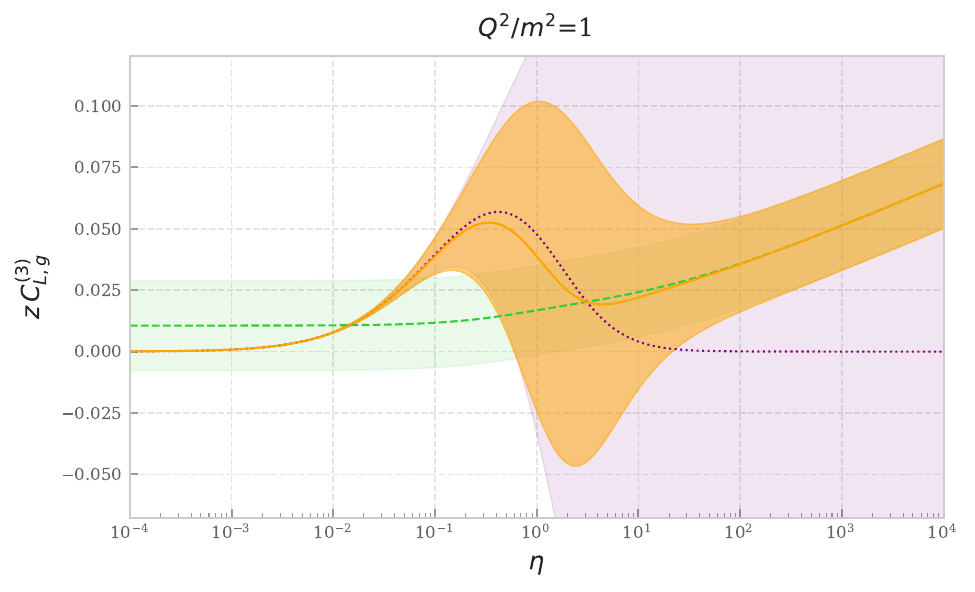}
  \includegraphics[width=0.49\textwidth]{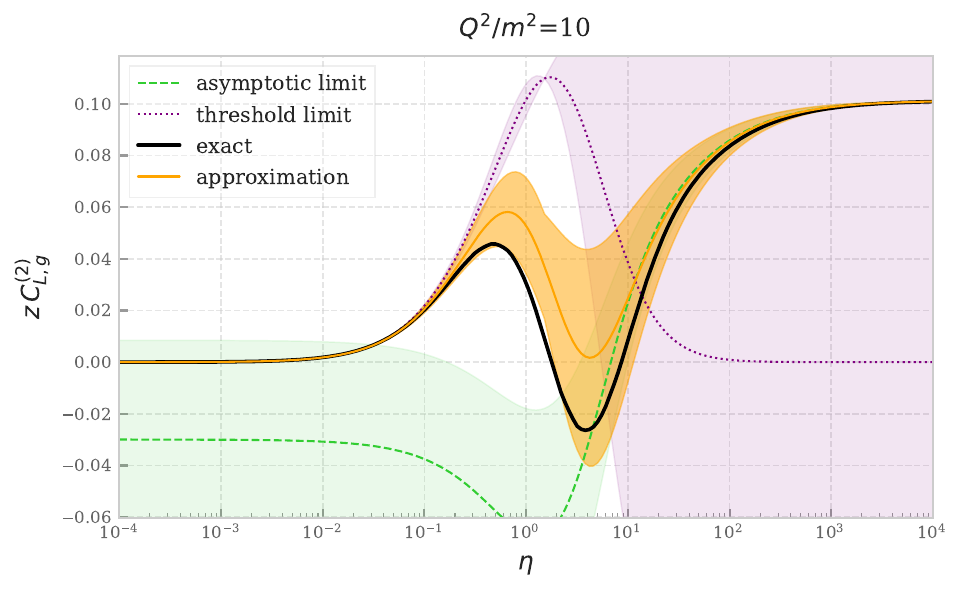}
  \includegraphics[width=0.49\textwidth]{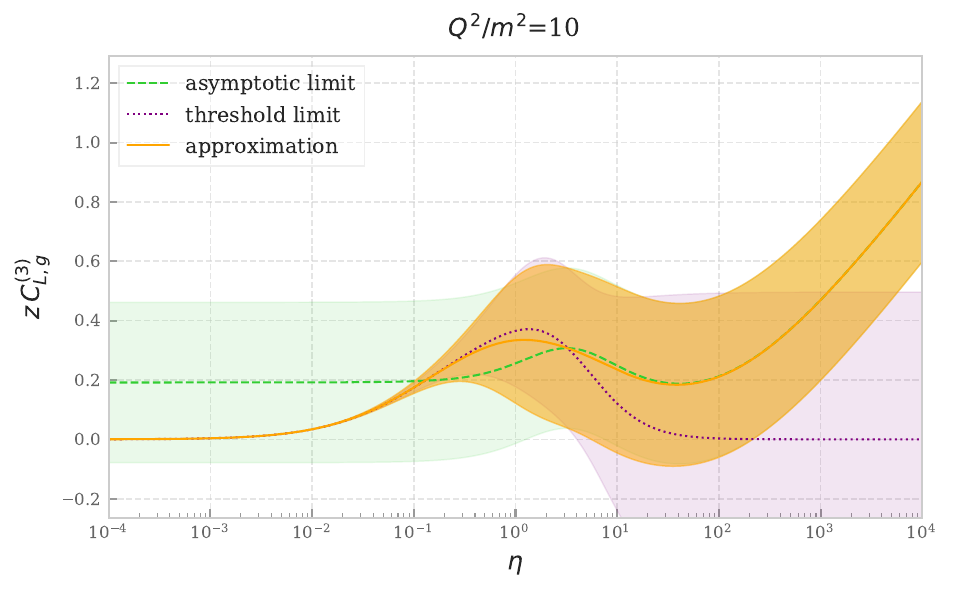}
  \includegraphics[width=0.49\textwidth]{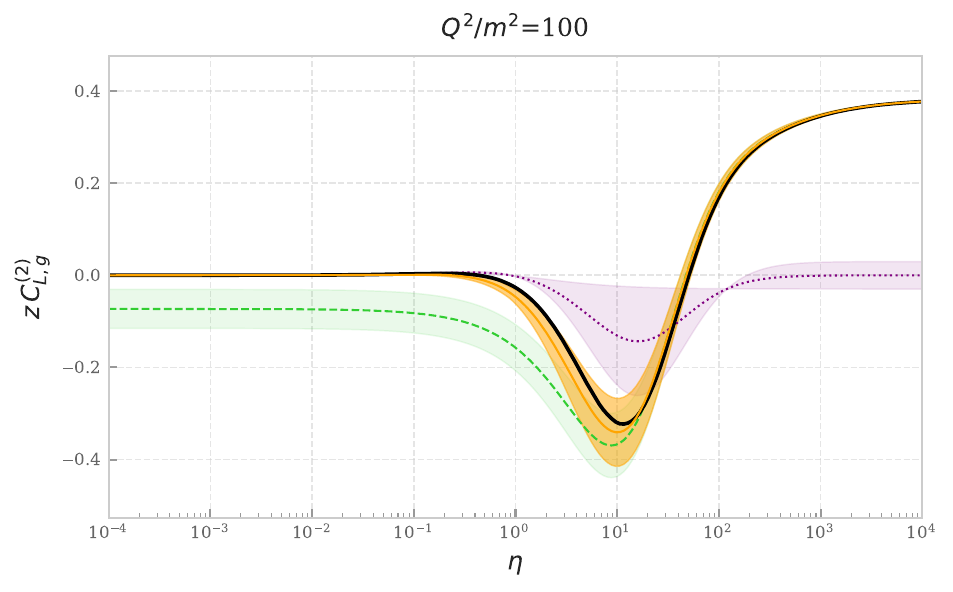}
  \includegraphics[width=0.49\textwidth]{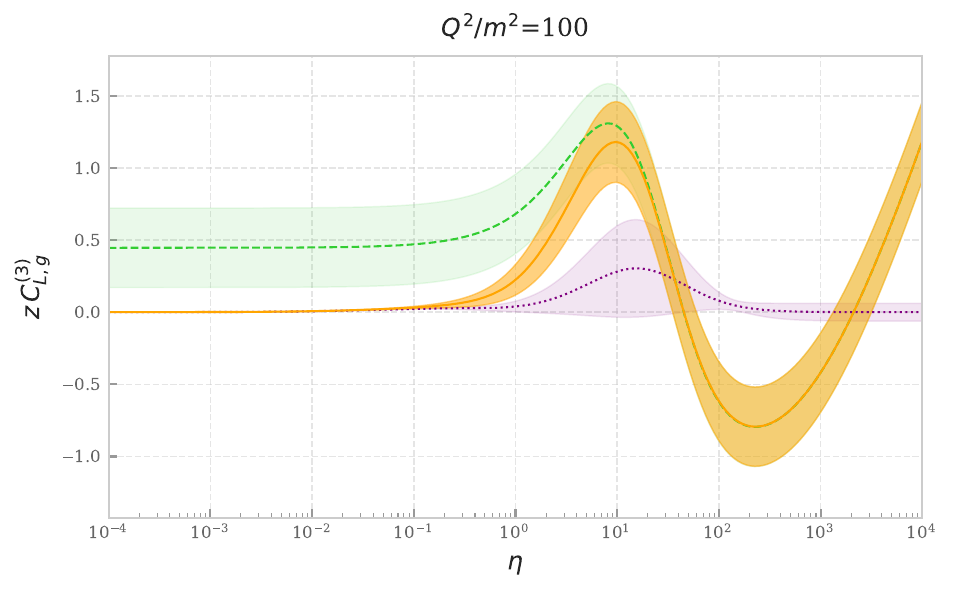}
  \includegraphics[width=0.49\textwidth]{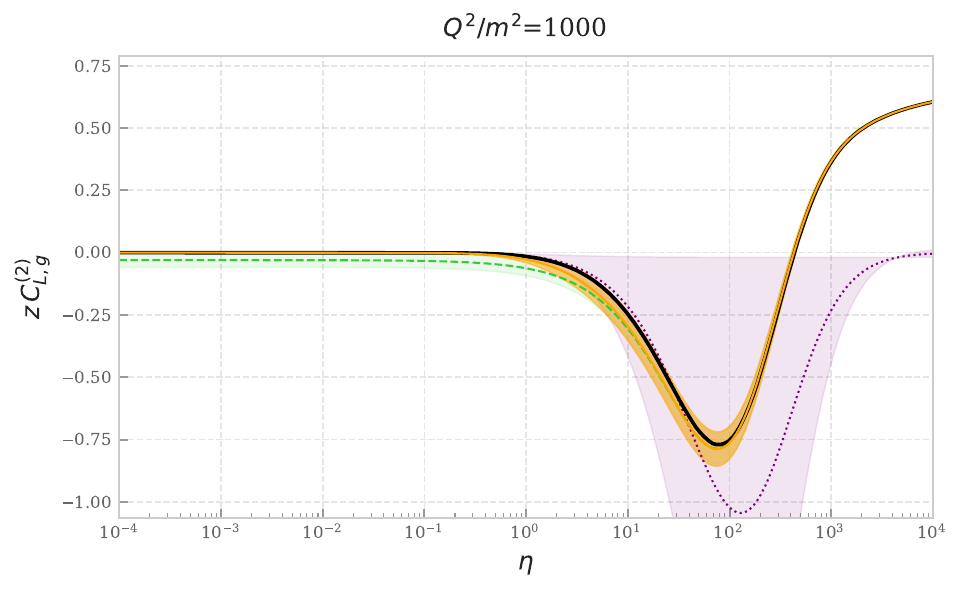}
  \includegraphics[width=0.49\textwidth]{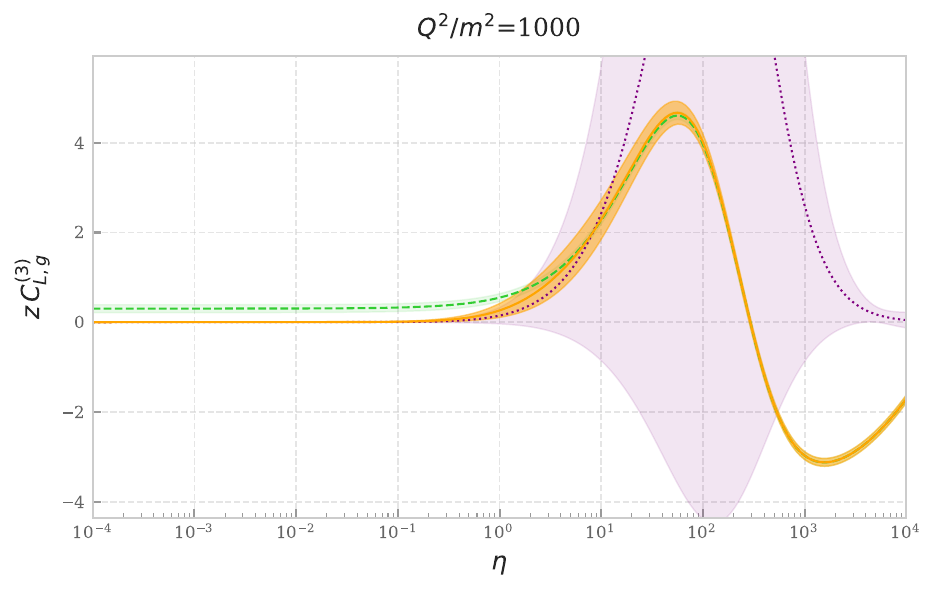}
  \caption{Left: Comparison between the exact longitudinal coefficient function $C_{L,g}^{[n_f]}$
    and its threshold and asymptotic limits, as well as the full approximation (with uncertainty), at $\ord2$.
    Right: Same as left, but at $\ord3$ and without showing the unknown exact result.
    All plots show the coefficient functions computed at $\mu=m$ as a function of $\eta$ for different values of $Q^2/m^2$.}
  \label{fig:appr:Lg}
\end{figure}
\begin{figure}[tp]
  \centering
  \includegraphics[width=0.49\textwidth]{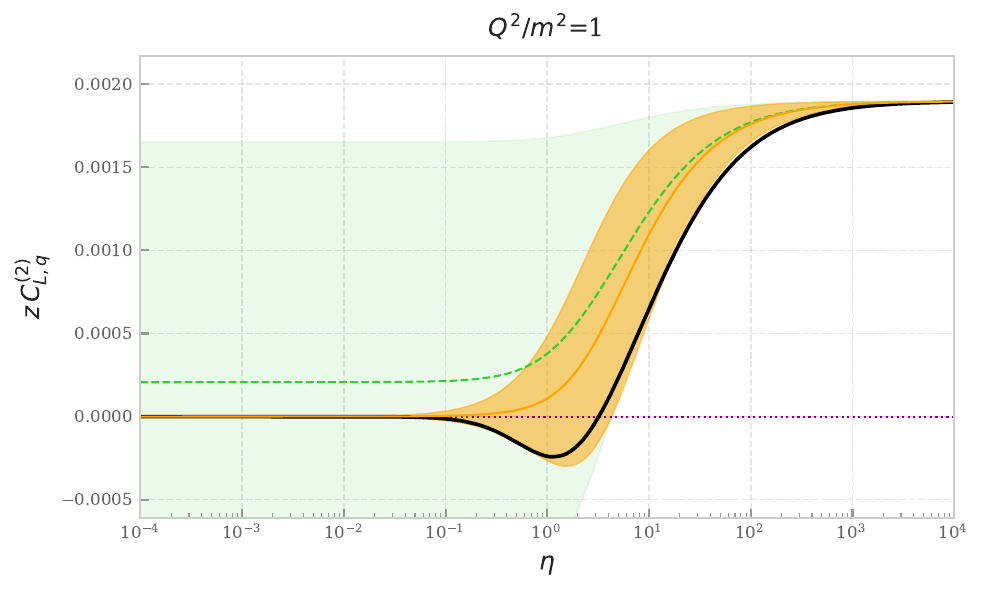}
  \includegraphics[width=0.49\textwidth]{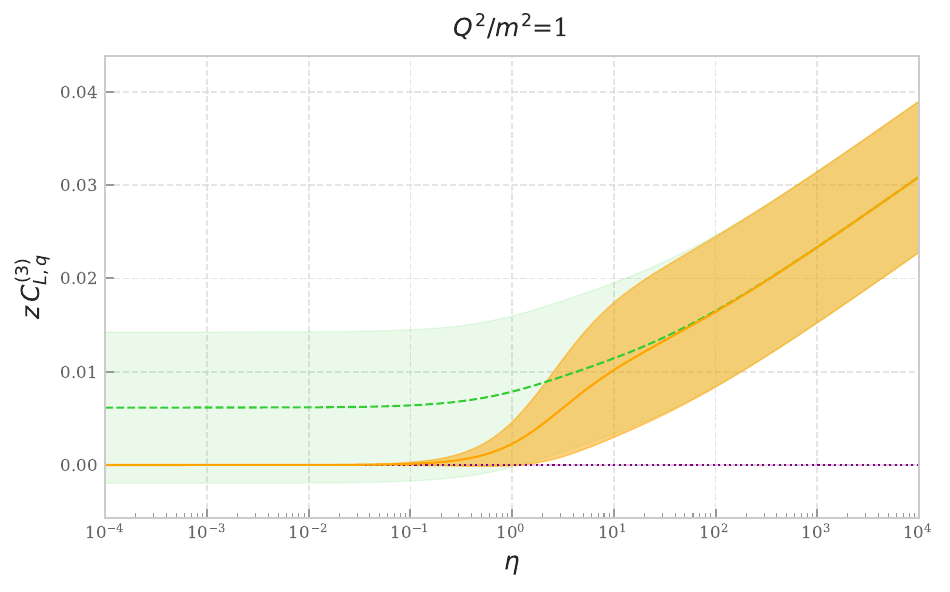}
  \includegraphics[width=0.49\textwidth]{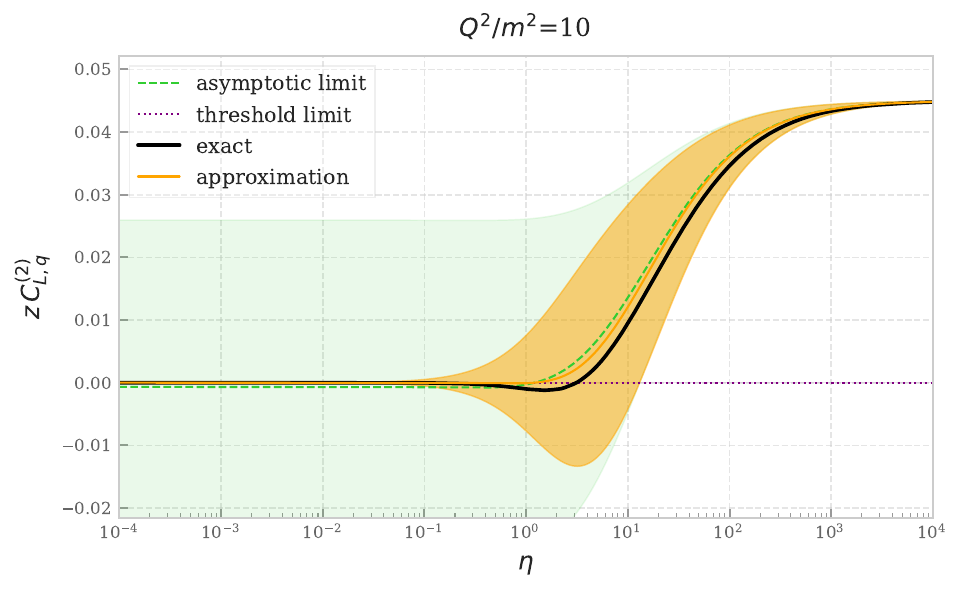}
  \includegraphics[width=0.49\textwidth]{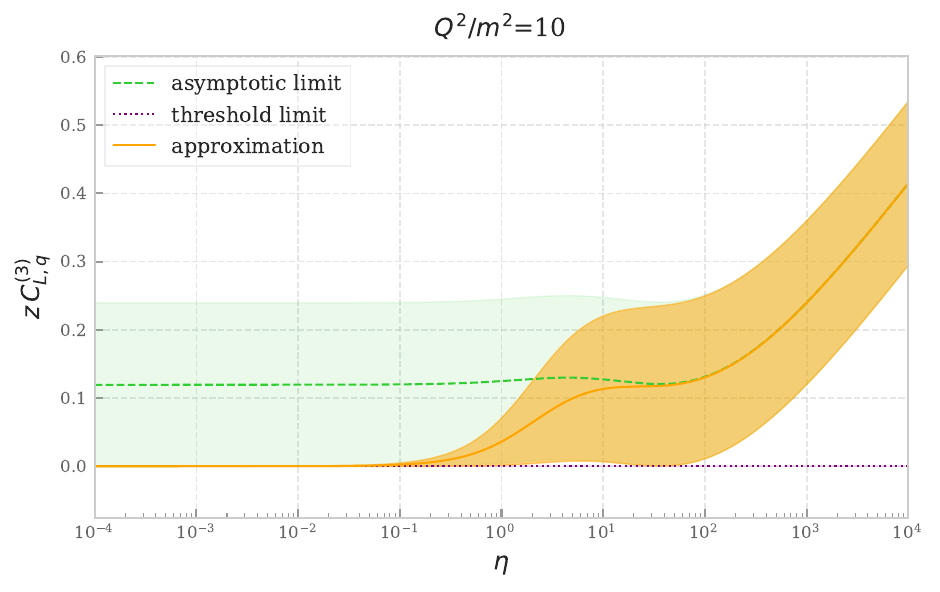}
  \includegraphics[width=0.49\textwidth]{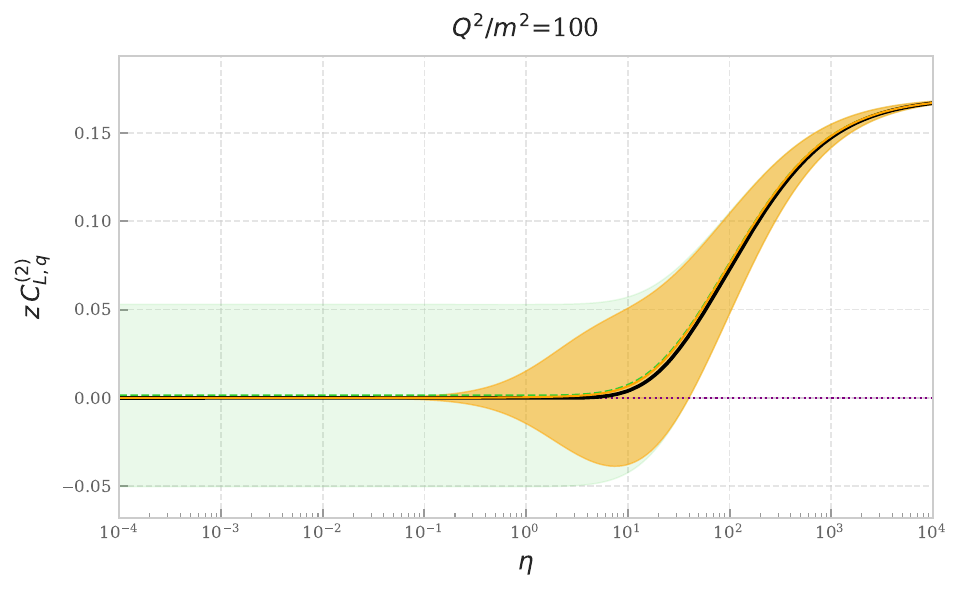}
  \includegraphics[width=0.49\textwidth]{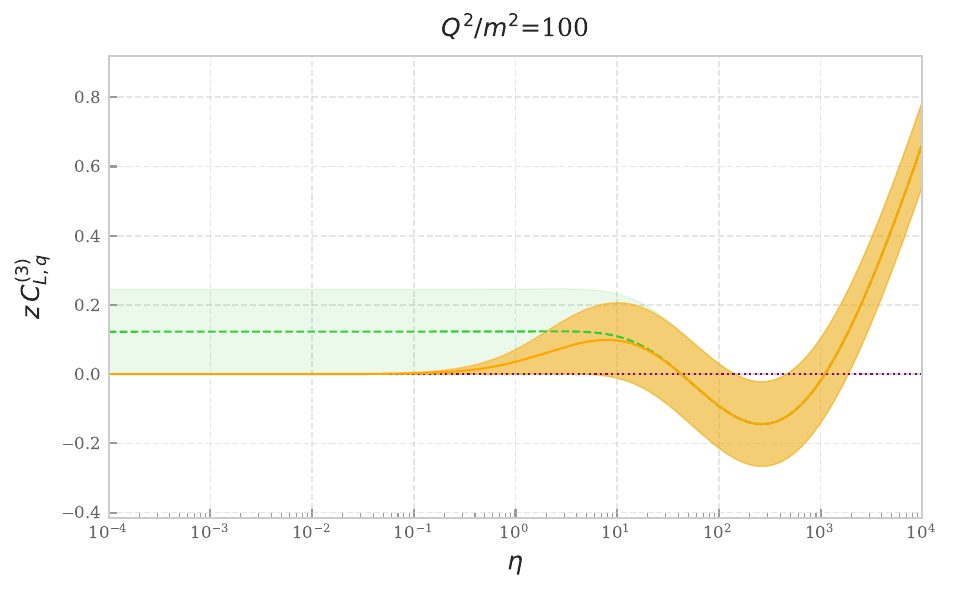}
  \includegraphics[width=0.49\textwidth]{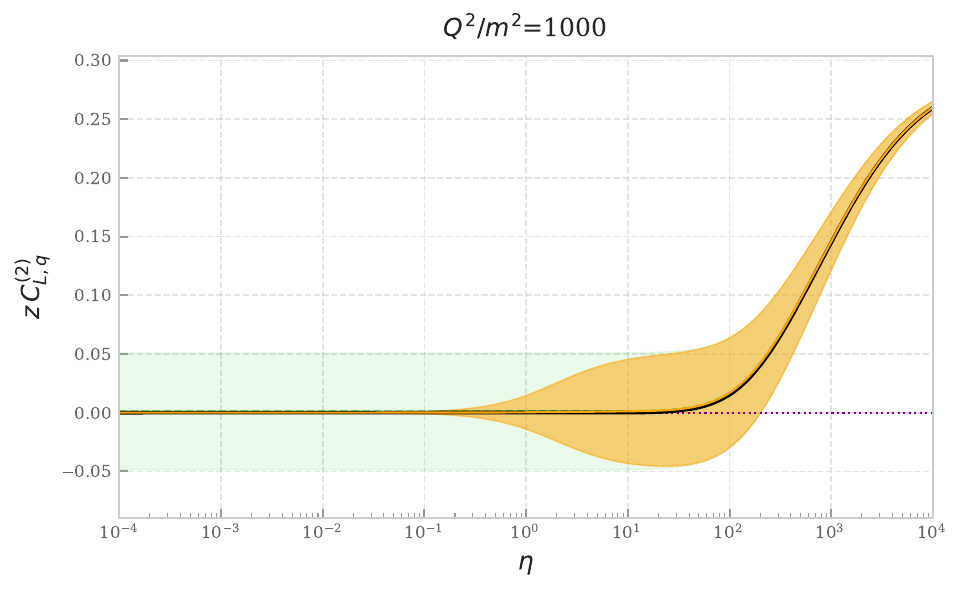}
  \includegraphics[width=0.49\textwidth]{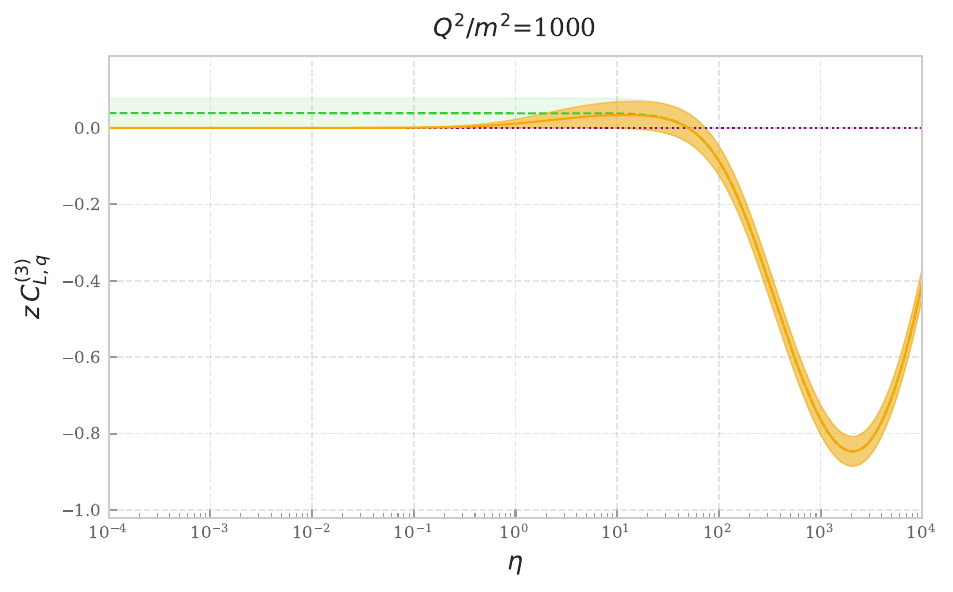}
  \caption{Same as Fig.~\ref{fig:appr:Lg} but for the quark channel longitudinal coefficient $C_{L,q}$.}
  \label{fig:appr:Lq}
\end{figure}

We now move to the longitudinal structure function $F_L$.
The construction of the approximation is morally the same as for $F_2$, with a practical difference in the asymptotic limit.
Indeed, we have noticed that at NNLO the additive matching for the asymptotic part, Eq.~\eqref{eq:Casy2},
is not so close to the exact result at mid values of $\eta$.
Conversely, the purely multiplicative matching, obtained by setting
$R_L^{(2)} = c_{L,i}^{(2,0)} / c_{L,i}^{(2,0)\,\rm h.s.}$ in Eq.~\eqref{eq:Casy2'}, is much closer
(at sufficiently high scales).
As there is no reason of principle for which one should be better than the other,
we can simply decide that for $F_L$ the multiplicative matching is used for the central value.
In principle we could use the additive matching for estimating the uncertainty,
but this would lead to a very large uncertainty band that we believe is not realistic.
Therefore, we estimate the uncertainty using a variation of the multiplicative matching
and symmetrising the difference with the central value to construct a band.
At NNLO, we achieve such variation by changing the form of $R_L^{(2)}$, as explained in appendix~\ref{sec:app:nll-approx},
see Eq.~\eqref{eq:RL} and discussion thereof.
At N$^3$LO, where we have introduced two variants of the modified multiplicative matching,
we use Eq.~\eqref{eq:Casy3'} for the central value and the variant without NLL terms, Eq.~\eqref{eq:Casy3''},
for the uncertainty, using in this case $R_L^{(3)} = c_{L,i}^{(3,0)} / c_{L,i}^{(3,0)\,\rm h.s.}$
(see again the discussion of appendix~\ref{sec:app:nll-approx} for a motivation of this choice).
As far as the threshold contribution is concerned, for $C_{L,g}$ we derive the threshold terms
in appendix~\ref{app:threshold}.
The uncertainty is constructed in the same way as for $F_2$, and in particular at N$^3$LO we include
the uncertainty due to the lack of knowledge of the $\beta^0$ term, by computing the corresponding $[0/1]$ Pad\'e approximant.
We recall that $F_L$ is more suppressed at threshold than $F_2$,
nevertheless we decided to include the information on this limit (for the gluon channel) in order to have a better description
of the transition from large $\eta$ to small $\eta$.

The results for $C_{L,g}$ are presented in figure~\ref{fig:appr:Lg} and for $C_{L,q}$ in figure~\ref{fig:appr:Lq},
at NNLO on the left and N$^3$LO on the right. Here we do not include a comparison with other results
as to our knowledge an approximation for this observable has never been constructed before.
At NNLO we observe excellent compatibility with the exact result,
with a reliable estimate of the uncertainty, in both channels.
We are thus confident that our N$^3$LO approximation be reliable.

\FloatBarrier
\section{The high-energy limit of the N$^3$LO coefficient functions}
\label{sec:app:nll-approx}

In this appendix we present the construction of the high-energy (small-$z$) approximation
for the massive coefficient functions.
We do this by expanding the resummed result in powers of the strong coupling $\as$.
Resummation at LL accuracy is performed according to Refs.~\cite{Catani:1990eg,Catani:1994sq} through
the Mellin-space expression (using a non-standard notation for the Mellin transform that is common in small-$z$ literature)
\begin{align}\label{eq:CresLLg}
C_{a,g}^{[n_f]}\(N,\mQ,\muQ\)
&\equiv \int_0^1 dz\, z^N\, C_{a,g}^{[n_f]}\(z,\mQ,\muQ\) \nonumber\\
&\overset{{\rm LL}}= R(N)\, {\cal C}_a\(\gamma(N),\mQ,\frac{m^2}{\mu^2}\),
\end{align}
where $N$ is the Mellin conjugate variable to $z$,
the function $R(N)=1+\ord3$ and thus it is of no interest for us,
and the functions ${\cal C}_a$ have been computed in Refs.~\cite{Catani:1990xk,Catani:1992zc,Catani:1990eg,Bonvini:2017ogt}
and are given by ($\xi\equiv m^2/Q^2$)
\begin{align}
  \label{eq:c2g_mell}
  {\cal C}_2\(\gamma, \xi, \frac{m^2}{\mu^2}\)
  &= \frac{\alpha_se_h^2}{3 \pi} \(\frac{m^2}{\mu^2}\)^\gamma \frac{3}{2}\,
    \frac{\Gamma(1-\gamma)^3\,\Gamma(1+\gamma)}{(3-2\gamma)(1+2\gamma)\Gamma(2-2\gamma)} \nonumber\\
  &\times \left[1+\gamma - \left( 1 + \gamma - \frac1{2\xi}(2+3\gamma - 3\gamma^2) \right)\,
    {}_2F_1\left(1-\gamma,1,\frac{3}{2}, -\frac1{4\xi}\right)\right],
  \\
  \label{eq:cLg_mell}
  {\cal C}_L\(\gamma, \xi, \frac{m^2}{\mu^2}\)
  &=\frac{\alpha_se_h^2}{3 \pi} \(\frac{m^2}{\mu^2}\)^\gamma \frac{3}{2}\,
    \frac{\Gamma(1-\gamma)^3\,\Gamma(1+\gamma)}{(3-2\gamma)(1+2\gamma)\Gamma(2-2\gamma)}\, \frac{4\xi}{1 + 4\xi} \nonumber\\
  &\times \left[3 + \frac1{2\xi}(1 - \gamma) - \left( 3 + \frac{1 - \gamma}\xi \left(1 - \frac\gamma{4\xi} \right)\right)\,
    {}_2F_1\left(1-\gamma,1,\frac{3}{2}, -\frac1{4\xi}\right)\right],
\end{align}
with $e_h$ the electric charge of the heavy quark $h$.
The function $\gamma(N)$ represents the resummed DGLAP anomalous dimension, which at LL reads
\beq
\gamma(N) \overset{{\rm LL}}= \frac{C_A\as}{\pi N} + \Ord\(\(\frac{\as}{N}\)^4\).
\eeq
We recall that in $N$ space the LL terms are those behaving as powers of $\as/N$.
The resummation for the quark coefficient functions $C_{a,q}^{[n_f]}$ are obtained from Eq.~\eqref{eq:CresLLg}
multiplying by the color factors $C_F/C_A$ and subtracting the lowest order in $\as$.

Expanding Eq.~\eqref{eq:CresLLg} in powers of $\as$ up to $\ord3$ leads to the LL result used in Ref.~\cite{Kawamura:2012cr}.
However, this does not predict any of the NLL contributions which are present at N$^3$LO.
In order to improve on this, we consider the resummation as implemented in the \texttt{HELL} code~\cite{Bonvini:2016wki,Bonvini:2017ogt},
where several NLL terms are included.
They have two origins: one is the use of $\gamma(N)$ accurate at NLL, and the other is the inclusion
of NLL running coupling effects whose resummation has been introduced in Ref.~\cite{Ball:2007ra}.
The latter effect can be incorporated in the fixed-order expansion of the resummed result
by expanding first in powers of $\gamma$ and then replacing the $k$-th power of $\gamma$
with $\[\gamma^k\]$ defined recursively by
\beq
\[\gamma^{k+1}\] = \(\gamma-k\as\beta_0\)\[\gamma^k\], \qquad \[\gamma\] = \gamma, \qquad \beta_0=\frac{11C_A-2n_f}{12\pi}.
\eeq
As we are interested in the expansion up to $\ord3$, and owing to the fact that the functions
${\cal C}_a$ are proportional to $\as$, we can expand in powers of $\gamma$ up to second order
(using $\zeta\equiv m^2/\mu^2$)
\begin{equation}
  {\cal C}_a(\gamma,\xi,\zeta) = {\cal C}_a^{(0)}(\xi,\zeta) + \gamma\, {\cal C}_a^{(1)}(\xi,\zeta)
  + \gamma^2 {\cal C}_a^{(2)}(\xi,\zeta) + \Ord\(\gamma^3\).
\end{equation}
After replacing $\gamma^k\to\[\gamma^k\]$ and expanding the anomalous dimension in powers of $\as$ as
\beq
\gamma(N) = \as\gamma_0(N) + \as^2\gamma_1(N) + \ord3,
\eeq
we have
\begin{align}\label{eq:calCexp}
  {\cal C}_a\(\gamma(N),\xi,\zeta\)
  &= {\cal C}_a^{(0)}(\xi,\zeta) + \as \gamma_0(N)\, {\cal C}_a^{(1)}(\xi,\zeta)\nonumber\\
  &+ \as^2\[ \(\gamma_0^2(N)-\beta_0\gamma_0(N)\)\, {\cal C}_a^{(2)}(\xi,\zeta) + \gamma_1(N)\, {\cal C}_a^{(1)}(\xi,\zeta)\] + \ord4.
\end{align}
The LL result used in Ref.~\cite{Kawamura:2012cr} is recovered by setting $\beta_0=0$ and using the LL values
for the expansion of $\gamma$, namely $\gamma_0=\frac{C_A}{\pi N}$ and $\gamma_1=0$.
Here instead we suggest to keep the $\beta_0$ term and to use the NLL expressions
\begin{align}
  \gamma_0(N) &= \frac{a_{11}}{N}+a_{10} + \Ord(N), \\
  \gamma_1(N) &= \frac{a_{21}}{N}+ \Ord\(N^0\),
\end{align}
with
\begin{align}
  a_{11}&=\frac{C_A}\pi, \\
  a_{10}&=-\frac{11C_A + 2n_f(1-2C_F/C_A)}{12\pi}, \\
  a_{21}&=n_f\frac{26C_F - 23C_A}{36\pi^2}.
\end{align}
Plugging this into Eq.~\eqref{eq:calCexp} and computing the inverse Mellin transform,
we can identify the coefficients of Eq.~\eqref{eq:Chighenergy} with
\begin{subequations}\label{eq:hec}
\begin{align}
 \as c_{a,g}^{(2,0)}\(\mQ,\muQ\) &= a_{11}\, {\cal C}_a^{(1)}\(\mQ,\frac{m^2}{\mu^2}\), \label{eq:hec20}\\
 \as c_{a,g}^{(3,0)}\(\mQ,\muQ\) &= -a_{11}^2\, {\cal C}_a^{(2)}\(\mQ,\frac{m^2}{\mu^2}\), \label{eq:hec30}\\
 \as c_{a,g}^{(3,1)}\(\mQ,\muQ\) &\simeq \(2a_{10}-\beta_0\)a_{11}\, {\cal C}_a^{(2)}\(\mQ,\frac{m^2}{\mu^2}\)
                                + a_{21}\, {\cal C}_a^{(1)}\(\mQ,\frac{m^2}{\mu^2}\). \label{eq:hec31}
\end{align}
\end{subequations}
While the LL terms Eq.~\eqref{eq:hec20} and \eqref{eq:hec30} are exact,
we stress once more that the NLL term Eq.~\eqref{eq:hec31} is an approximation (which is why we used a $\simeq$ symbol).

To complete the derivation, we need to write explicitly the functions ${\cal C}_a^{(k)}$ for $a=2,L$ and $k=1,2$.
We start observing that the dependence on $\mu$ is fully dictated by the term $(m^2/\mu^2)^\gamma$
in Eq.~\eqref{eq:c2g_mell} and \eqref{eq:cLg_mell}.
In other words, both equations can be written as
\beq
{\cal C}_a\(\gamma, \xi, \frac{m^2}{\mu^2}\) = \(\frac{m^2}{\mu^2}\)^\gamma  {\cal C}_a\(\gamma, \xi, 1\),
\eeq
in terms of a $\mu$-dependent factor and the same function computed at $\mu=m$.
Therefore, the scale dependence of the expansion coefficients can be written
in terms of the $\mu=m$ coefficients as
\begin{subequations}\label{eq:calCmu}
\begin{align}
  {\cal C}_a^{(1)}\(\xi,\frac{m^2}{\mu^2}\)
  &= {\cal C}_a^{(1)}\(\xi,1\) + {\cal C}_a^{(0)}\(\xi,1\) \log\frac{m^2}{\mu^2}, \\
  {\cal C}_a^{(2)}\(\xi,\frac{m^2}{\mu^2}\)
  &= {\cal C}_a^{(2)}\(\xi,1\)  + {\cal C}_a^{(1)}\(\xi,1\) \log\frac{m^2}{\mu^2} + \frac12{\cal C}_a^{(0)}\(\xi,1\) \log^2\frac{m^2}{\mu^2}.
\end{align}
\end{subequations}
Expanding Eq.~\eqref{eq:c2g_mell} and \eqref{eq:cLg_mell} in powers of $\gamma$ at $\mu=m$,
we find
\begin{subequations}\label{eq:calCxi}
\begin{align}
  {\cal C}_2^{(0)}\(\xi,1\)
  &= \frac{\as e_h^2}{6\pi}\[1 + \frac{1-\xi}{2\xi} J(\xi)\],
  \\
  {\cal C}_2^{(1)}\(\xi,1\)
  &=  \frac{\as e_h^2}{6\pi}\[ \frac53 + \frac{1-\xi}{2\xi} I(\xi) + \frac{13-10\xi}{12\xi} J(\xi)\],
  \\
  {\cal C}_2^{(2)}\(\xi,1\)
  &=  \frac{\as e_h^2}{6\pi}\[ \frac{46}9 - \frac{1-\xi}{4\xi} K(\xi) + \(\frac{13-10\xi}{12\xi} - \frac{1-\xi}{4\xi}\log\frac{4\xi}{1+4\xi}\) I(\xi)
    + \frac{71-92\xi}{36\xi} J(\xi)\],
  \\
  {\cal C}_L^{(0)}\(\xi,1\)
  &= \frac{\as e_h^2}{6\pi}\frac{2(1 + 6\xi) - 2(1+3\xi) J(\xi)}{1+4\xi},
  \\
  {\cal C}_L^{(1)}\(\xi,1\)
  &=  \frac{\as e_h^2}{6\pi} \frac{-\frac23 +8\xi -2(1+3\xi) I(\xi) + \(\frac1{2\xi}+\frac23-4\xi\) J(\xi)}{1+4\xi},
  \\
  {\cal C}_L^{(2)}\(\xi,1\)
  &=  \frac{\as e_h^2}{6\pi} \frac{\frac{68}9 +\frac{160}3\xi +(1+3\xi) K(\xi)
    + \(\frac1{2\xi}+\frac23-4\xi + (1+3\xi)\log\frac{4\xi}{1+4\xi}\) I(\xi)
    - \(\frac1{6\xi}+\frac{68}9+\frac{80}3\xi\)J(\xi)}{1+4\xi},
\end{align}
\end{subequations}
with
\begin{align}
  I(\xi) &= \frac{4\xi}{\sqrt{1+4\xi}} H_{-,+}\(\frac1{\sqrt{1+4\xi}}\), \\
  J(\xi) &= \frac{4\xi}{\sqrt{1+4\xi}} \log\frac{\sqrt{1+4\xi}+1}{\sqrt{1+4\xi}-1}, \\
  K(\xi) &= \frac{4\xi}{\sqrt{1+4\xi}} H_{-,+,-}\(\frac1{\sqrt{1+4\xi}}\),
\end{align}
where $H_{-,+}(x)$ and $H_{-,+,-}(x)$ are harmonic polylogarithms~\cite{Kawamura:2012cr,Maitre:2007kp}.
The results for $F_2$ have already been presented in Ref.~\cite{Kawamura:2012cr} and they do coincide,
while the results for $F_L$ are new.

For completeness, we also report the high-scale (small $\xi$) limit of the coefficients, which is found to be
\begin{subequations}
\begin{align}
  {\cal C}_2^{(0)}\(\xi,1\)
  &= \frac{\as e_h^2}{6\pi}\[1 - 2\log\xi + \Ord(\xi)\],
  \\
  {\cal C}_2^{(1)}\(\xi,1\)
  &=  \frac{\as e_h^2}{6\pi}\[ \frac53 -2\zeta_2 - \frac{13}{3} \log\xi + \log^2\xi + \Ord(\xi)\],
  \\
  {\cal C}_2^{(2)}\(\xi,1\)
  &=  \frac{\as e_h^2}{6\pi}\[ \frac{46}9 - \frac{13}3\zeta_2 + 4\zeta_3
    + \(- \frac{71}{9} + 2\zeta_2\) \log\xi
    +\frac{13}{6}\log^2\xi -\frac13\log^3\xi
    + \Ord(\xi)\],
  \\
  {\cal C}_L^{(0)}\(\xi,1\)
  &= \frac{\as e_h^2}{6\pi} \[2 + \Ord(\xi)\],
  \\
  {\cal C}_L^{(1)}\(\xi,1\)
  &=  \frac{\as e_h^2}{6\pi} \[-\frac23 -2\log\xi + \Ord(\xi)\],
  \\
  {\cal C}_L^{(2)}\(\xi,1\)
  &=  \frac{\as e_h^2}{6\pi} \[\frac{68}9 -2\zeta_2 + \frac23 \log\xi + \log^2\xi + \Ord(\xi)\].
\end{align}
\end{subequations}
Out of these expansions, using Eqs.~\eqref{eq:hec} and \eqref{eq:calCmu}
it is possible to construct the high-scale limit of the LL and NLL coefficients.

\begin{figure}[t]
  \centering
  \includegraphics[width=0.49\textwidth, page=1]{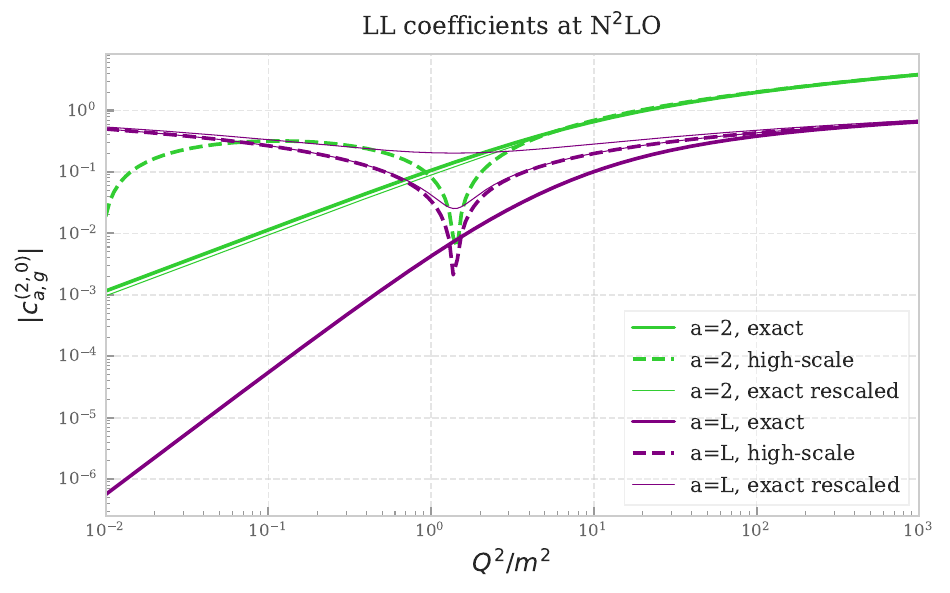}
  \includegraphics[width=0.49\textwidth, page=1]{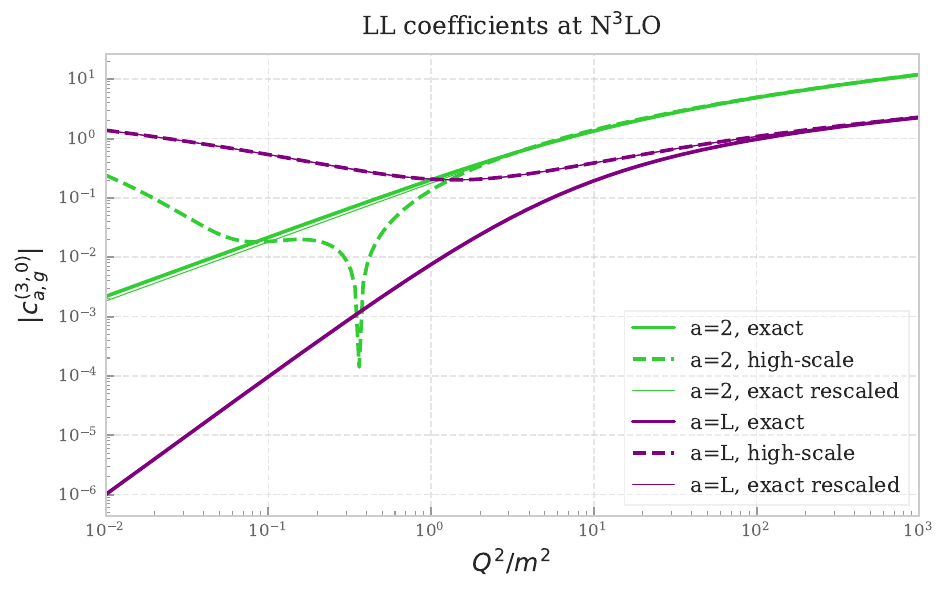}
  \caption{The absolute value, shown in logarithmic scale, of the LL coefficient $c_{a,g}^{(n,0)}(m^2/Q^2,\mu^2/Q^2)$
    computed at $\mu^2=m^2$ as a function of $Q^2/m^2$.
    In green the results for $F_2$ and in purple for $F_L$.
    The thick solid line represents the exact result, the dashed line its high-scale limit,
    and the thin solid lines the rescaled result $c_{a,g}^{(n,0)}/R_a^{(n)}$ as described in the text.
    The left plot shows these curves at order $\as^2$ and the right plot at order $\as^3$.}
  \label{fig:LLcoeff}
\end{figure}

We can now investigate the behaviour of the high-energy coefficients as a function of the scales.
In particular, as we construct our approximation at $\mu^2=m^2$, we fix this scale and present the LL
coefficients of the gluon contribution to $F_2$ and $F_L$ as a function of $Q^2/m^2$.
In order to be general, we do not include the quark-dependent factor $e_h^2$ in the results.
In figure~\ref{fig:LLcoeff} we show the absolute value of the LL coefficient $c_{a,g}^{(n,0)}(m^2/Q^2,1)$
at NNLO ($n=2$, left plot) and N$^3$LO ($n=3$, right plot) for $F_2$ (green) and $F_L$ (purple).
The exact coefficient is given by a thick solid line, while its high-energy limit uses a dashed style.
We observe that the exact coefficient keeps its sign (which is positive at NNLO and negative at N$^3$LO)
throughout the whole scale range shown, and it approaches zero at small $Q^2$,
as one can indeed verify analytically by expanding Eqs.~\eqref{eq:calCxi} at large $\xi$.
Conversely, the high-scale limit of the LL coefficients departs from the exact value as $Q^2$ gets smaller,
eventually changing sign at some scale (except for $F_L$ at N$^3$LO, where in any case the high-scale coefficient departs significantly from the exact one).
As a result, the behaviour of the exact LL coefficients and their high-scale limit is very different
far from the high-scale region.
While this is not problematic for the additive matching Eq.~\eqref{eq:Casy},
it would generate a crazy behaviour at low scales for the multiplicative matching Eq.~\eqref{eq:Casy'}
if we defined $R_a^{(n)}=c_{a,g}^{(n,0)}/c_{a,g}^{(n,0)\,\rm h.s.}$.

In order to avoid this, we construct the rescaling function $R_a^{(n)}$ introduced in section~\ref{sec:asy}
in a way that equals $c_{a,g}^{(n,0)}/c_{a,g}^{(n,0)\,\rm h.s.}$ at high scales but stays reasonable at smaller scales.
What ``reasonable'' means depends on which structure function we look at.
Starting from $F_2$, we have noticed in section~\ref{sec:asy} that the additive matching works great at NNLO,\footnote
{The observation was done at the level of the gluon channel, but holds the same for the quark channel.}
so when transitioning to the modified multiplicative matching Eq.~\eqref{eq:Casy'} we shall avoid
a large rescaling that would ruin the agreement.
For this reason, we shall construct $R_2^{(n)}$ such that it never gets bigger than $A_2$ or smaller than $1/A_2$,
with $A_2>1$ a parameter that we can choose/tune at our will.
Rather than a sharp cut, we prefer to implement these limits through a smooth transition, e.g.\ with the function
\beq\label{eq:R2}
R_2^{(n)} = \exp\[ -\frac{2\log A_2}\pi \tan^{-1}\(\frac{\pi\(c_{2,g}^{(n,0)\,\rm h.s.}/c_{2,g}^{(n,0)}-1\)}{2\log A_2} \)\].
\eeq
For practical applications, we suggest to use $A_2=1.2$. Larger values are possible,
but they lead to enlarged uncertainty bands without increasing the accuracy of the approximation in a significant way.
A curve corresponding to the exact coefficient rescaled by this function, namely $c_{2,g}^{(n,0)}/R_2^{(n)}$,
is also shown in figure~\ref{fig:LLcoeff} (thin solid green). As one can see, as soon as the high-scale limit
of the coefficient departs from the exact one, this rescaled curve stays close to the exact.

For $F_L$, the situation is different. We have already commented in appendix~\ref{sec:app:coeff-func}
that at NNLO a purely multiplicative matching seems to work best, while the additive matching is inaccurate.
In this case, a value for $R_L^{(n)}$ which is close to $c_{L,g}^{(n,0)}/c_{L,g}^{(n,0)\,\rm h.s.}$ even at small scales
is advisable. However, if the denominator changes sign there is still a problem.
In order to avoid it, we suggest to construct $R_L^{(n)}$ as
\beq\label{eq:RL}
R_L^{(n)} = \frac{\left|c_{L,g}^{(n,0)}\right|}{\sqrt{\(c_{L,g}^{(n,0)\,\rm h.s.}\)^2 + A_L^2}},
\eeq
where the denominator is essentially an absolute value smoothened by the presence of the parameter $A_L\neq0$,
such that it can never be zero.
In fact, this is unnecessary at N$^3$LO, where there is no sign change, but it is important at NNLO,
where the sign changes at $Q^2\sim m^2$.
A value of $A_L$ that allows to be pretty close to the exact NNLO result is $A_L=1/(4\pi^2)$,
that we use as our central choice. At this order, since there are no NLL terms to vary and the additive matching is too
far from the exact result, we construct the uncertainty band considering another value of this parameter,
namely $A_L=2/\pi^2$.
The two curves corresponding to $c_{L,g}^{(n,0)}/R_L^{(n)}$ are shown in figure~\ref{fig:LLcoeff}, both in thin solid purple style.
The one corresponding to the smaller value of $A_L$ is the one closer to the high-scale curve, while the variation is the other curve.
In the left panel (NNLO) these curves are distinct, while in the right panel (N$^3$LO) they are indistinguishable,
and actually they are overlapping perfectly with the high-scale curve.
This shows that this construction is effectively equivalent to setting $A_L=0$ in the N$^3$LO case,
which is perfectly acceptable as we already commented.

\section{Details on the threshold approximation}
\label{app:threshold}

Following Ref.~\cite{Kawamura:2012cr},
we can write the threshold contribution, which is present in the gluon channel only, as
(we omit the scale arguments which are not essential here)\footnote
{Note that this definition of the Mellin transform differs from the one
used in the previous section; therefore, the variable $N$ is not the same,
but we use the same name as these notations are confined to each respective appendix.}
\begin{align}\label{eq:CthrRes}
  C_{a,g}^{[n_f]\,\rm thr}(N)
  &= \int_0^1 d\rho\, \rho^{N-1}  C_{a,g}^{[n_f]\,\rm thr}(z) &a=2,L\nonumber\\
  &= \as C_{a,g}^{[n_f](1)}(N)\, g_{0,a}(\as, N) \,\exp G(\as, N)
\end{align}
where
\beq
\rho \equiv \frac{4m^2}s = \frac{4m^2}{Q^2}\frac z{1-z} = 1-\beta^2.
\eeq
The functions $g_{0,a}$, which are observable dependent, and $G$,
which is not, admit the expansions
\begin{align}
  g_{0,a}(\as,N) &= g_{0,a}^h(\as)\, g_{0,a}^c(\as,N),\\
  g_{0,a}^h(\as) &= 1 +\as g_{01,a}^h +\as^2g_{02,a}^h + \Ord(\as^3),\\
  g_{0,a}^c(\as,N) &= 1 +\as g_{01,a}^c(N) +\as^2g_{02,a}^c(N) + \Ord(\as^3),\\
  G(\as,N) &= \sum_{n=1}^\infty \as^n\sum_{k=1}^{n+1} t_{nk}\log^k\tilde N,
\end{align}
with $\tilde N=Ne^{\gammae}$, with $\gammae$ the Euler-Mascheroni constant.
The function $g_{0,a}^c(\as,N)$ represents Coulomb contributions~\cite
{Hagiwara:2008df,Kiyo:2008bv,Beneke:2009rj,Beneke:2011mq,Czarnecki:2001gi,Pineda:2006ri,Beneke:1999qg},
and it depends on $N$ in the following way,
\begin{align}
  g_{01,a}^c(N) &= g_{01,a}^{c,\sqrt N} \sqrt N, \\
  g_{02,a}^c(N) &= g_{02,a}^{c,N} N + g_{02,a}^{c,\sqrt N\log}\sqrt N \log\tilde N +g_{02,a}^{c,\sqrt N} \sqrt N
                  + g_{02,a}^{c,\log} \log\tilde N +g_{02,a}^{c,\rm const},
\end{align}
where, in the case of $F_2$, the coefficients are~\cite{Kawamura:2012cr}
\begin{subequations}
\begin{align}
  g_{01,2}^{c,\sqrt N} &= \(C_F-\frac{C_A}2\)\sqrt\pi, \\
  g_{02,2}^{c,N} &=\(C_F-\frac{C_A}2\)^2\zeta_2, \\
  g_{02,2}^{c,\sqrt N\log} &= \(C_F-\frac{C_A}2\)\frac{\pi\beta_0}{\sqrt\pi}, \\
  g_{02,2}^{c,\sqrt N} &= \(C_F-\frac{C_A}2\) \frac1{\sqrt\pi} \(\frac{31}{36}C_A -\frac5{18}n_f -\pi\beta_0 \log\frac{4m^2}{\mu^2}\),\\
  g_{02,2}^{c,\log} &= \(C_F-\frac{C_A}2\) C_F,
\end{align}
\end{subequations}
 and
$g_{02,2}^{c,\rm const}= \(C_F-\frac{C_A}2\) 2C_F (\log2-1) + \(C_F-\frac{C_A}2\)^2\frac{\zeta_2}2$
is fixed such that in physical space no $\beta^0$ terms are produced (we will come back later on this).\footnote
{Note that the term proportional to $\zeta_2$ in $g_{02,2}^{c,\rm const}$ is not reproduced by Eq.~(3.13) of Ref.~\cite{Kawamura:2012cr}.
  This is due to a typo in that equation, confirmed by the authors: the term proportional to $(2N)$ is in fact proportional to $(2N+1)$.
  Eq.~(3.14) of Ref.~\cite{Kawamura:2012cr} is consistent with the corrected version of Eq.~(3.13).}
For completeness and for later use, we also report the first two orders of the function $G(\as,N)$,
\begin{align}
  G(\as, N)
  &= \frac{\as}\pi\[C_A\log^2\tilde N + C_A\(1-\log\frac{4m^2}{\mu^2}\)\log\tilde N\] \nonumber\\
  &\quad+ \frac{\as^2}{\pi^2}\bigg[\frac23C_A \pi\beta_0\log^3\tilde N
    + \(A_2+C_A \pi\beta_0\(1-\log\frac{4m^2}{\mu^2}\)\)\log^2\tilde N
    \nonumber\\
  &\qquad\qquad + \(-D_2 + 2 A_1 \pi\beta_0 \zeta_2 - (A_2+C_A \pi\beta_0) \log\frac{4m^2}{\mu^2} + 
 \frac12 C_A \pi\beta_0 \log^2\frac{4m^2}{\mu^2}\)\log\tilde N\bigg] \nonumber\\
  &\quad + \Ord(\as^3)
\end{align}
with
\begin{align}
  A_2 &= \frac{C_A}2 \(\frac{67}{18}C_A - \frac59 n_f - C_A\zeta_2\),\\
  D_2 &= \frac{C_A}4 \[\(-\frac{547}{27} + \frac{28}3\zeta_2 + 5 \zeta_3\) C_A + \(\frac{94}{27} - \frac43\zeta_2\)n_f\],
\end{align}
from which we can easily identify the various $t_{nk}$ for $n=1,2$: $t_{12}=C_A/\pi$, etc.

Now, we recall that, following Ref.~\cite{Kawamura:2012cr}, we have written the threshold approximation Eq.~\eqref{eq:Cthreshold}
in factorised form in physical space,
\beq
  C_{a,g}^{[n_f]\,\rm thr}(z)
  \equiv \as C_{a,g}^{[n_f](1)}(z)\, T_a^{\rm thr}(\beta).
\eeq
By comparison with Eq.~\eqref{eq:CthrRes} we find that
\beq
T_a^{\rm thr}(\beta) = \frac{\Mell^{-1}\[C_{a,g}^{[n_f](1)}(N)\, g_{0,a}(\as, N) \,\exp G(\as, N)\]}{C_{a,g}^{[n_f](1)}(z)},
\eeq
where $\Mell^{-1}$ denotes the inverse Mellin transform.
What appears clear here is that the function $T_a^{\rm thr}(\beta)$ depends not only on the resummation functions $g_{0,a}$ and $G$,
but also on the lowest order coefficient. This induces a further difference between the two structure functions $F_2$ and $F_L$,
because the NLO terms behave as
\begin{subequations}\label{eq:C1betaexp}
\begin{align}
  C_{2,g}^{[n_f](1)}(z) &\propto \beta^{\phantom3} \[1+\Ord(\beta^2)\], \\
  C_{L,g}^{[n_f](1)}(z) &\propto \beta^3 \[1+\Ord(\beta^2)\],
\end{align}
\end{subequations}
 where the correction to the leading small-$\beta$ term is suppressed in both cases by two powers of $\beta$.
To simplify the following discussion, we will consider an alternative definition of the function $T_a^{\rm thr}(\beta)$
based only on the leading threshold behaviour of the NLO coefficients,
\beq
\tilde T_a^{\rm thr}(\beta) \equiv \frac{\Mell^{-1}\[C_{a,g}^{[n_f](1)\,\rm thr}(N)\, g_{0,a}(\as, N) \,\exp G(\as, N)\]}{C_{a,g}^{[n_f](1)\,\rm thr}(z)},
\eeq
where $C_{a,g}^{[n_f](1)\,\rm thr}(\beta)$ is defined by neglecting the $\Ord(\beta^2)$ corrections in Eqs.~\eqref{eq:C1betaexp}.
Note that, ignoring terms that vanish in the threshold ($\beta\to0$) limit, $T_a^{\rm thr}(\beta)$ and $\tilde T_a^{\rm thr}(\beta)$
start differing at $\Ord(\as^2\beta^0)$, which corresponds to the constant term at N$^3$LO
(i.e., a term of the same kind of $g_{02,a}^h$), whose exact value is currently unkonwn.
Therefore, at the level of accuracy that we can reach at N$^3$LO, the difference between the two definitions is irrelevant.
We thus focus on $\tilde T_a^{\rm thr}(\beta)$ from now on for simplicity.

The Mellin transform of an odd power of $\beta$ is given by
\begin{align}
  \Mell[\beta^{1+2k}]
  &= \frac{\Gamma\(\frac32+k\) \Gamma(N)}{\Gamma\(\frac32+k+N\)} \nonumber\\
  &= \frac{\Gamma\(\frac32+k\)}{N^{\frac32+k}}\[1-\frac{3+8k+4k^2}8\frac1N+\Ord\(\frac1{N^2}\)\].
\end{align}
For most of the terms we are interested in, the first term of the expansion is sufficient to achieve the desired accuracy.
The only exception is the leading Coulomb term at N$^3$LO, proportional to $N$, which by interference with the $1/N$ correction
produces a constant term at the same order.
We can thus keep the two leading terms only and get the expressions
\begin{align}
  \tilde T_2^{\rm thr}(\beta)
  &= \frac1{\beta}\Gamma\(\frac32\)\, \Mell^{-1}\[\frac{\Gamma(N)}{\Gamma\(\frac32+N\)}\, g_{0,a} (\as, N) \,\exp G(\as, N)\]\nonumber\\
  &\simeq \frac{\Gamma(\frac32)\, \Mell^{-1}\[\(N^{-\frac32}-\frac38N^{-\frac52}\)\, g_{0,a} (\as, N) \,\exp G(\as, N)\]}{\beta},\\
  \tilde T_L^{\rm thr}(\beta)
  &= \frac1{\beta^3}\Gamma\(\frac52\)\, \Mell^{-1}\[\frac{\Gamma(N)}{\Gamma\(\frac52+N\)}\, g_{0,a} (\as, N) \,\exp G(\as, N)\]\nonumber\\
  &\simeq \frac{\Gamma(\frac52)\, \Mell^{-1}\[\(N^{-\frac52}-\frac{15}8N^{-\frac72}\)\, g_{0,a} (\as, N) \,\exp G(\as, N)\]}{\beta^3},
\end{align}
where $\Gamma(\frac32)=\sqrt\pi/2$ and $\Gamma(\frac52)=\frac32\Gamma(\frac32)=3\sqrt\pi/4$.
The most general form of a term in $g_{0,a}\exp G$ is
\beq
N^p\log^jN = \frac{d^j}{dp^j}N^p,
\eeq
whose inverse Mellin transform is thus easy to compute.
For the terms appearing up to N$^3$LO we get the results of table~\ref{tab:thresholdMellin}.
Apart from the obvious result in the first line, we immediately see that the leading term
is the same for $F_2$ and $F_L$ in the purely logarithmic contributions, while there is a rescaling factor
when there is a positive power of $N$: $\frac32$ for $\sqrt N$ and $3$ for $N$.
The differences in subleading terms are less trivial.

\begin{table}[t]
  \centering
  {\renewcommand{\arraystretch}{1}
  \begin{tabular}{lcc}
    term & $\tilde T_2^{\rm thr}(\beta)$ & $\tilde T_L^{\rm thr}(\beta)$ \\
    \midrule
    $1$ & $1$ & $1$\\
    \cmidrule(lr){1-3}
    $N$ & $\frac12\(\frac1{\beta^2}-1\)$ & $\frac32\(\frac1{\beta^2}-1\)$ \\
    \cmidrule(lr){1-3}
    $\sqrt N$ & $\frac{\Gamma(\frac32)}{\beta}$ & $\frac{\Gamma(\frac52)}{\beta}$ \\
    \cmidrule(lr){1-3}
    $\sqrt N \log\tilde N$ & $-\frac{\Gamma(\frac32)}{\beta}2\log\beta$ & $-\frac{\Gamma(\frac52)}{\beta}(2\log\beta-1)$ \\
    \cmidrule(lr){1-3}
    $\sqrt N \log^2\tilde N$ & $\frac{\Gamma(\frac32)}{\beta}(4\log^2\beta-\zeta_2)$ & $\frac{\Gamma(\frac52)}{\beta}(4\log^2\beta -4\log\beta-\zeta_2+2)$ \\
    \cmidrule(lr){1-3}
    $\log\tilde  N$ & $-2\log(2\beta)+2$ & $-2\log(2\beta)+\frac83$ \\
    \cmidrule(lr){1-3}
    $\log^2\tilde N$ & $4\log^2(2\beta)-8\log(2\beta)+8-3\zeta_2$ & $4\log^2(2\beta)-\frac{32}3\log(2\beta)+\frac{104}9-3\zeta_2$ \\
    \cmidrule(lr){1-3}
    $\log^3\tilde N$ & $\begin{gathered}-8\log^3(2\beta) +24\log^2(2\beta)\\
      -(48-18\zeta_2)\log(2\beta)\\ +48-18\zeta_2 -14\zeta_3\end{gathered}$
         & $\begin{gathered}-8\log^3(2\beta) +32\log^2(2\beta)\\
           -\(\tfrac{208}3-18\zeta_2\)\log(2\beta)\\
           +\tfrac{640}9-24\zeta_2-14\zeta_3\end{gathered}$ \\
    \cmidrule(lr){1-3}
    $\log^4\tilde N$ & $\begin{gathered} 16 \log^4(2\beta) - 64\log^3(2\beta)\\
      + (192 - 72\zeta_2)\log^2(2\beta)\\
      - (384 - 144\zeta_2 - 112 \zeta_3)\log(2\beta)\\
      +384 - 144\zeta_2 - 112 \zeta_3- 9\zeta_2^2 \end{gathered}$
         & $\begin{gathered} 16 \log^4(2\beta) - \frac{256}3\log^3(2\beta)\\
      + \(\tfrac{832}3 - 72\zeta_2\)\log^2(2\beta)\\
      - \(\tfrac{5120}9 - 192\zeta_2 - 112 \zeta_3\)\log(2\beta)\\
      +\tfrac{15488}{27} - 208\zeta_2 - \frac{448}3 \zeta_3- 9\zeta_2^2 \end{gathered}$
  \end{tabular}}
  \caption{Table of inverse Mellin transforms for $F_2$ and $F_L$, in the limit $\beta\to0$. Here, $\tilde N=N e^{\gammae}$.}
  \label{tab:thresholdMellin}
\end{table}

Having clarified this aspect of the way threshold terms are written, we now want to focus on the construction
of the threshold limit for $F_L$ at N$^3$LO.
The idea is to use the knowledge from $F_2$ at this order~\cite{Kawamura:2012cr} to determine the form of $\tilde T_L^{\rm thr}(\beta)$.
To better understand how the conversion works, let us consider the NNLO first, where the threshold limit
is known both for $F_2$ and $F_L$~\cite{Hekhorn:2018ywm}.
For this, we write
\begin{align}
g_{0,a} (\as, N) \,\exp G(\as, N)
  &= 1+\as\[t_{12}\log^2\tilde N + t_{11}\log\tilde N + g_{01,a}^{c,\sqrt N}\sqrt N + g_{01,a}^h\]
       + \Ord(\as^2),
\end{align}
which in physical space becomes
\begin{align}
  \tilde T_2^{\rm thr}(\beta) &= 1+\as\[4t_{12}\log^2(2\beta) - (2t_{11}+8t_{12})\log(2\beta) + g_{01,2}^{c,\sqrt N}\frac{\sqrt\pi}{2\beta} + \text{const}\] +\Ord(\as^2), \\
  \tilde T_L^{\rm thr}(\beta) &= 1+\as\[4t_{12}\log^2(2\beta) - \(2t_{11}+\frac{32}3t_{12}\)\log(2\beta) + g_{01,L}^{c,\sqrt N}\frac{3\sqrt\pi}{4\beta} + \text{const}\] +\Ord(\as^2),
\end{align}
where we did not write the constant term explicitly because that is observable dependent and we do not learn anything by looking it.
It turns out that the only $\beta$-dependent term that differs between $F_2$ and $F_L$ is the logarithmic term $\log(2\beta)$,
see Ref.~\cite{Hekhorn:2018ywm}.
This difference is completely captured by the different constant multiplying $t_{12}=C_A/\pi$,
which is in turn due to the different NLO $\beta$ dependence appearing in the definition
of $\tilde T_2^{\rm thr}(\beta)$ and $\tilde T_L^{\rm thr}(\beta)$.
In other words, this difference reflects the fact that factorization of threshold logarithms happens
in $N$ space, where the only observable dependence is governed by $g_0$,
and when trying to write a factorized form in $\beta$ space the universality of the logarithmic terms is lost
(except for the highest power at each order).

This expected behavior is contrasted by what happens to the Coulomb term.
Being proportional to $\sqrt N$, from table~\ref{tab:thresholdMellin} we know that the inverse Mellin
in the $F_2$ and $F_L$ cases differs by a factor $3/2$. However, this difference is not seen in
the $\beta$-space results of Ref.~\cite{Hekhorn:2018ywm}, where the $1/\beta$ Coulomb term
has the same coefficient for both structure functions, thus implying that the coefficients in $N$ space are not the same:
\beq
g_{01,2}^{c,\sqrt N} = \frac32 g_{01,L}^{c,\sqrt N}.
\eeq
It thus seems that the Coulomb terms naturally factorize in physical space, where they are equal for $F_2$ and $F_L$,
as opposed to threshold logarithms that factorize in Mellin space.

Even though we do not have a proof for this, we proceed assuming that the Coulomb terms
in $\tilde T_2^{\rm thr}(\beta)$ and $\tilde T_L^{\rm thr}(\beta)$ be the same,
except those coming from the interference with the hard function $g_{0,a}^h$ which are naturally observable dependent,
and infer from this assumption the values of the corresponding coefficients in Mellin space at N$^3$LO.
To this end, we focus on those terms generated at N$^3$LO that contain Coulomb coefficients, which are
\begin{align}
\left. g_{0,a}^c (\as, N) \,\exp G(\as, N)\right|_{\as^2}^c
  &= \as^2\bigg[\(t_{12}\log^2\tilde N + t_{11}\log\tilde N \) g_{01,a}^{c,\sqrt N}\sqrt N + g_{02,a}^{c,N} N\\
  &\qquad\quad+g_{02,a}^{c,\sqrt N\log}\sqrt N \log\tilde N +g_{02,a}^{c,\sqrt N} \sqrt N +g_{02,a}^{c,\log} \log\tilde N +g_{02,a}^{c,\rm const}\bigg].
    \nonumber
\end{align}
Their inverse Mellin transform reads\footnote
{The notation is a bit misleading because at this order there are Coulomb terms coming from the interference
  between $g_{0,a}^c$ and $g_{0,a}^h$, which we have omitted as they are assumed to factorize.}
\begin{align}
  \left.\tilde T_2^{\rm thr}(\beta)\right|_{\as^2}^c
  &= \as^2
    \bigg[g_{02,2}^{c,N}\frac1{2\beta^2}
    +t_{12}g_{01,2}^{c,\sqrt N}\frac{2\sqrt\pi\log^2\beta}\beta
    -\(t_{11}g_{01,2}^{c,\sqrt N}+g_{02,2}^{c,\sqrt N\log}\) \frac{\sqrt\pi\log\beta}\beta \nonumber\\
  &\qquad\quad
    +\( -\zeta_2 t_{12}g_{01,2}^{c,\sqrt N}+ g_{02,2}^{c,\sqrt N}\)\frac{\sqrt\pi}{2\beta}
    - 2 g_{02,2}^{c,\log}\log(2\beta)
    -\frac12 g_{02,2}^{c,N} + 2 g_{02,2}^{c,\log}+g_{02,2}^{c,\rm const} \bigg],
  \\
  \left.\tilde T_L^{\rm thr}(\beta) \right|_{\as^2}^c
  &= \as^2
    \bigg[g_{02,L}^{c,N}\frac3{2\beta^2}
    +t_{12}g_{01,L}^{c,\sqrt N}\frac{3\sqrt\pi\log^2\beta}\beta
    -\(\(t_{11}+2t_{12}\)g_{01,L}^{c,\sqrt N}+g_{02,L}^{c,\sqrt N\log}\) \frac{3\sqrt\pi\log\beta}{2\beta} \nonumber\\
  &\qquad\quad
    +\(\(t_{11}+(2-\zeta_2) t_{12}\)g_{01,L}^{c,\sqrt N}+ g_{02,L}^{c,\sqrt N}+g_{02,L}^{c,\sqrt N\log}\)\frac{3\sqrt\pi}{4\beta}
    - 2 g_{02,L}^{c,\log}\log(2\beta) \nonumber\\
  &\qquad\quad
    -\frac32 g_{02,L}^{c,N} + \frac83 g_{02,L}^{c,\log}+g_{02,L}^{c,\rm const} \bigg].
\end{align}
Imposing that these terms in $\beta$ space are the same for the two structure functions leads to
\begin{subequations}\label{eq:CoulombCoeffsFL}
\begin{align}
  g_{02,L}^{c,N}
  &= \frac13 g_{02,2}^{c,N} = \(C_F-\frac{C_A}2\)^2\frac{\zeta_2}3, \\
  g_{02,L}^{c,\sqrt N\log}
  &= \frac23 \(g_{02,2}^{c,\sqrt N\log} -2t_{12}g_{01,2}^{c,\sqrt N}\) \nonumber\\
  &= \(C_F-\frac{C_A}2\)\frac{2 \pi\beta_0-4C_A}{3\sqrt\pi},\\
  g_{02,L}^{c,\sqrt N}
  &= \frac23\(g_{02,2}^{c,\sqrt N} -t_{11} g_{01,2}^{c,\sqrt N}- g_{02,2}^{c,\sqrt N\log}\) \nonumber\\
  &= \(C_F-\frac{C_A}2\)  \frac2{3\sqrt\pi} \(-\frac{19}{18}C_A -\frac{n_f}9 +\frac{C_A+2n_f}{12} \log\frac{4m^2}{\mu^2}\),\\
  g_{02,L}^{c,\log}
  &= g_{02,2}^{c,\log} = \(C_F-\frac{C_A}2\) C_F,
\end{align}
\end{subequations}
 where we have also reported the values of these coefficients.
As far as the constant terms are concerned, imposing that there are no constant terms in
$\left.\tilde T_a^{\rm thr}(\beta) \right|_{\as^2}^c$, leads to
\begin{align}
  g_{02,2}^{c,\rm const}
  &= 2 g_{02,2}^{c,\log}(\log2-1) + \frac12 g_{02,2}^{c,N}\nonumber\\
  &= \(C_F-\frac{C_A}2\) 2C_F (\log2-1) + \(C_F-\frac{C_A}2\)^2\frac{\zeta_2}2,\\
  g_{02,L}^{c,\rm const} &= 2 g_{02,L}^{c,\log}\(\log2-\frac43\) + \frac32 g_{02,L}^{c,N}\nonumber\\
  &= \(C_F-\frac{C_A}2\) 2C_F \(\log2-\frac43\) + \(C_F-\frac{C_A}2\)^2\frac{3\zeta_2}2.
\end{align}
The former reproduces the result already mentioned for $F_2$, while the latter,
together with the coefficients in Eqs.~\eqref{eq:CoulombCoeffsFL}, is new.

\phantomsection
\addcontentsline{toc}{section}{References}

\bibliographystyle{jhep}
\bibliography{references}

\end{document}